\documentclass[12pt]{article}
\usepackage{amsmath,amssymb}
\usepackage{tikz}
\usepackage{graphicx}

\parskip=\medskipamount

\title{B\"acklund Transformations for the Boussinesq Equation and Merging Solitons}
\author{Alexander G. Rasin  \\
Department of Mathematics,\\ 
Ariel University, Ariel 40700, Israel \\
{E-mail: rasin@ariel.ac.il}\and  Jeremy Schiff \\
Department of Mathematics,\\
Bar-Ilan University, Ramat Gan, 52900, Israel \\
{E-mail: schiff@math.biu.ac.il}}

\begin{document}
\maketitle
\begin{abstract}{
    The B\"acklund transformation (BT) for the ``good'' Boussinesq equation and its
    superposition principles are presented and applied. Unlike many other standard integrable
    equations, the Boussinesq equation does not have a strictly algebraic superposition principle
    for 2 BTs, but it does for 3. We present associated lattice systems.
    Applying the BT to the trivial solution generates standard solitons but also what we
    call ``merging solitons'' --- solutions in which two solitary waves (with related speeds)
    merge into a single one. We use the superposition principles to generate a variety of interesting solutions,
    including superpositions of a merging soliton with $1$ or $2$ regular solitons, and solutions that
    develop a singularity in finite time which then disappears at some later finite time. We prove a
    Wronskian formula for the solutions obtained by applying a general sequence of BTs on the trivial solution. 
    Finally, we show how to obtain the standard conserved quantities of the Boussinesq equation from the BT, and 
    how the hierarchy of local symmetries follows in a simple manner from the superposition principle for 3 BTs.}
\end{abstract}

\section{Introduction} 

In this paper we explore the B\"acklund transformation (BT) of the Boussinesq equation (BEq)
\begin{equation}
  U_{tt} - 4\beta U_{xx} + {\textstyle{\frac13}} U_{xxxx} - 2(U^2)_{xx} = 0   \label{be} 
\end{equation}   
where $\beta$ is a positive constant. The BEq is one of the oldest of the classical 
integrable nonlinear partial differential equations (PDE) \cite{b1,b2}, and its BT 
was given in bilinear form by Hirota and Satsuma \cite{beq17} and in standard form by Chen \cite{beq45}. 
who also gave a superposition principle 
(see also \cite{beq21, beq18, beq16}). However, certain aspects seem not to have been discussed. There is a
second superposition principle, and using this it is possible to give a superposition principle for
3 BTs that is algebraic (as opposed to the superposition principle of \cite{beq45} that involves derivatives). 
In addition, there does not seem to be a systematic study of solutions generated by the BT, and this involves
several surprises, as we shall shortly explain. 

Our original motivation  for looking at the BT of the BEq was connected with lattice versions of the equation.
In recent years there has been substantial interest in integrable lattice equations, and one of the origins of these is as
superposition principles of BTs of integrable PDE (for example, the Q4 equation in the ABS classification \cite{beq48}
was originally discovered by Adler as the superposition principle for the Krichever-Novikov equation \cite{beq47}).
Discrete versions of the BEq have been given by Nijhoff {\em et al.} \cite{beq53}  as a scalar equation on a large
stencil (see also \cite{beq54}), and by Tongas and Nijhoff \cite{beq28} as a system of equations for 3 fields on a
rectangular plaquette. These, along with related ``modified'' and ``Schwarzian''  systems,  have attracted much
attention recently \cite{beq56,beq14,beq25,beq26,beq27,beq24,beq31,beq57,beq55,beq22}. Having fully understood the
superposition principle for the continuum BEq, we present two associated lattice systems.
One is a system of 2 equations for 2 fields on a rectangular plaquette
(like the discrete modified and Schwarzian BEqs, as introduced in \cite{beq54,beq56}),  
the other is a system of 2 equations for a single field on a cube.

However, it seems there is much to be learnt from simply applying the BT.  Note that in (\ref{be})
we have written the ``good'' version of the BEq, in which the signs of the $U_{tt}$ and $U_{xxxx}$
terms are the same. For the ``bad'' version, in which the signs are opposite, the $N$-soliton solutions
of the BEq equation were given by Hirota \cite{beq34}, using his eponymous method. For the good BEq there is
a subtlety in applying the Hirota method, and there are a variety of interesting, non-standard, soliton-type solutions as
discovered by Manoranjan {\em et.al.} \cite{beq58,beq59} and 
Bogdanov and Zakharov \cite{beq11}, citing unpublished work of Orlov. (Similar phenomena were
observed by Hietarinta and Zhang \cite{beq27} in their study of solitons in a modified discrete BEq.)
We show that applying the BT to the trivial solution can generate standard solitons, but also what we 
call ``merging solitons'' --- solutions in which two solitary waves (with related speeds)
merge into a single one. The superposition principle enables us to superpose a merging soliton with a
standard 1-soliton or a standard 2-soliton. We have not succeeded so far to obtain a nonsingular solution 
involving the superposition of 2 or more merging solitons, but from the superposition of 3 merging solitons
we find a solution which initially describes 6 solitary waves, becomes singular in finite time, but then becomes regular
again, leaving 3 solitary waves. The possibility of finite time singularities forming in the BEq is well-known,
originating, we believe,  in \cite{Kalantarov}. 

We also use the BT to prove a Wronskian formula for
the general soliton solution of BEq, a generalization of the formula given in \cite{beq20} for the bad BEq. 

Finally, we show how to use the BT of the BEq to generate its conservation laws and symmetries.
The idea of using a BT to generate
conservation laws of an integrable PDE is very old, see for example \cite{beq44}. In \cite{rs1} we showed how the
superposition principles of BTs of a number of integrable PDEs can be used to generate their symmetries. This
works for the BEq, but it is necessary to use the superposition principle for 3 BTs. This is a
consequence of the fact that the BEq is associated with the Lie group $SL(3)$, while 
equations such as Korteweg-de Vries, Sine-Gordon and Camassa-Holm are associated with $SL(2)$. We show how
to use the 3 BT superposition principle to obtain the local symmetries of the BEq, and also obtain the recursion operator 
and some nonlocal symmetries. 

This paper is structured as follows: In Section 2 we give the BT of the BEq and its superposition principles.
In Section 3 we discuss associated lattice equations. In Section 4 we describe solutions of the BEq generated
by the BT. In Section 5 we use the BT to generate the symmetries and conservation of the BEq. In Section 6
we conculde and indicate areas for further study. 

\section{\bf B\"acklund Transformation and Superposition} 
 
We work with the potential BEq in the form 
\begin{eqnarray}
f_t &=& \left( f_{x} - f^2 - 2h \right)_x \ ,  \label{feq}\\
h_t &=& \left( {\textstyle{\frac23}} f_{xx} - h_{x} + {\textstyle{\frac23}}f^3 - 2ff_{x} \right)_x  + 2fh_x \ .
\label{heq}
\end{eqnarray}
This arises from the consistency of the Lax pair $Y_x = A Y,\  Y_t=B Y$  where 
\begin{eqnarray*}
A  &=& \left( \begin{array}{ccc}
  f & 1 & 0  \\
  f_x - f^2 - h & 0 & 1 \\
  \lambda + f_{xx} + f^3 - 3ff_x  - 2h_x + 2fh  &   h &  -f  
     \end{array} \right) \ , \\ 
B  &=& \left( \begin{array}{ccc}
  -h & f & 1  \\
  \lambda+fh - h_x & f_x-f^2 & -f  \\
  B_{31}  & \lambda + f_{xx} +f^3 -3ff_x - h_x + fh &  h+f^2-f_x 
\end{array} \right) \ , \\
B_{31} &=& -{\textstyle{\frac13}} f_{xxx} + f_x^2 + ff_{xx} - f^2f_x - h^2 + h(f_x-f^2) \ . 
\end{eqnarray*}
The reason to take the equation in this apparently complicated form is that it simplifies the action of
the BT, as we shall see shortly.  It is easy to check that (\ref{feq})-(\ref{heq}) imply  
\begin{equation} f_{tt} = -{\textstyle{\frac13}}  f_{xxxx} + 4f_x f_{xx}  \label{pb}\end{equation} 
which is the standard form of the potential BEq. If we define  $w = h + {\textstyle{\frac12}} f^2 -f_x$ then
the above system simplifies to
\begin{eqnarray*}
  f_t &=&  \left(-2 w - f_x \right)_x \ , \\
  w_t &=&  w_{xx} + {\textstyle{\frac23}} f_{xxx} - f_x^2  \ .
\end{eqnarray*}
So if $u=f_x$, $v=w_x$ then 
\begin{eqnarray}
  u_t &=&  \left(-2v-u_x \right)_x \ ,  \label{ueq}\\
  v_t &=&  \left(v_x + {\textstyle{\frac23}} u_{xx} -u^2 \right)_x  \ . \label{veq}
\end{eqnarray}
This is the two component form of the BEq that we use (it is maybe more standard to replace the field
$v$ by $\tilde{v}=-2v-u_x$ to simplify the first equation, but we find our form marginally more convenient). 
By eliminating $v$ we obtain the scalar form of the BEq: 
\begin{equation}
u_{tt} = -{\textstyle{\frac13}} u_{xxxx} + (2u^2)_{xx}  \ .
\label{usceq}\end{equation}
Finally, if we write $u=U+\beta$ we recover the familiar form (\ref{be}). However, we will work  with
the form (\ref{usceq}), and just remember, when looking at explicit solutions, that we are interested
in solutions with $u\rightarrow \beta$ at spatial infinity. We will focus on
the case $\beta>0$, in which case (\ref{be}) is a linearly stable perturbed wave equation. 

It is straightforward to verify that 
the potential BEq in the $f,h$ form (\ref{feq})-(\ref{heq}) has a BT 
\begin{equation}
  f \rightarrow f_{\rm new} = f - s \ , \qquad  h \rightarrow h_{\rm new} = h - f_x + f s
\label{BT}  \end{equation}
where $s$ satisfies the equations 
\begin{eqnarray}
  s_{xx} &=&  \theta  -  3ss_x  - s^3 + 3f_x s  + 3f_{xx} - 3ff_x - 3h_x\ ,  \label{sxx}\\
  s_{t}  &=&  \theta  -   ss_x  - s^3 + 3f_x s  + f_{xx} - 3ff_x - 3h_x\ ,   \label{st}
\end{eqnarray}
or, equivalently,
\begin{eqnarray}
  s_{xx} &=&  \theta  -  3ss_x  - s^3 + 3u s  -3v\ ,  \label{sxx1}\\
  s_{t}  &=&  \theta  -   ss_x  - s^3 + 3u s  - 2u_x - 3v\ .   \label{st1} 
\end{eqnarray}
This is a BT in the sense that if $f,h$ satisfy the potential BEq system (\ref{feq})-(\ref{heq}), then so do
$f_{\rm new},h_{\rm new}$ given by (\ref{BT}). Furthermore, the equations for $s$, (\ref{sxx1})-(\ref{st1}), 
are consistent if and only $u,v$ satsify the BEq system (\ref{ueq})-(\ref{veq}). 

Denote by $f_1,h_1$ ($f_2,h_2$) the solution obtained from $f,h$ using a BT with parameter $\theta_1$ ($\theta_2$),
and by $f_{12},h_{12}$ ($f_{21},h_{21}$) the solution obtained from $f_1,h_1$ ($f_2,h_2$)  using a BT with parameter $\theta_2$
($\theta_1$). Assuming commutativity of BTs gives
$$  f_{12} = f_{21}\ , \qquad h_{12} = h_{21}\ . $$
Eliminating $s$ from 4 copies of (\ref{BT}) we have
\begin{eqnarray*}
h_1  &=& h - f_x + f(f-f_1) \ ,   \\
h_2  &=& h - f_x + f(f-f_2) \ ,   \\
h_{12} &=& h_1 - f_{1x} + f_1(f_1-f_{12}) \ , \\ 
h_{21} &=& h_2 - f_{2x} + f_2(f_2-f_{21}) \ . 
\end{eqnarray*}
Using commutativity and
taking the obvious linear combination of these equations to eliminate all $h$ fields, we arrive at the 
superposition principle, as given by Chen \cite{beq45}
\begin{equation} 
  f_{2x}-f_{1x} + (f + f_{12} - f_1 - f_2)(f_2-f_1)  = 0\ , \label{pl2} 
\end{equation}
which can be solved for $f_{12}$: 
\begin{equation} 
f_{12}  =  f_1 + f_2 - f - \frac{f_{2x}-f_{1x}}{f_2-f_1}\ .   \label{pl3} 
\end{equation}
In place of giving a proof for commutativity, it is possible to directly verify that the new solution
given by (\ref{pl3}) is a solution of the potential BEq (\ref{pb}). 

But in fact there is also a second superposition formula.
Assuming the commutativity of 2 BTs, we have four versions of equation (\ref{sxx}): 
\begin{eqnarray*}
  (f-f_1)_{xx} &=&  \theta_1  -  3(f-f_1)(f-f_1)_x  - (f-f_1)^3 + 3f_x (f-f_1)  + 3f_{xx} - 3ff_x - 3h_x\ , \\
  (f-f_2)_{xx} &=&  \theta_2  -  3(f-f_2)(f-f_2)_x  - (f-f_2)^3 + 3f_x (f-f_2)  + 3f_{xx} - 3ff_x - 3h_x \ ,\\
  (f_1-f_{12})_{xx} &=&  \theta_2  -  3(f_1-f_{12})(f_1-f_{12})_x  - (f_1-f_{12})^3 + 3f_{1x} (f_1-f_{12})  + 3f_{1xx} - 3f_1f_{1x} - 3h_{1x}\ ,  \\
  (f_2-f_{12})_{xx} &=&  \theta_1  -  3(f_2-f_{12})(f_2-f_{12})_x  - (f_2-f_{12})^3 + 3f_{2x} (f_2-f_{12})  + 3f_{2xx} - 3f_2f_{2x} - 3h_{2x}\ ,  
\end{eqnarray*}
where $h_1 = h-f_x+f(f-f_1)$, $h_2 = h-f_x+f(f-f_2)$. Taking a suitable linear combination of these equations eliminates
the second derivatives of $f,f_1,f_2,f_{12}$ and the function $h$, giving the result 
\begin{eqnarray*}
0 &=& \left(f_1-f_2 \right)
\left( f_{12x}+f_x -2f^2 -f_1^2 -f_2^2 -f_{12}^2  +2f(f_1+f_2)  
-f_1f_2+ f_{12}(f_1+f_2)  \right) \nonumber \\ 
&& + \theta_2-\theta_1 + (f_1-f)f_{1x} - (f_2-f)f_{2x}\ .  
\end{eqnarray*}
Finally, adding $f$ times equation (\ref{pl2})  gives 
\begin{eqnarray}
0 &=& \left(f_1-f_2 \right)
\left( f_x+f_{12x}  - f^2 - f_1^2 - f_2^2  - f_{12}^2  +  (f+f_{12})(f_1+f_2) + ff_{12} -  f_1f_2  \right)
\nonumber \\ 
&& +  \theta_2-\theta_1  + f_1f_{1x} - f_2f_{2x} 
  \ .  \label{pl1}
\end{eqnarray}
This can be solved for $f_{12x}$:
\begin{equation} 
f_{12x}  =   f^2 + f_1^2 +f_2^2  + f_{12}^2  - (f+f_{12})(f_1+f_2) - ff_{12} +  f_1f_2
  +  \frac{ \theta_1-\theta_2  + f_2f_{2x} - f_1f_{1x} }{f_1-f_2}   - f_x  \ .
    \label{pl4}  
\end{equation}
Equation (\ref{pl4}) does not follow directly from (\ref{pl3}). Differentiating the right
hand side of (\ref{pl3}) will include second derivatives of $f_1,f_2$. However note that if these are eliminated 
by using the first two of the 4 versions of (\ref{sxx}) above, then (\ref{pl4}) can be proved directly,
without needing to assume commutativity. 

\section{Lattice Equations}

Writing $u=f_x$ in (\ref{pl2}) and (\ref{pl1}) we obtain the pair of quad-graph  equations 
\begin{eqnarray}
0  &=&  u_2-u_1 + (f + f_{12} - f_1 - f_2)(f_2-f_1)\ ,  \label{fu1} \\
0 &=& \left(f_1-f_2 \right)\left( u+u_{12}  - f^2 - f_1^2 - f_2^2  - f_{12}^2  +  (f+f_{12})(f_1+f_2) + ff_{12} -  f_1f_2  \right)
\nonumber \\ 
&& + \theta_2-\theta_1  + f_1u_{1} - f_2u_{2} \ . \label{fu2}
\end{eqnarray}
(Here we are thinking of $f,f_1,f_2,f_{12}$ as $4$ values of the field $f$ around the vertices of a rectangle. Other
notations common in the literature are $f,\tilde{f},\hat{f},\hat{\tilde{f}}$ and
$f_{n,m},f_{n+1,m},f_{n,m+1},f_{n+1,m+1}$.) These equations are somewhat simplfied by introducing the field  $g=u-f^2$: 
\begin{eqnarray}
0 &=&  g_2-g_1 + (f + f_{12})(f_2-f_1) \ .  \label{fg1} \\
0 &=& \left(f_1-f_2 \right)\left( g+g_{12}  +  (f+f_{12})(f_1+f_2) + ff_{12}  \right) + \theta_2-\theta_1  + f_1g_1 - f_2g_2 \ .
  \label{fg2}
\end{eqnarray}
It is straightforward to check that these equations 
have the consistency around the cube (CAC) property \cite{beq48}, and also arise as the 
consistency conditions for the following Lax pair: 
\begin{eqnarray*}
  Y_1 &=& \left( \begin{array}{ccc}
    f_1 &  -1 & 0  \\
    -(ff_1+g_1) & f & 1  \\
    \theta_1 + f^2f_1 +fg_1 - f_1g  & g+g_1-f^2 & -(f+f_1)  
    \end{array}\right)  Y \ , \\ 
  Y_2 &=& \left( \begin{array}{ccc}
    f_2 &  -1 & 0  \\
    -(ff_2+g_2) & f & 1  \\
    \theta_2 + f^2f_2 +fg_2 - f_2g  & g+g_2-f^2 & -(f+f_2)  
    \end{array}\right)  Y  \ .
\end{eqnarray*}

The lattice potential Boussinesq system (\ref{fg1})-(\ref{fg2}) should be compared with
the lattice Boussinesq system of Tongas and Nijhoff \cite{beq28}. 
Their system involves 3 fields $u,v,w$, satisfying $5$ equations on an elementary plaquette, $4$ of which
being the ``same'' equation on the $4$ sides of the plaquette. Using $u,v,w$ for the fields, as in \cite{beq28}
(and not as in the rest of this paper), the equations are 
\begin{eqnarray*}
  w_1 &=& u u_1 - v \ ,\\
  w_2 &=& u u_2 - v \ ,\\
  w_{12} &=& u_2 u_{12} - v_2\ , \\
  w_{12} &=& u_1 u_{12} - v_1 \ ,\\
  w   &=& u u_{12} - v_{12}  + \frac{\theta_2-\theta_1}{u_2-u_1}\ .
\end{eqnarray*}
We would argue that since the Tongas-Nijhoff system involves $5$ relations between $12$ quantities
on an elementary plaquette ($3$ fields at each of $4$ vertices), whereas our system involves $2$
relations between $8$ quantites, there is a fundamental difference. However, we suspect there may be
relations between solutions of the two systems. Likewise, we suspect there are relations with the
lattice modified Boussinesq system introduced in \cite{beq54} and the lattice Schwarzian Boussinesq system 
that appears in \cite{beq56}, both of which are systems of 2 equations for 2 fields on an elementary
plaquette. 

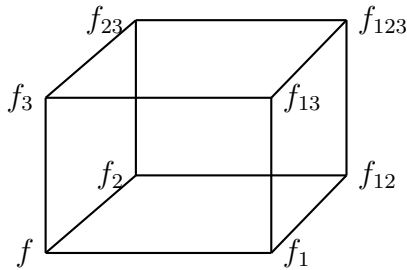
\begin{figure}
\centerline{
    \begin{tikzpicture}[side/.style={thick}]
      \draw[side] (1cm,-10em) node[left]{$f$} -- (4cm,-10em)  node[right]{$f_1$};
      \draw[side] (1cm,-10em) -- (1cm,-5em) ;
      \draw[side] (1cm,-10em) -- (2.2cm,-7.5em) ;      
      \draw[side] (4cm,-10em) -- (4cm,-5em) ;
      \draw[side] (4cm,-10em) -- (5cm,-7.5em) ;
      \draw[side] (5cm,-7.5em) -- (5cm,-2.5em) ;            
      \draw[side] (1cm,-5em) -- (4cm,-5em) node[right]{$f_{13}$};
      \draw[side] (1cm,-5em) node[left]{$f_3$} -- (2.2cm,-2.5em) ;
      \draw[side] (2.2cm,-7.5em) node[left]{$f_{2}$} -- (5cm,-7.5em) node[right]{$f_{12}$};
      \draw[side] (2.2cm,-7.5em) -- (2.2cm,-2.5em)   ;
      \draw[side] (4cm,-5em) -- (5cm,-2.5em)   ;    
      \draw[side] (2.2cm,-2.5em) node[left]{$f_{23}$} -- (5cm,-2.5em)  node[right]{$f_{123}$}; 
    \end{tikzpicture}
}
\caption{8 solutions around a cube}
\end{figure}

In checking the CAC property for (\ref{fg1})-(\ref{fg2}) it emerges that it is possible to eliminate
the field $g$ when considering the equations on a cube. So we introduce a third BT, with parameter
$\theta_3$, and denote by $f_3$ the solution obtained from $f$ via this BT, and consider the
set of $8$ solutions $f,f_1,f_2,f_3,f_{12},f_{13},f_{23},f_{123}$ associated with vertices of a cube, as
indicated in Figure 1. These solutions satisfy the equations 
\begin{equation} 
    f_{12} (f_2-f_1)  
  + f_{23} (f_3-f_2)  
  + f_{13} (f_1-f_3)  = 0   \label{latt}
\end{equation}
and
\begin{equation}
f_{123} = f + \frac{(\theta_3-\theta_2)f_1+(\theta_1-\theta_3)f_2+(\theta_2-\theta_1)f_3}
{  (f_2-f_3)f_1f_{23}   +(f_3-f_1)f_2f_{13}   +(f_1-f_2)f_3f_{12}}\ . 
\label{latt2}
\end{equation}
Once again, we have $2$ relations between $8$ quantities. 
The first of these equations has a superficial similarity to the Hirota DAGTE equation \cite{beq29},
which was used as a starting point to find discrete Boussinesq systems in \cite{beq24}. We wonder in what sense 
the system (\ref{latt})-(\ref{latt2}) is integrable.

\section{Solitons of the Boussinesq Equation} 

As explained in section 2, we wish to look at solutions of the BEq with (\ref{usceq}) with $u\rightarrow\beta$
at spatial infinity, with $\beta>0$. Equivalently, we want solutions of the potential BEq (\ref{pb}) for
which $f\sim\beta x + \gamma_+$ as $x\rightarrow\infty$
and $f\sim\beta x + \gamma_-$ as $x\rightarrow-\infty$, where $\gamma_{\pm}$ are constants. We
obtain such solutions by applying the BT to the starting solution $f = \beta x$. 
Applying the BT once gives new solutions
\begin{equation}
  f  =  \beta x - \frac{y_x}{y} \quad {\rm where} \quad 
y  =  C_1 e^{\lambda_1 x + \lambda_1^2 t} + C_2 e^{\lambda_2 x + \lambda_2^2 t} + C_3 e^{\lambda_3 x + \lambda_3^2 t}\ .
\label{yeq}
\end{equation} 
Here
$\lambda_1,\lambda_2,\lambda_3$ are the three roots of the cubic equation $\lambda^3 = 3 \beta \lambda + \theta$,
and
$C_1,C_2,C_3$ are constants, not all zero, which can be jointly rescaled without changing the solution. There are two
main situations to look at: the case $\theta^2<4\beta^3$ when $\lambda_1,\lambda_2,\lambda_3$ are all real and distinct,
and  the case  $\theta^2>4\beta^3$ when one is real and the other two are a complex conjugate pair. (Note
that the first situation can only happen if $\beta>0$.) As we wish to focus on soliton-type solutions, we look
only at the first case, when there are $3$ real, distinct roots. 


The case when two of the constants $C_1,C_2,C_3$ are zero is trivial. If one is zero, say $C_3$, then the new
solution is  
$$
f =  \beta x + \frac{c}{2}-  p {\rm tanh}\left( p(x-ct) + \alpha \right) 
$$  
or 
$$
  f =  \beta x + \frac{c}{2}  -  p {\rm coth}\left( p(x-ct) + \alpha \right)  
$$
where $c = -(\lambda_1+\lambda_2)=\lambda_3$, $p=\frac12(\lambda_1-\lambda_2)$ and $\alpha$ is an
arbitrary constant. The corresponding solutions of the B. equation are 
$$
u =  \beta -  p^2 {\rm sech}^2\left( p(x-ct) + \alpha \right) \ ,\qquad 
u =  \beta +  p^2 {\rm csch}^2\left( p(x-ct) + \alpha \right) \ . 
$$  
These are the standard soliton and singular soliton solutions. 
A direct calculation confirms that these are solutions provided
\begin{equation}   \frac{c^2}{4} + \frac{p^2}{3} = \beta\ .   \label{speedamp} \end{equation} 
From this we again deduce the need for $\beta$ to be positive, and obtain bounds on both the velocity $c$ and the
amplitude parameter $p$ for fixed $\beta$.  Note that there are solutions with both positive and negative
velocity, but that the solutions do not depend on the sign of $p$. Note also that for solitons $u<\beta$ and for singular
solitons $u>\beta$. 

Proceeding to the case where all three constants $C_1,C_2,C_3$ are nonzero, we need to distinguish between the
case that all have the same sign, in which case the solution will be nonsingular, and the case that there are
differing signs, in which case there is singularity. We start with the former. Looking at the expression for $y$ in 
(\ref{yeq}), at a given time $t$ and position $x$ we will ``see'' a soliton if two of the terms balance and
are much bigger than the third term.  So for example, we will see a soliton determined by the first two terms
at position 
$$
x \approx  \frac{1}{\lambda_1-\lambda_2} \log\left( \frac{C_2}{C_1} \right)  - (\lambda_1+\lambda_2) t   
$$
(this is obtained from balance between the first two terms), provided
$$
t(\lambda_1 - \lambda_3)(\lambda_2 - \lambda_3) \ll  K
$$
where $K$ is some constant (this being the condition that at the given $x$, the first two terms are much
bigger than the third one).  Clearly a critical role is played by the sign of the product
$(\lambda_1 - \lambda_3)(\lambda_2 - \lambda_3)$. If this is positive we ``see'' a soliton determined
by the first two terms for large negative times, if it is negative we will see the soliton for large positive times.
It is straightforward to check that this product is  positive for two of the three possible pairs of
terms in $y$ and negative for the other pair. Thus the solutions describe the merger of two
solitary waves  into a single one. Furthermore, if we choose, without loss of generality,
$\lambda_1<\lambda_2<\lambda_3$, then the incoming solitary waves are those of velocity $\lambda_1$ and $\lambda_3$,
one of which is negative and one positive, and the outgoing one has velocity $\lambda_2$.
Since $\lambda_1+\lambda_2+\lambda_3=0$ there is a ``law of conservation of speed''. See Figure 2 (compare
with Figures 3.1 and 5.1 in \cite{beq59} and 
Figure 11 in \cite{beq11}). In this plot, as in all subsequent plots in this section, $u$ is plotted as a function of $x$. 
We call this solution a ``merging soliton''. 

\begin{figure}
      \centerline{
        \includegraphics[width=3.6cm]{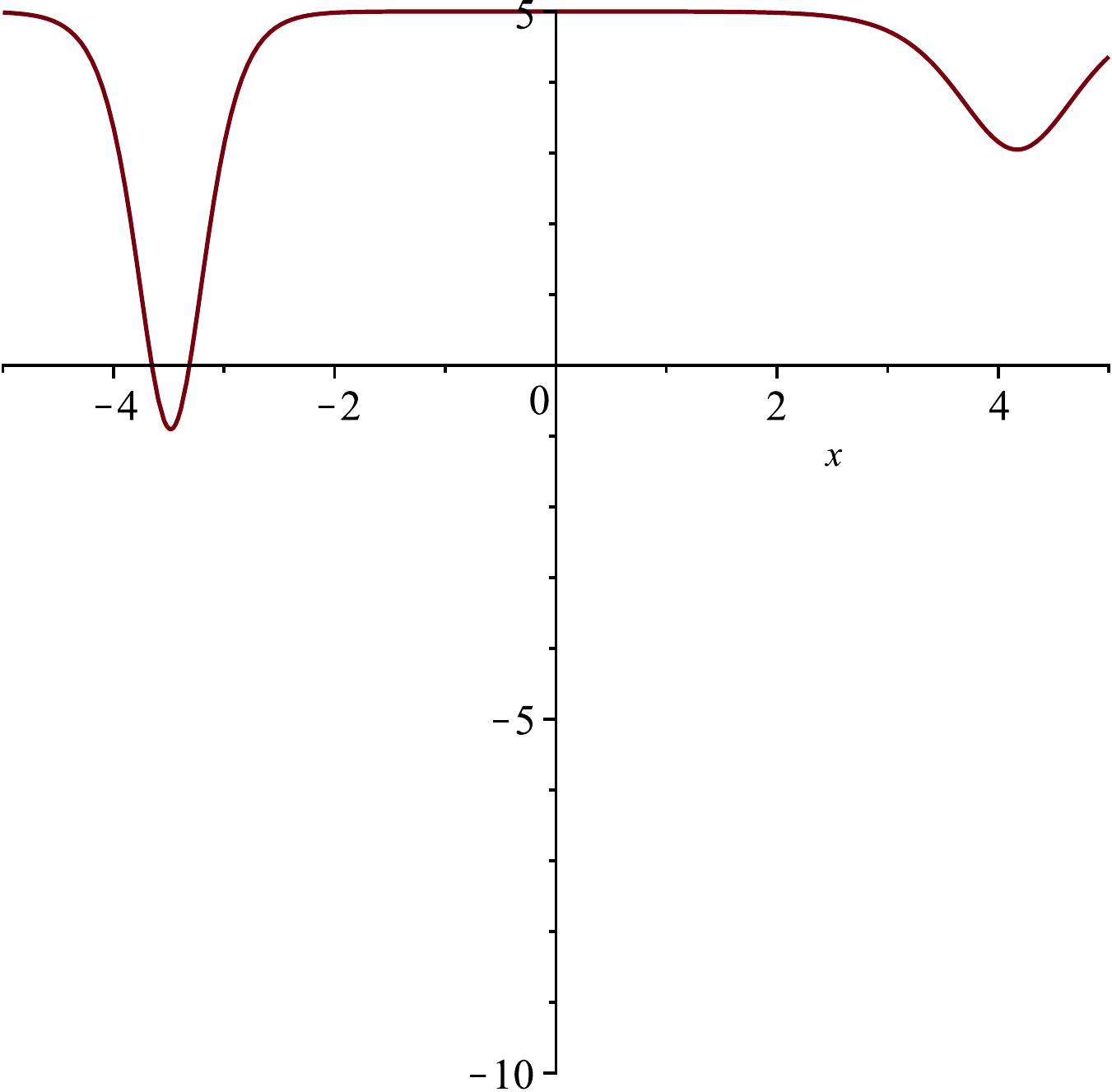} ~~~
        \includegraphics[width=3.6cm]{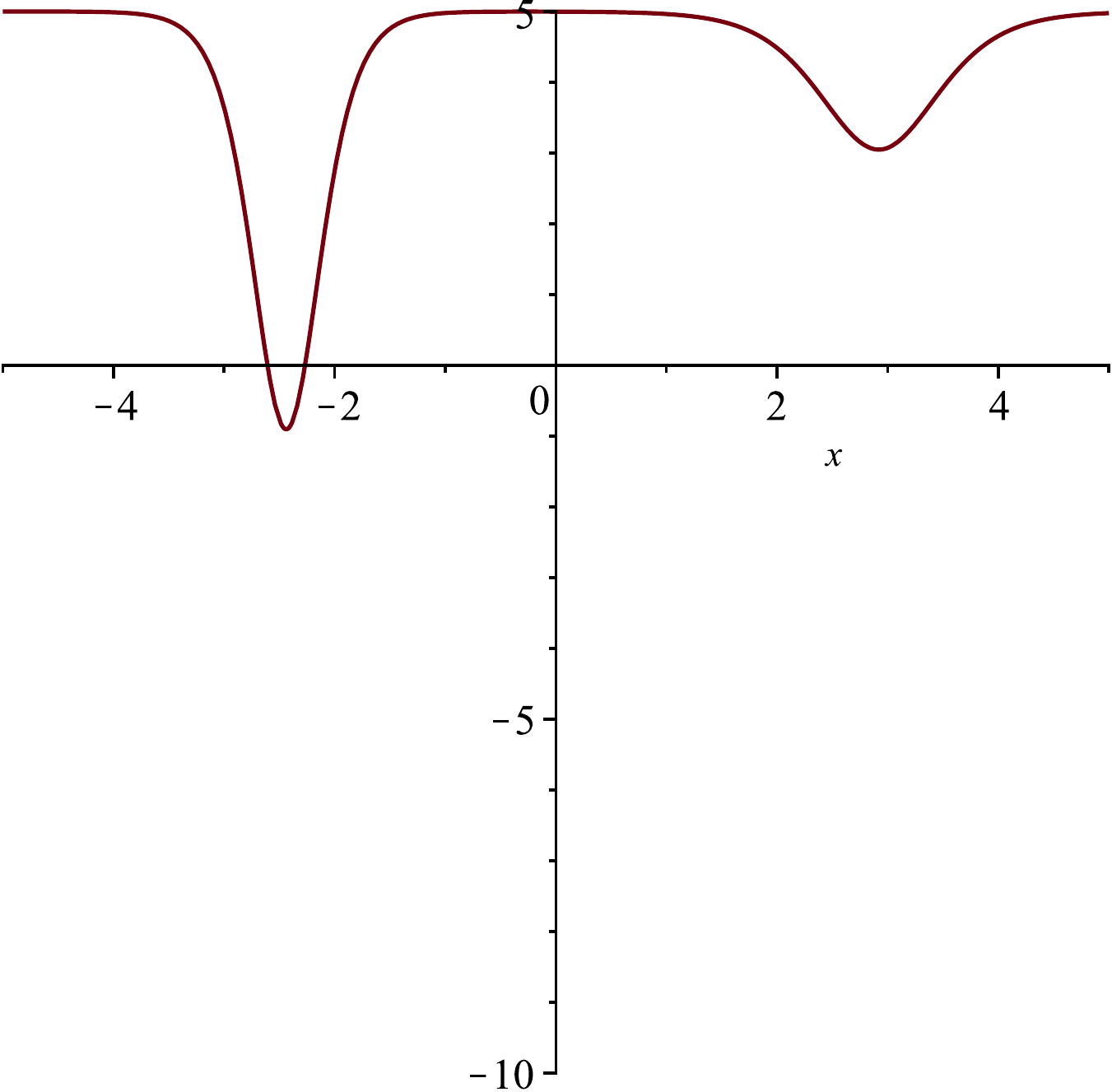} ~~~
        \includegraphics[width=3.6cm]{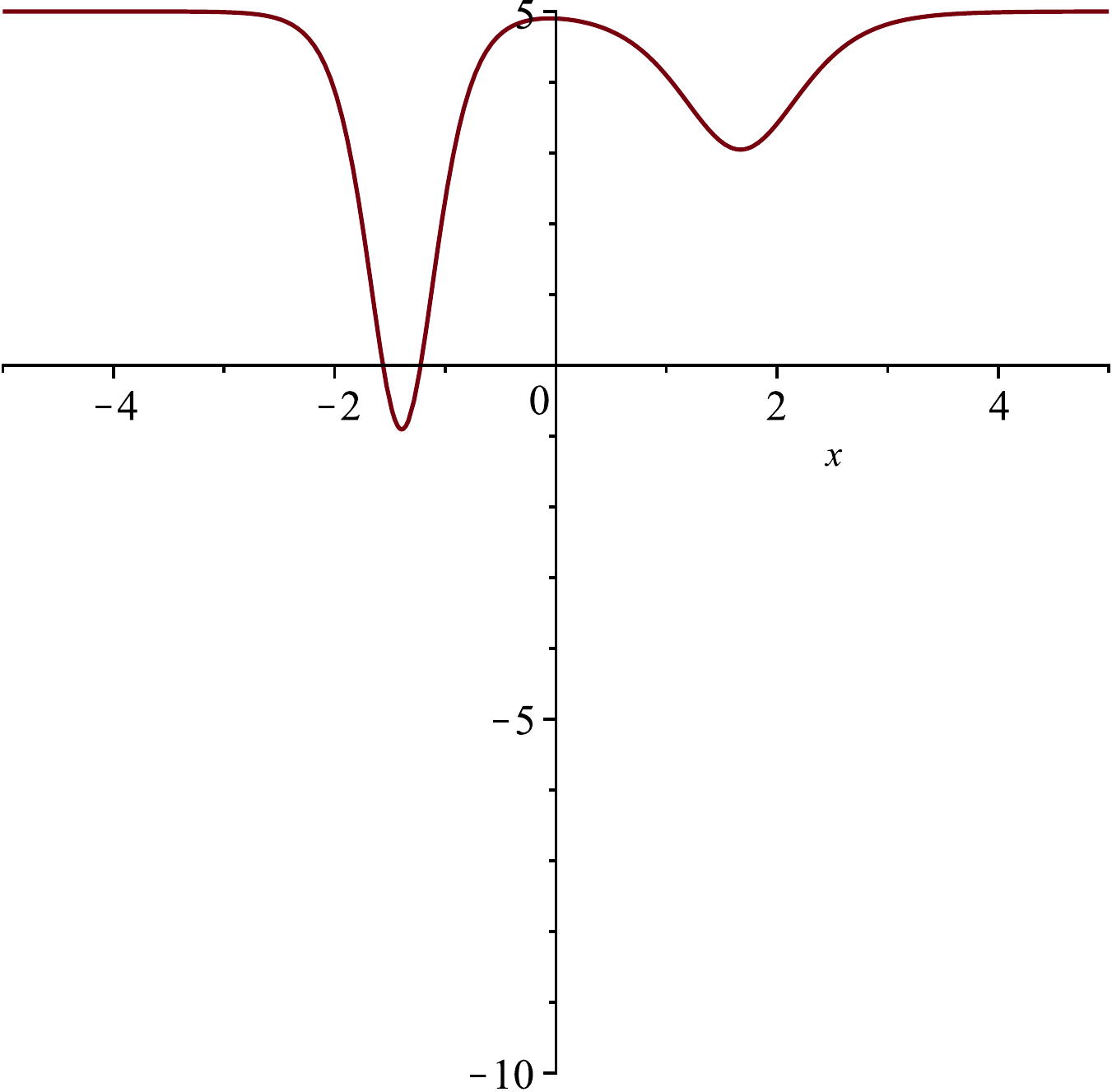} ~~~
        \includegraphics[width=3.6cm]{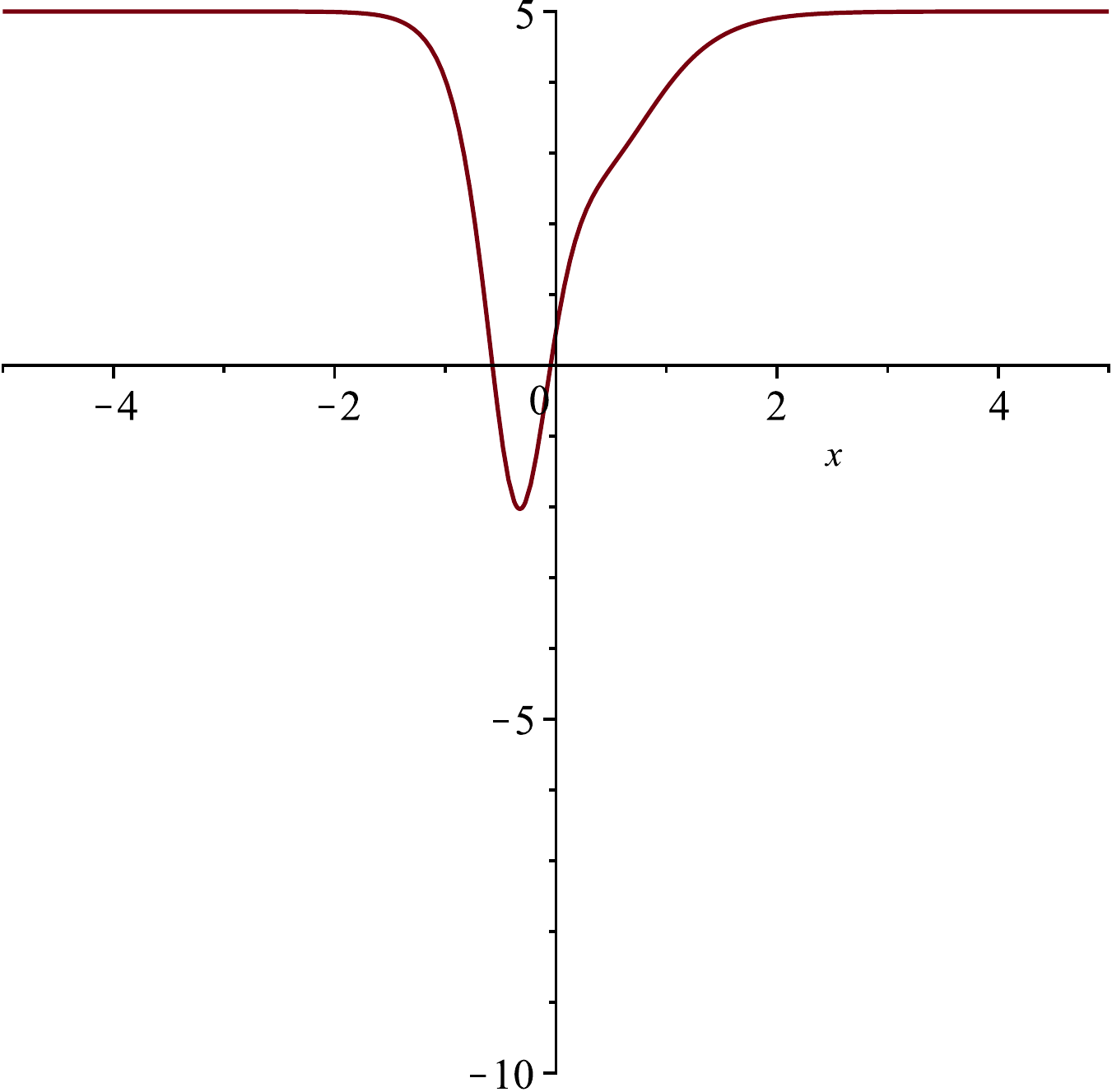} 
        }
      \centerline{
        \includegraphics[width=3.6cm]{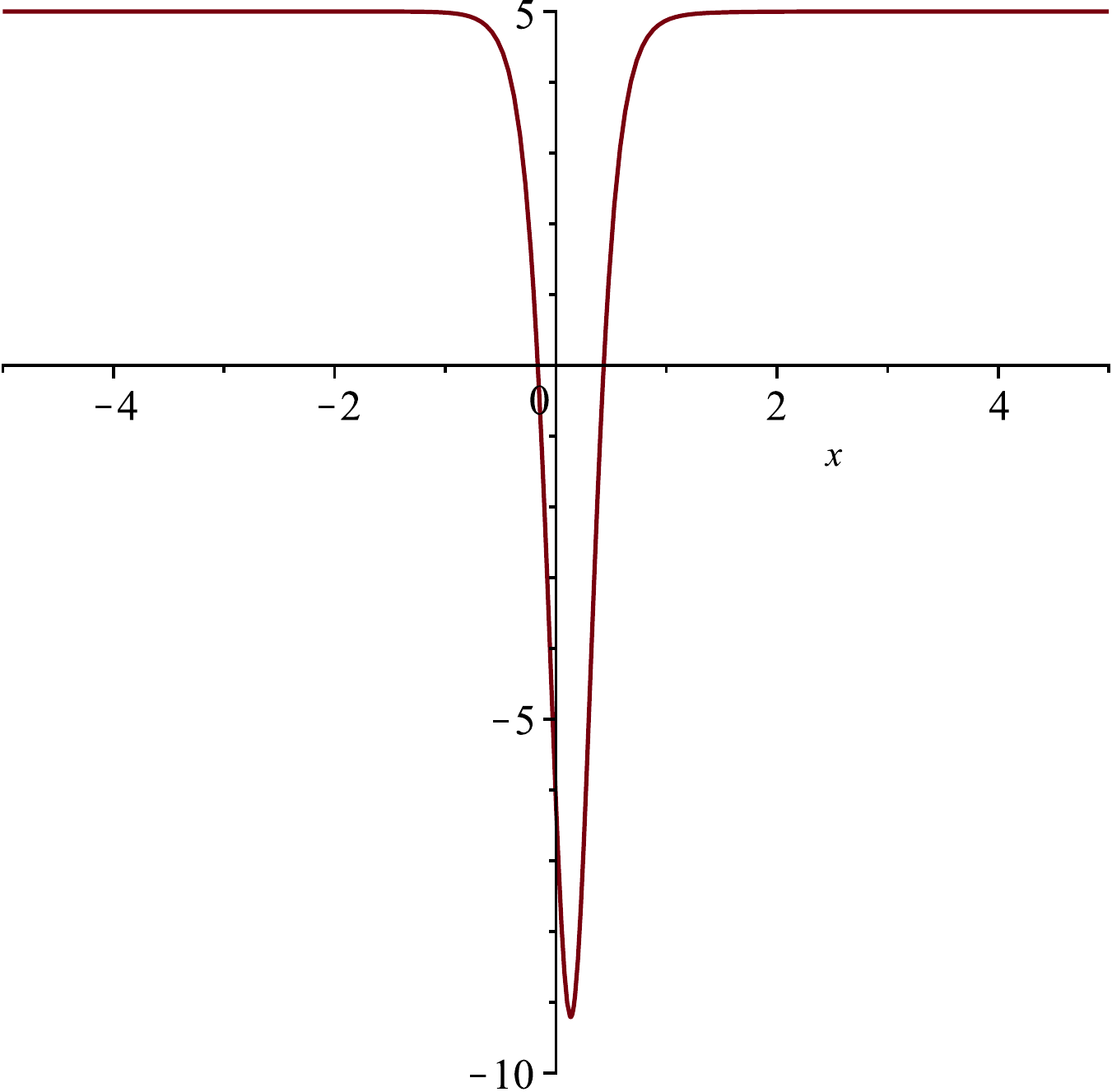} ~~~
        \includegraphics[width=3.6cm]{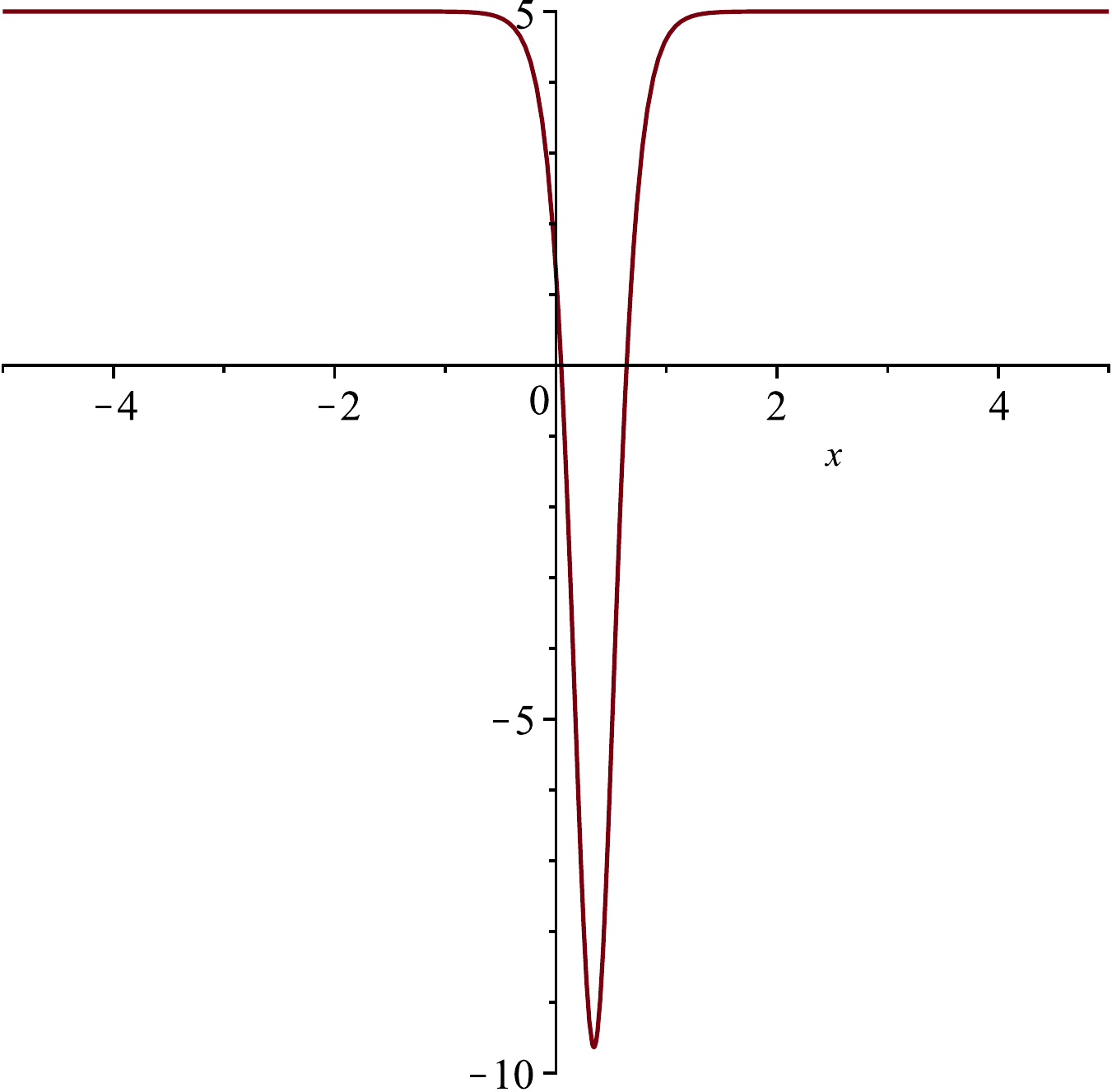} ~~~
        \includegraphics[width=3.6cm]{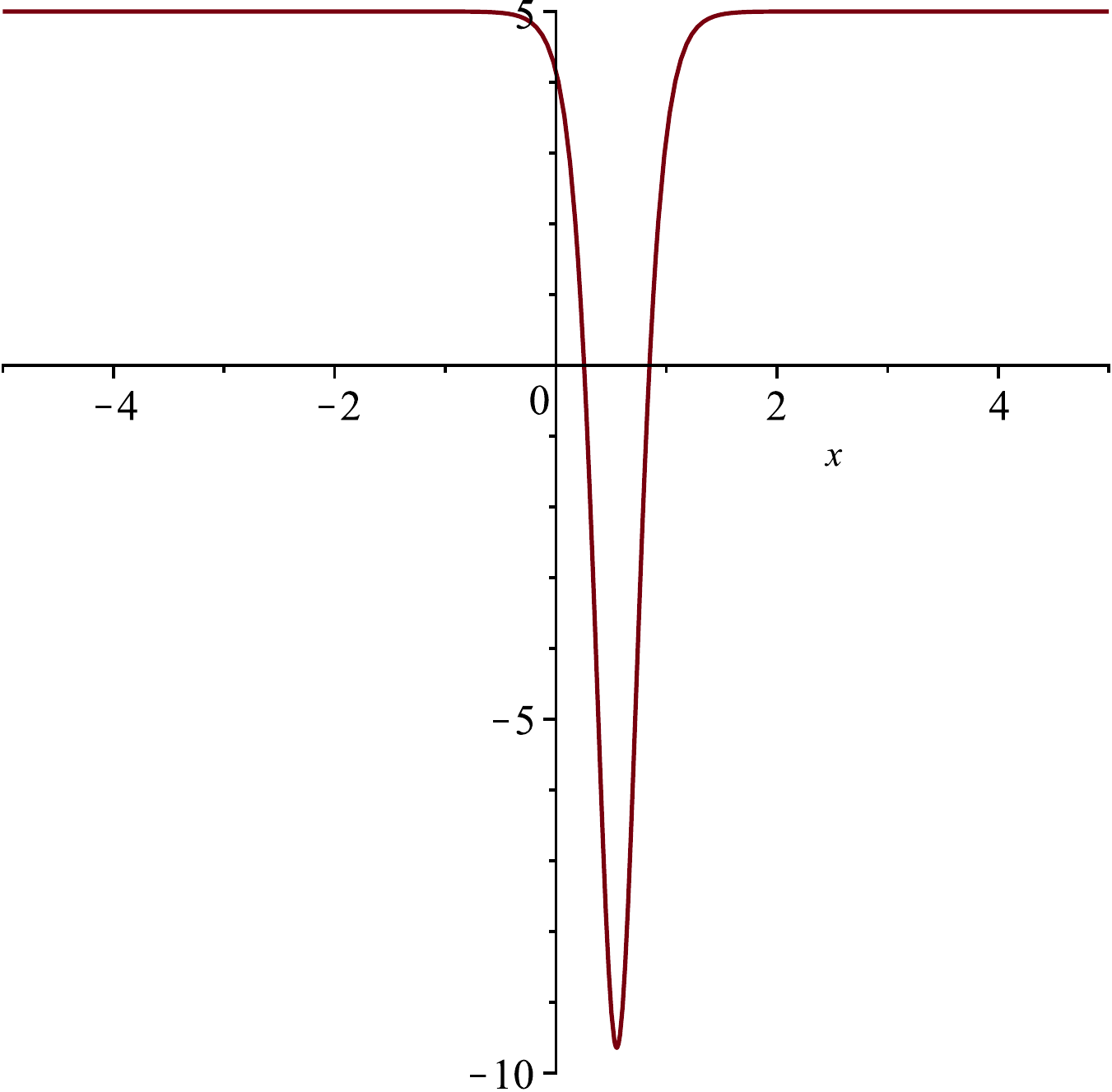} 
        }
      \caption{The merging soliton.  Parameter values $\beta=5$ and $\theta=-10$, so $\lambda_1,\lambda_2,\lambda_3 \approx -4.17,0.69,3.48$.
        The constants $C_1,C_2,C_3$ are all taken to be $1$.
        Plots of $u=f_x$ (with $f$ given by (\ref{yeq})),  displayed for times
      $t=-1,-0.7,-0.4,-0.1,0.2,0.5,0.8$. Note  smaller solitons are faster (see Equation (\ref{speedamp})).} 
\end{figure}

Similar considerations can be applied in the case that all three constants $C_1,C_2,C_3$ in (\ref{yeq}) are nonzero, but their signs
differ. There will be one pair with the same sign and two pairs with opposite signs. The solutions can describe the absorption of a
standard soliton by a singular soliton, see Figure 3,  or the merger of two singular solitons to a standard soliton, see Figures 4
and 5 (compare Figure 12 in \cite{beq11}). 

\begin{figure}
      \centerline{
        \includegraphics[width=3.6cm]{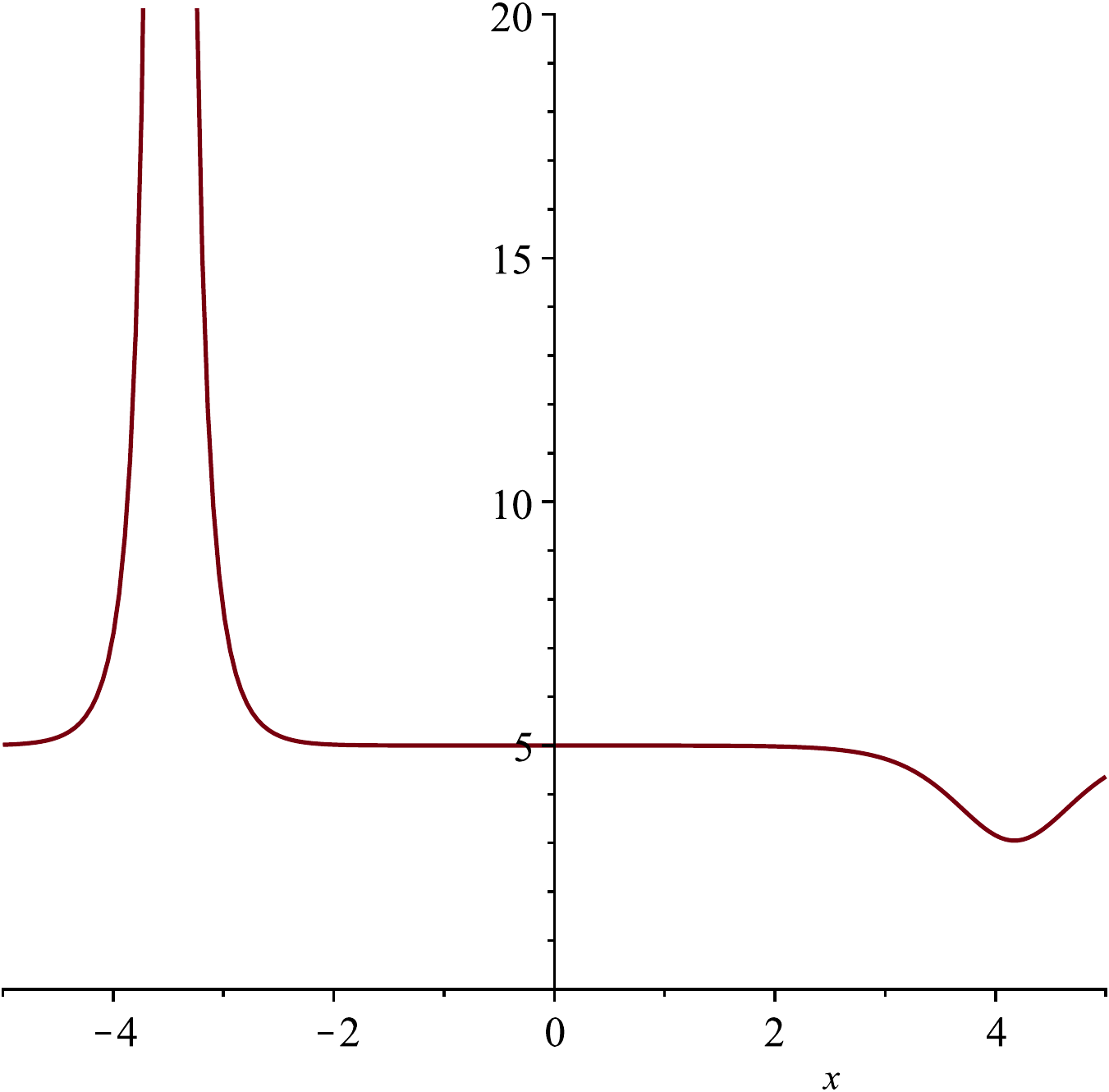} ~~~
        \includegraphics[width=3.6cm]{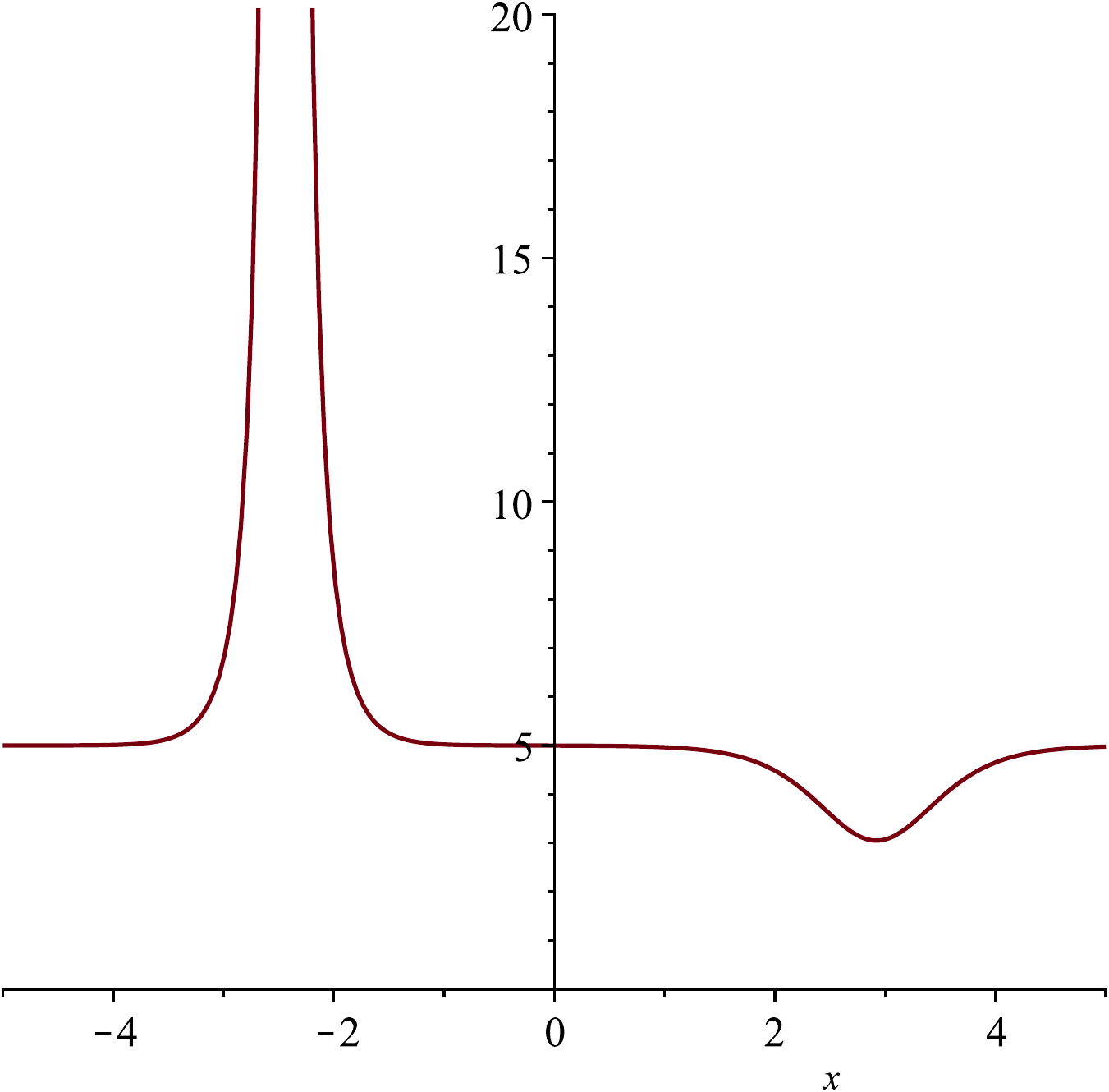} ~~~
        \includegraphics[width=3.6cm]{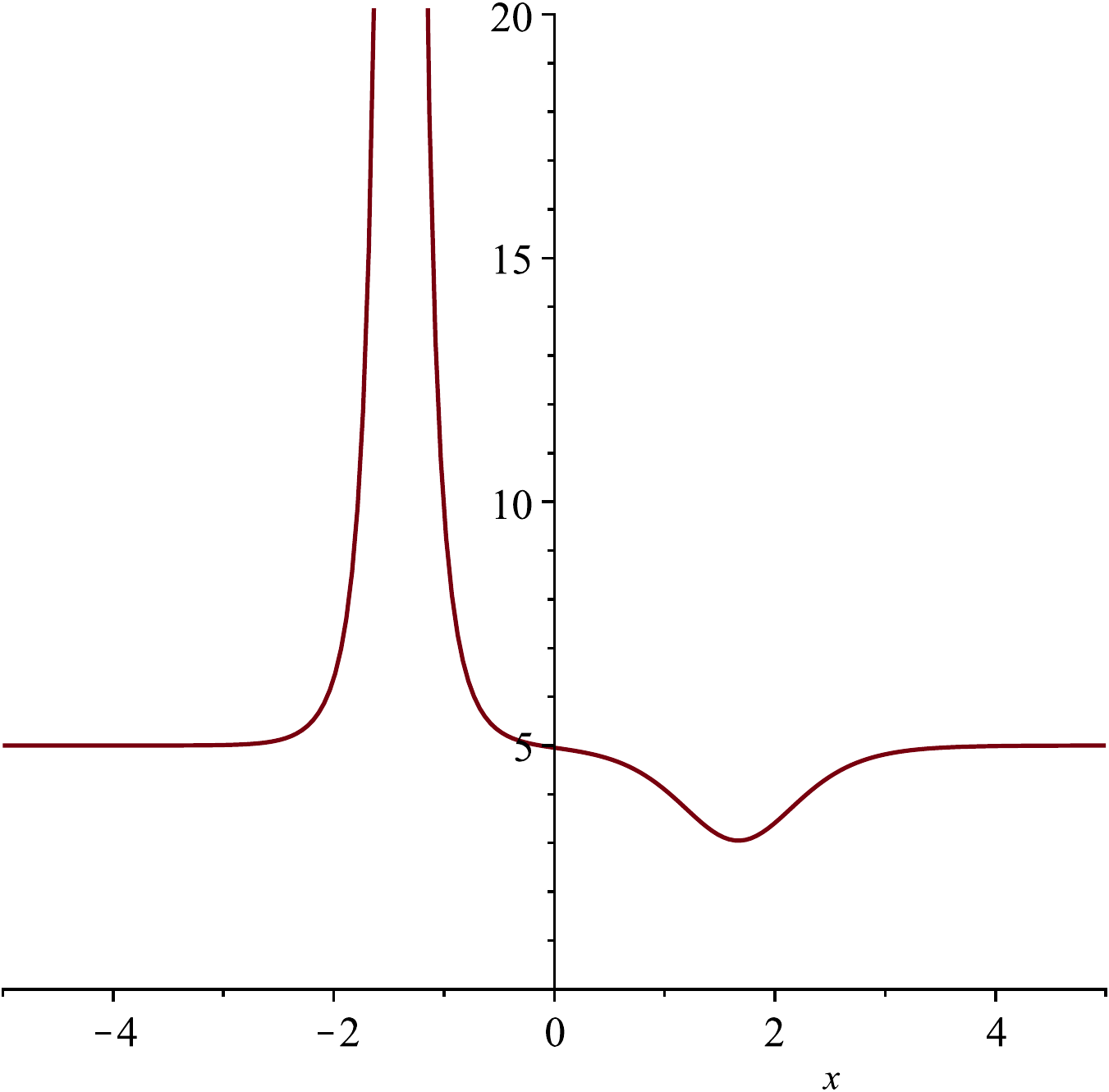} ~~~
        \includegraphics[width=3.6cm]{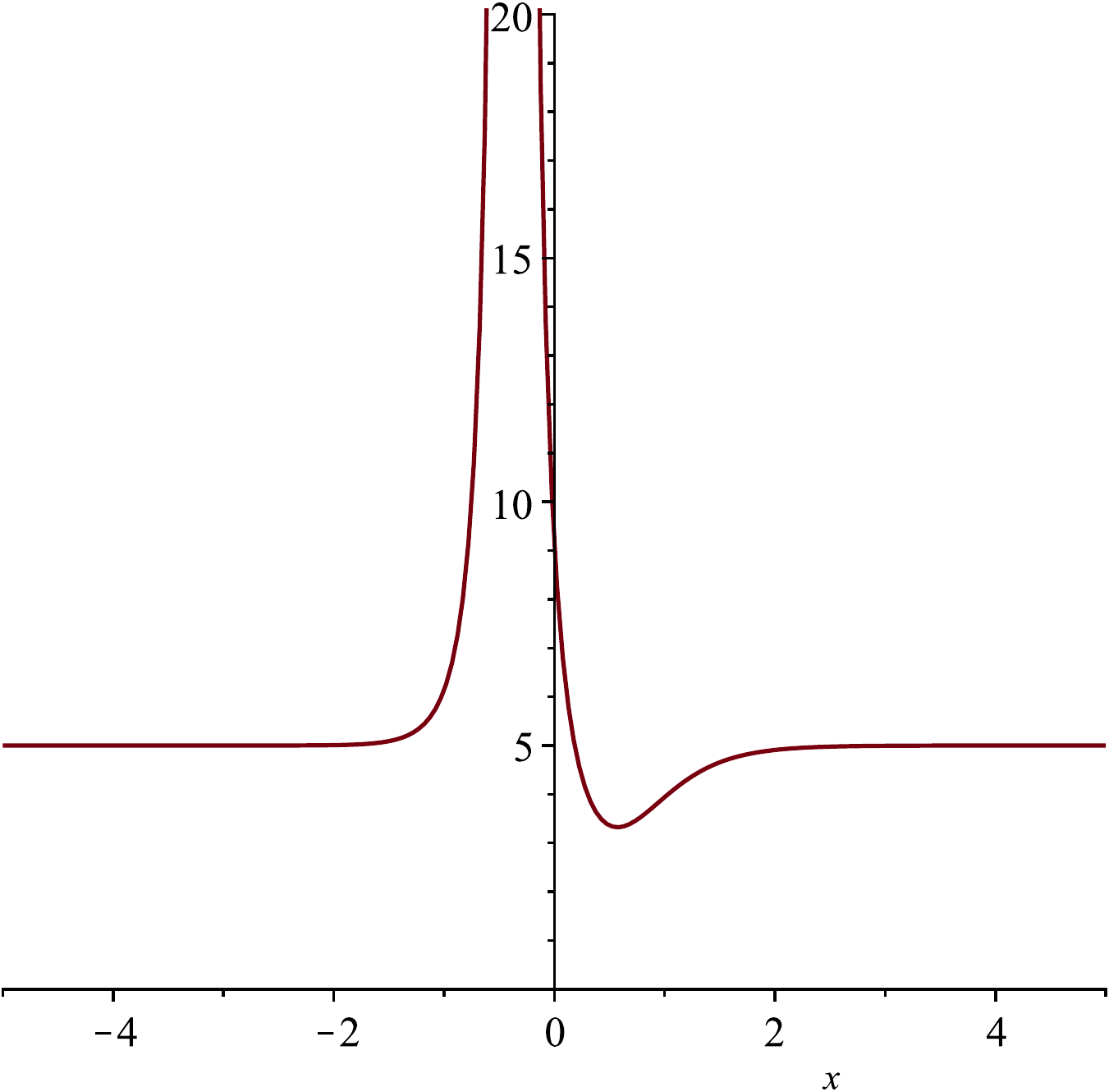} 
        }
      \centerline{
        \includegraphics[width=3.6cm]{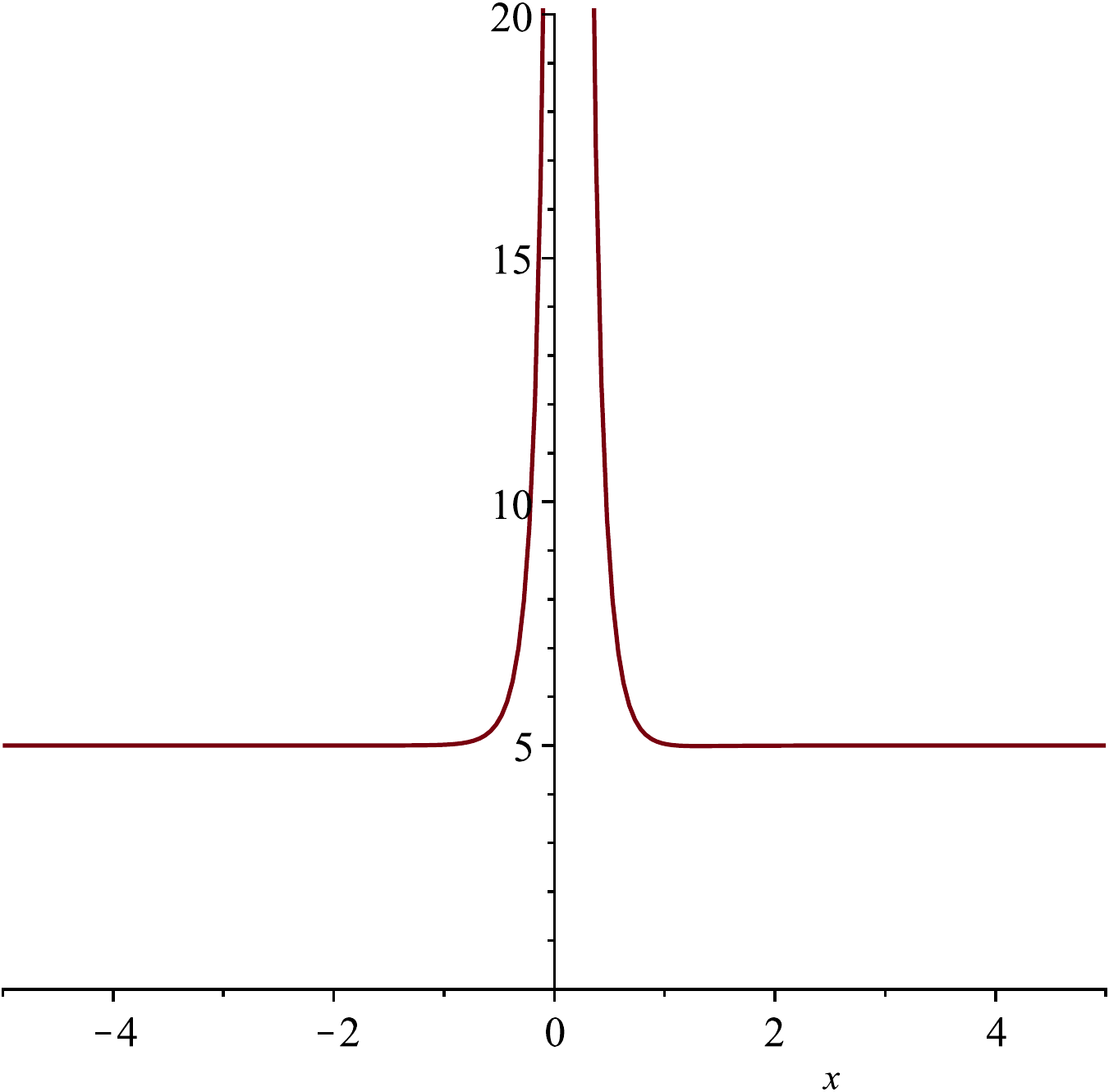} ~~~
        \includegraphics[width=3.6cm]{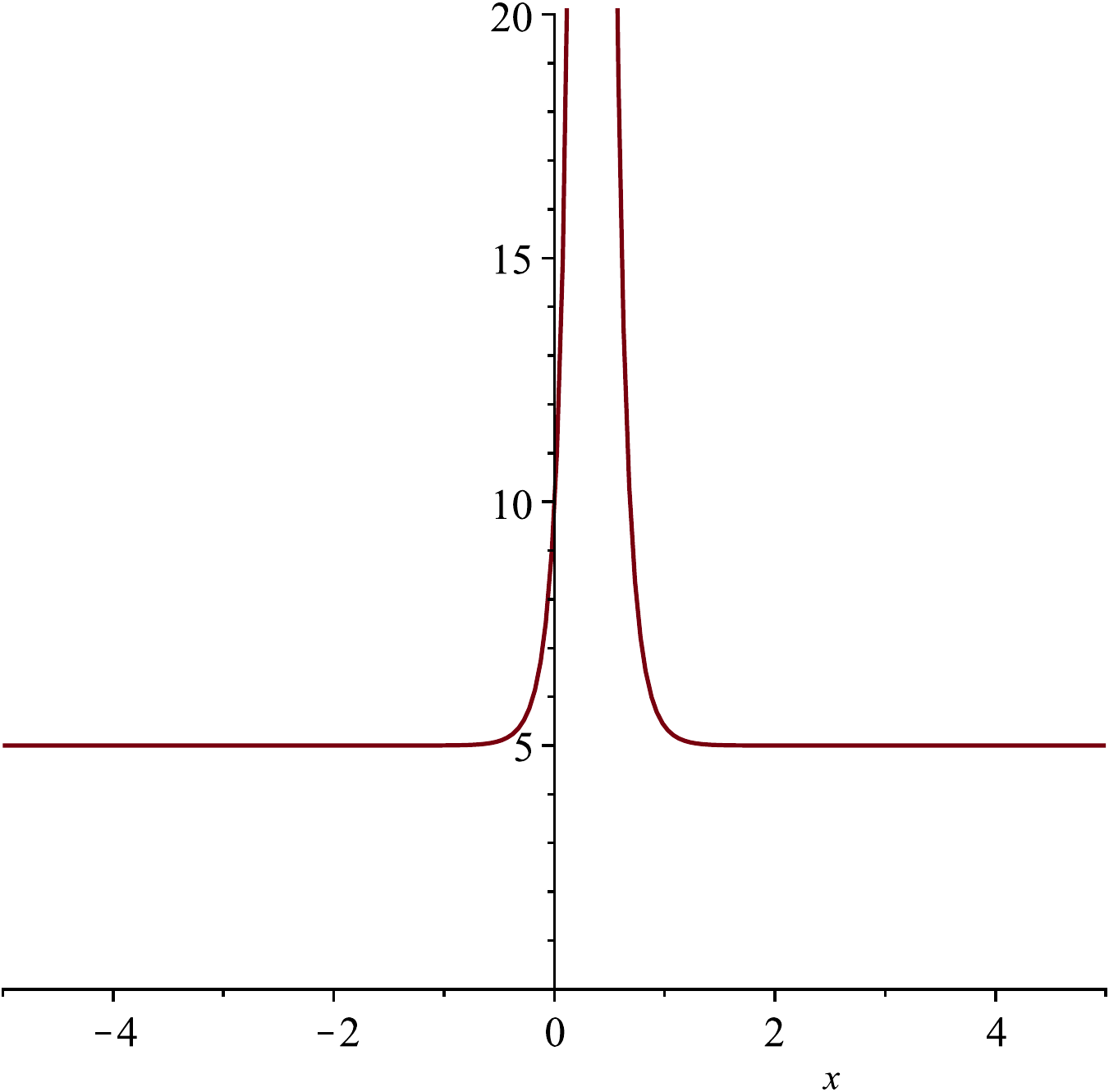} ~~~
        \includegraphics[width=3.6cm]{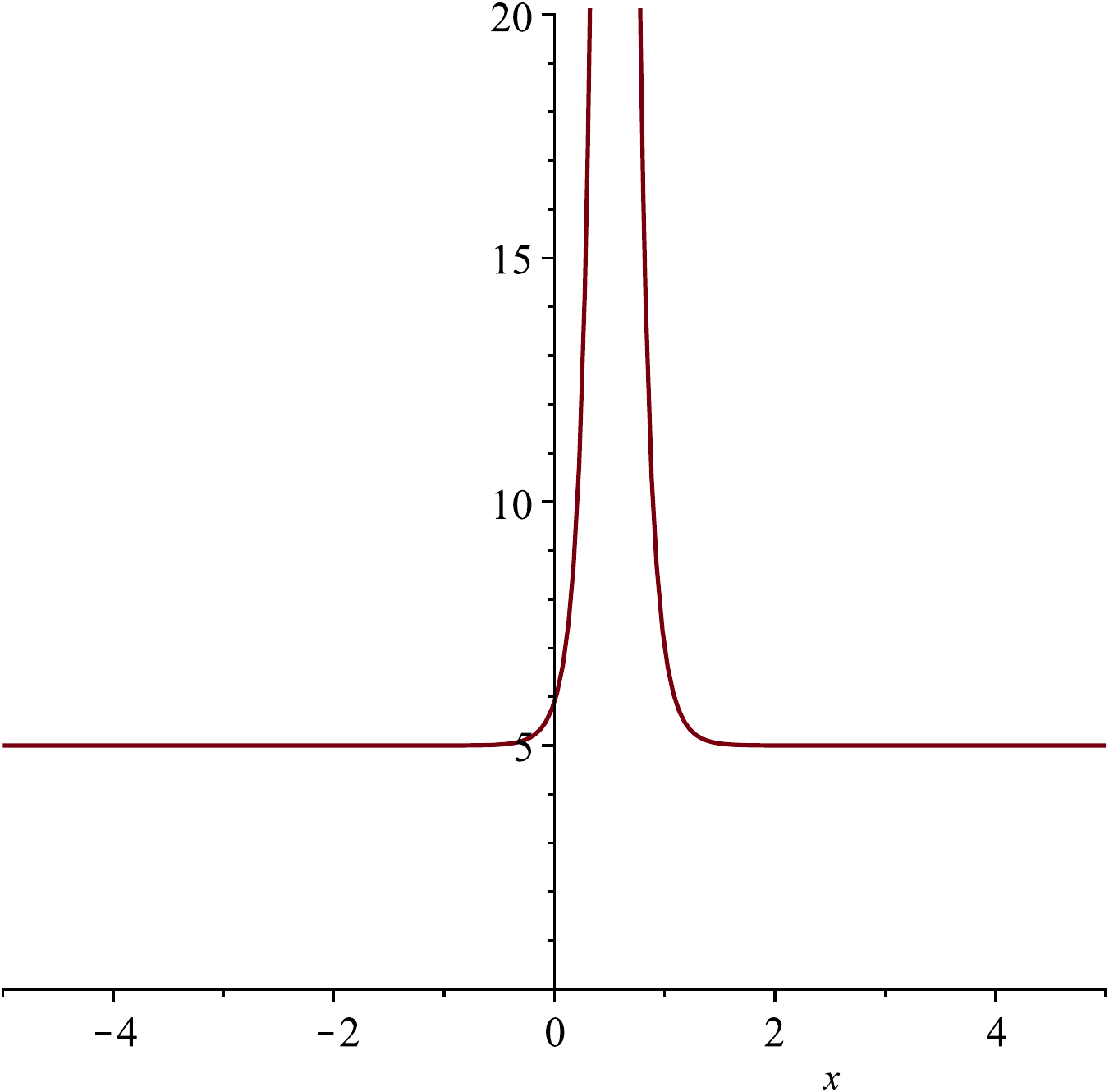} 
        }
      \caption{Absorption of a soliton by singular soliton. Parameters and times identical to Figure $2$ except $C_1=-1$.} 
\end{figure}

\begin{figure}
      \centerline{
        \includegraphics[width=3.6cm]{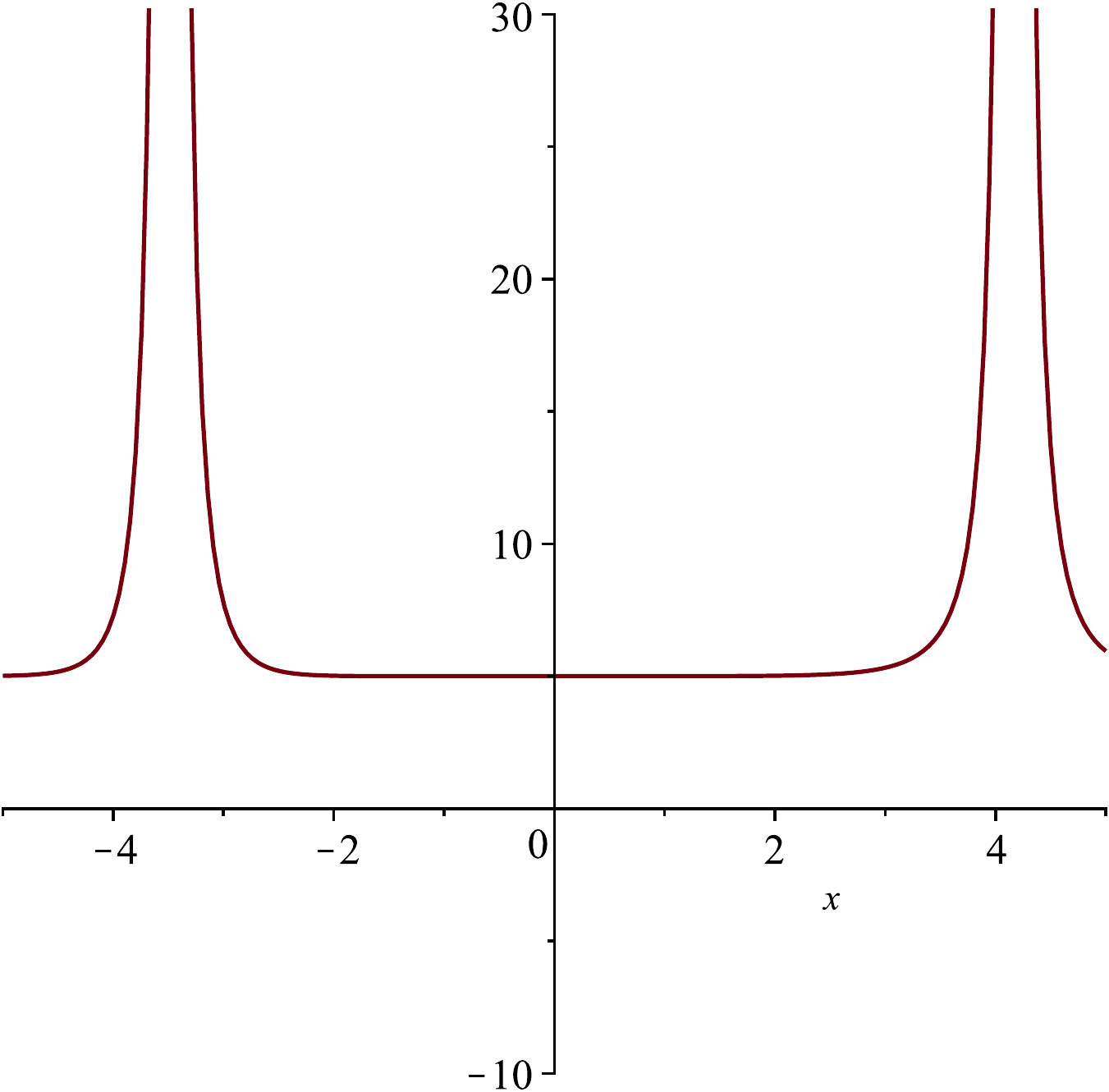} ~~~
        \includegraphics[width=3.6cm]{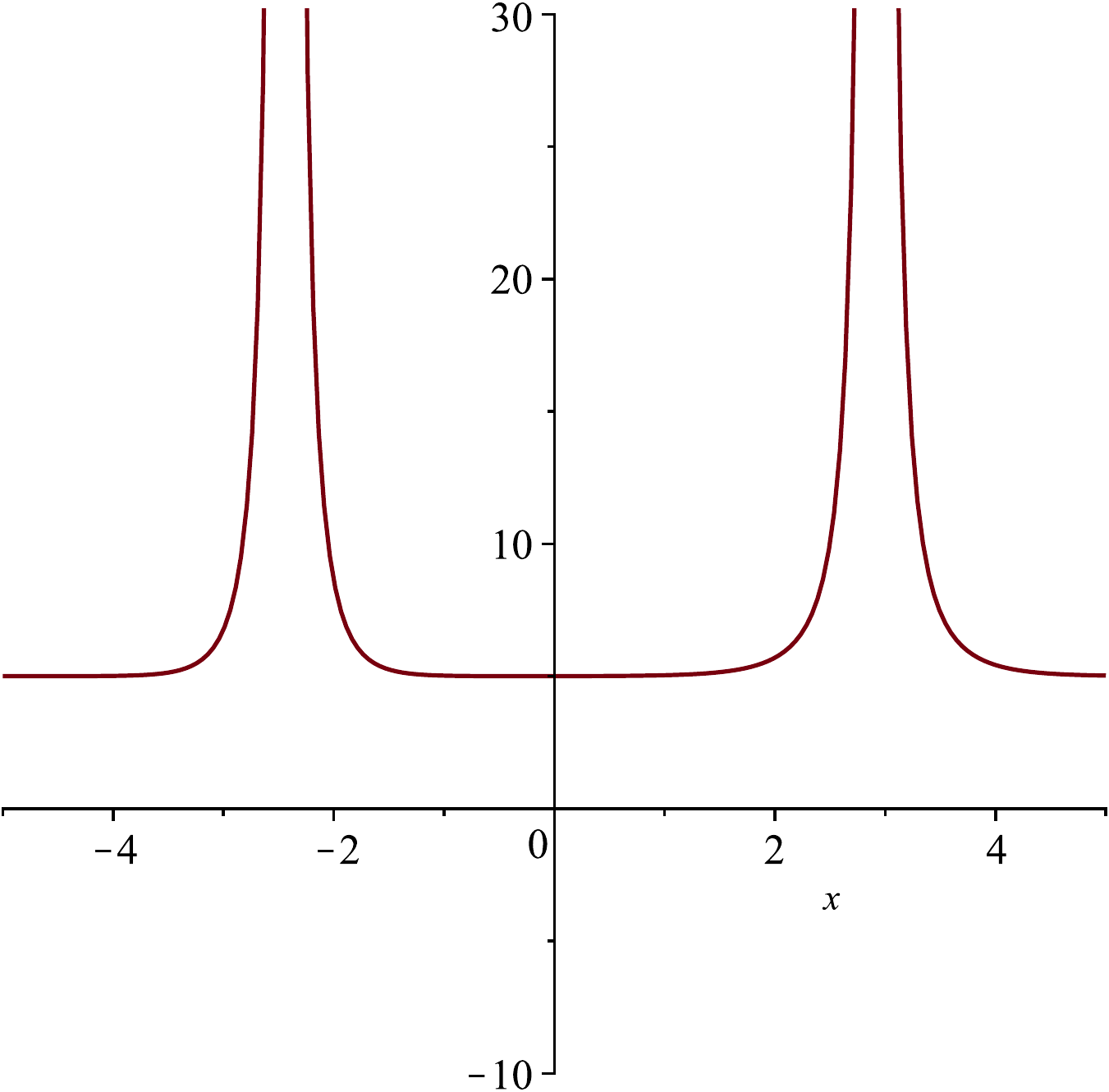} ~~~
        \includegraphics[width=3.6cm]{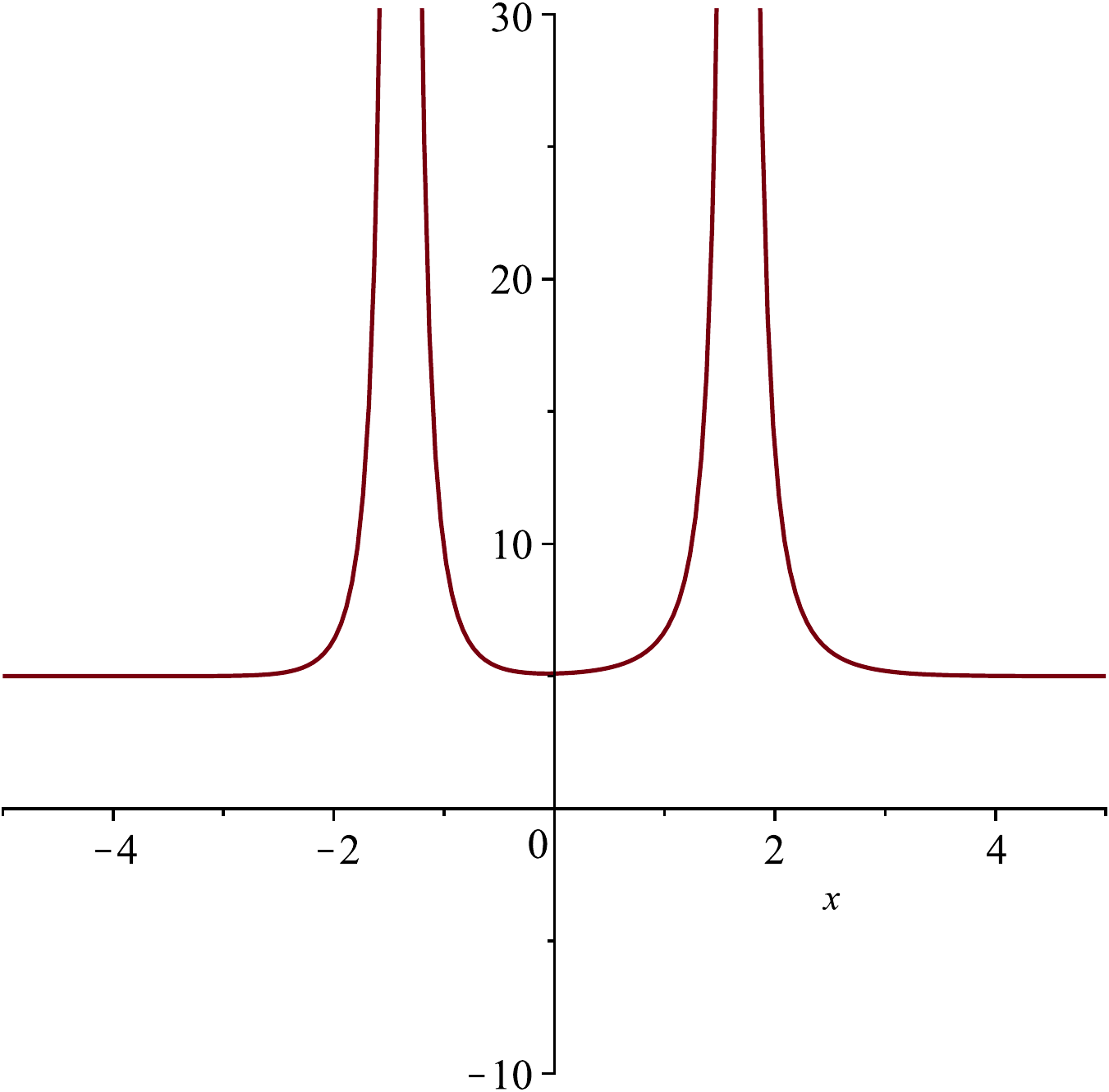} ~~~
        \includegraphics[width=3.6cm]{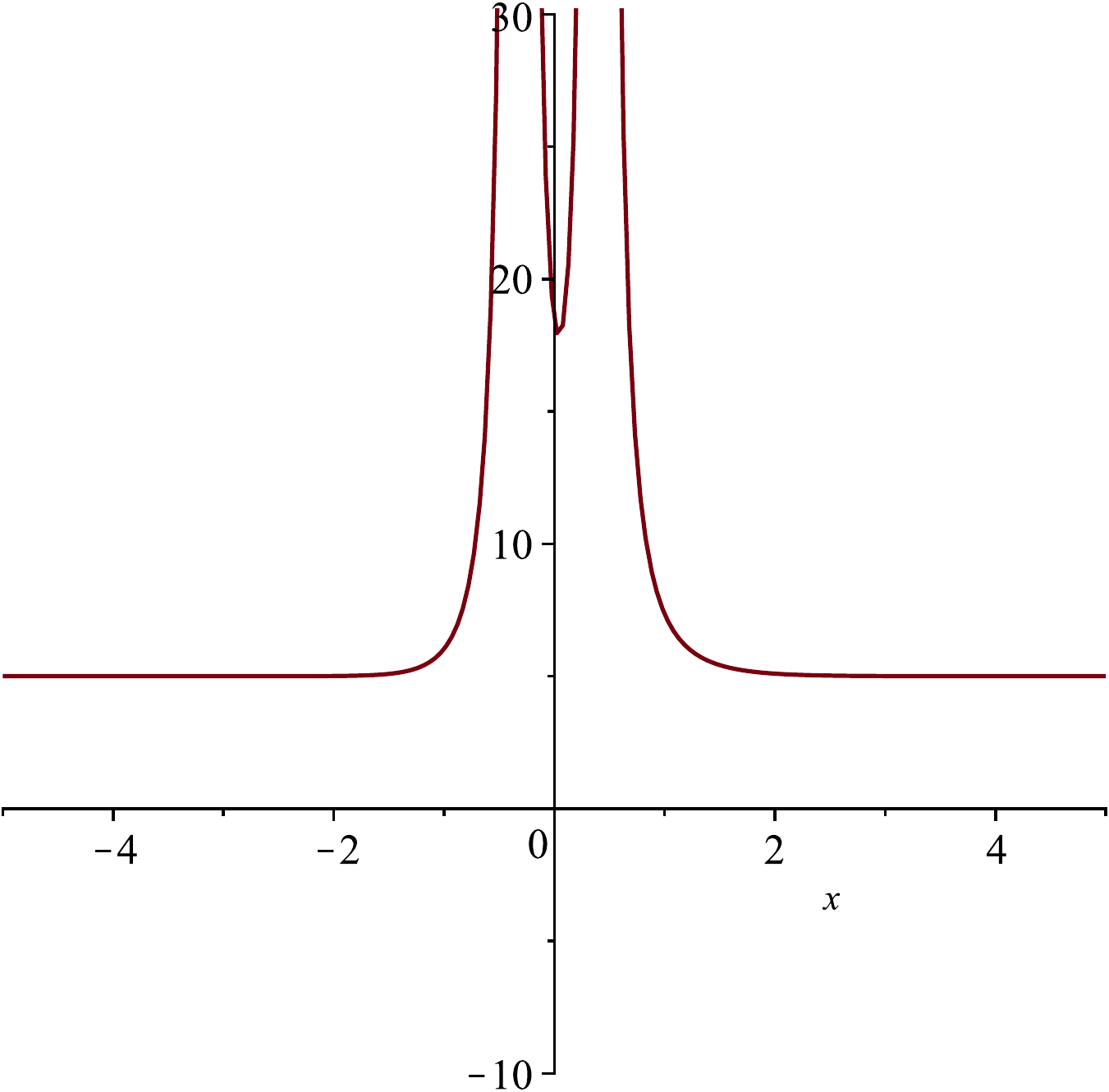} 
        }
      \centerline{
        \includegraphics[width=3.6cm]{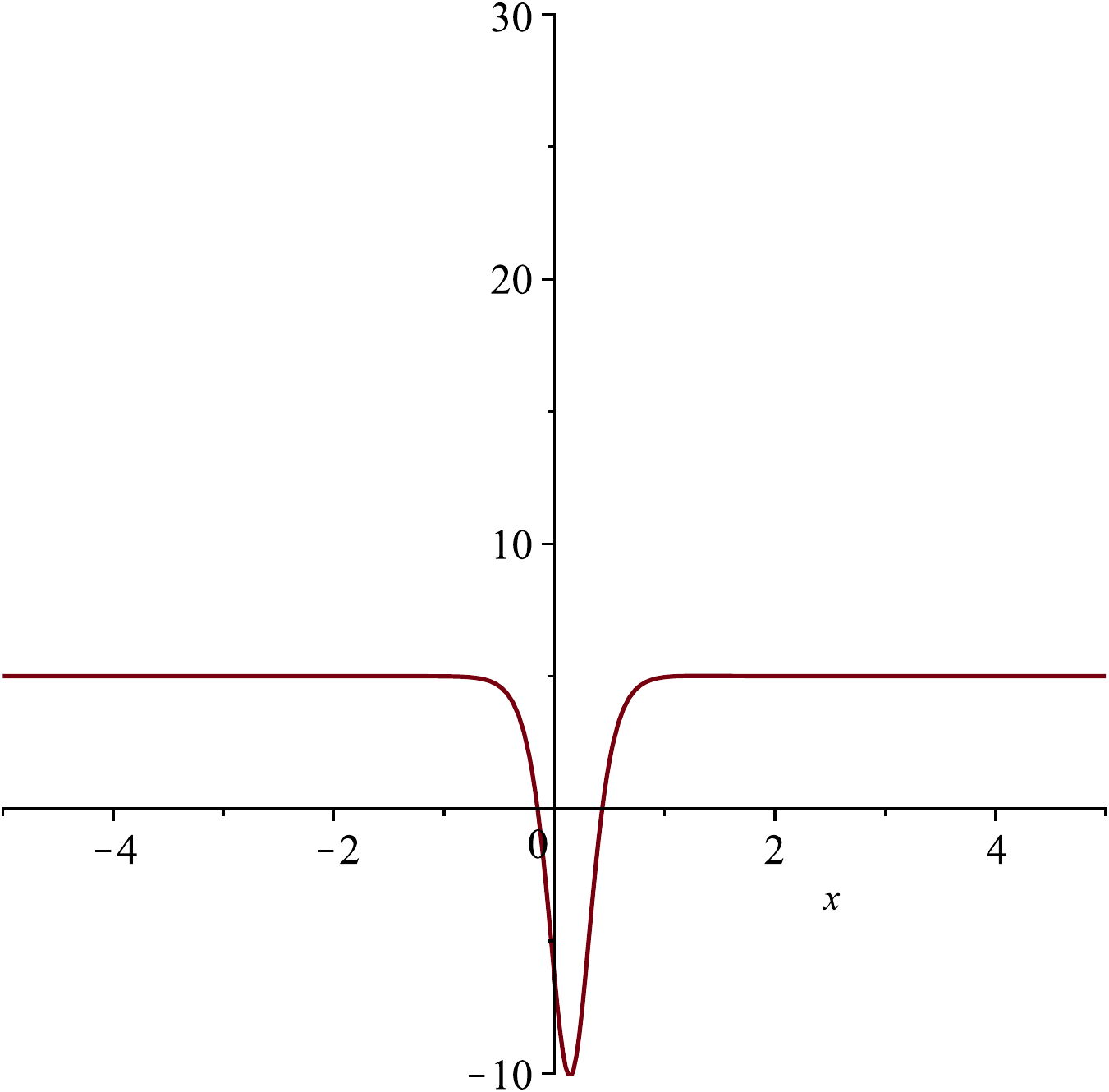} ~~~
        \includegraphics[width=3.6cm]{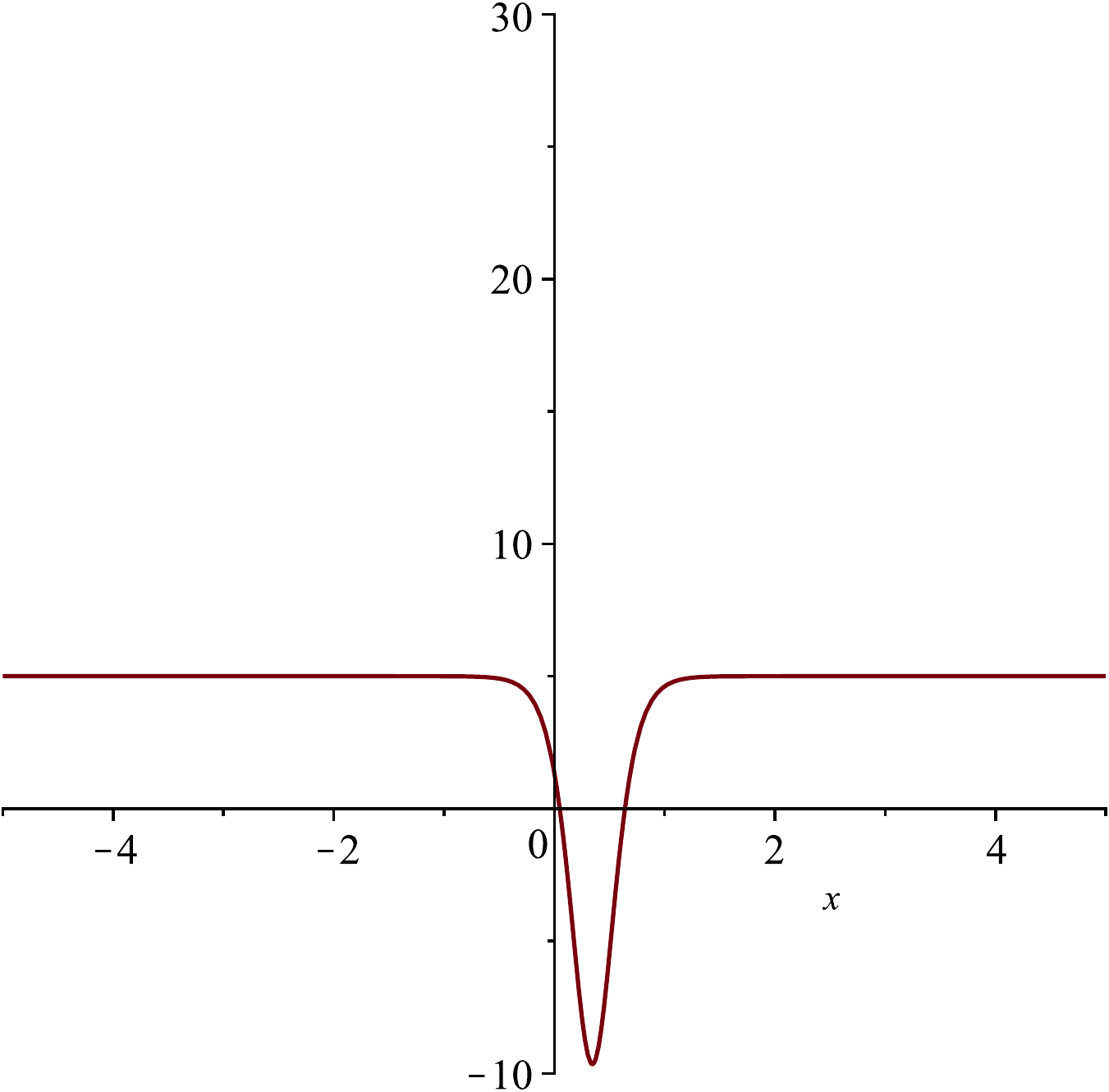} ~~~
        \includegraphics[width=3.6cm]{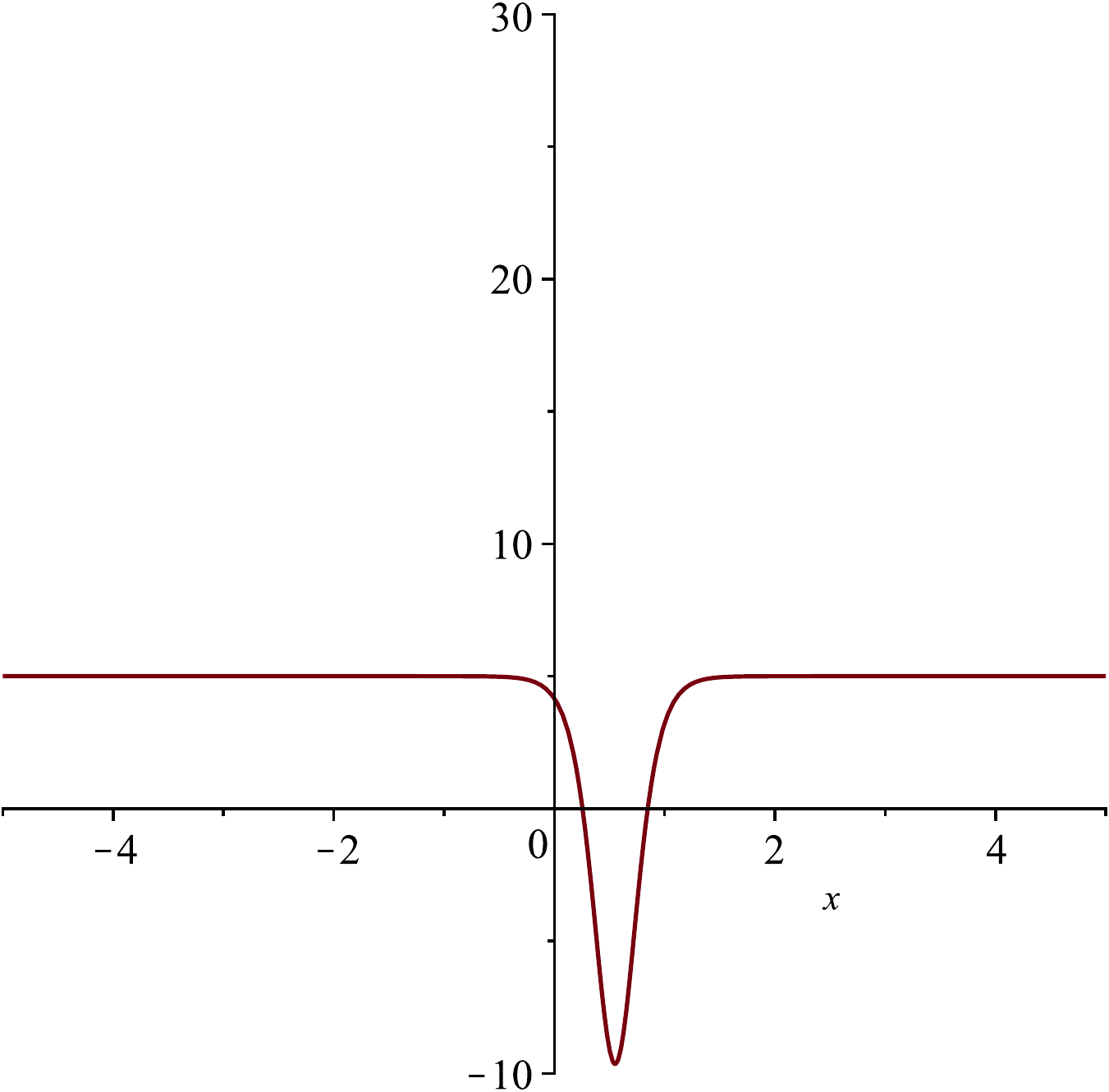} 
        }
      \caption{Singular soliton merger. Parameters and times identical to Figure $2$ except $C_2=-1$.} 
\end{figure}

\begin{figure}
      \centerline{
        \includegraphics[width=3.6cm]{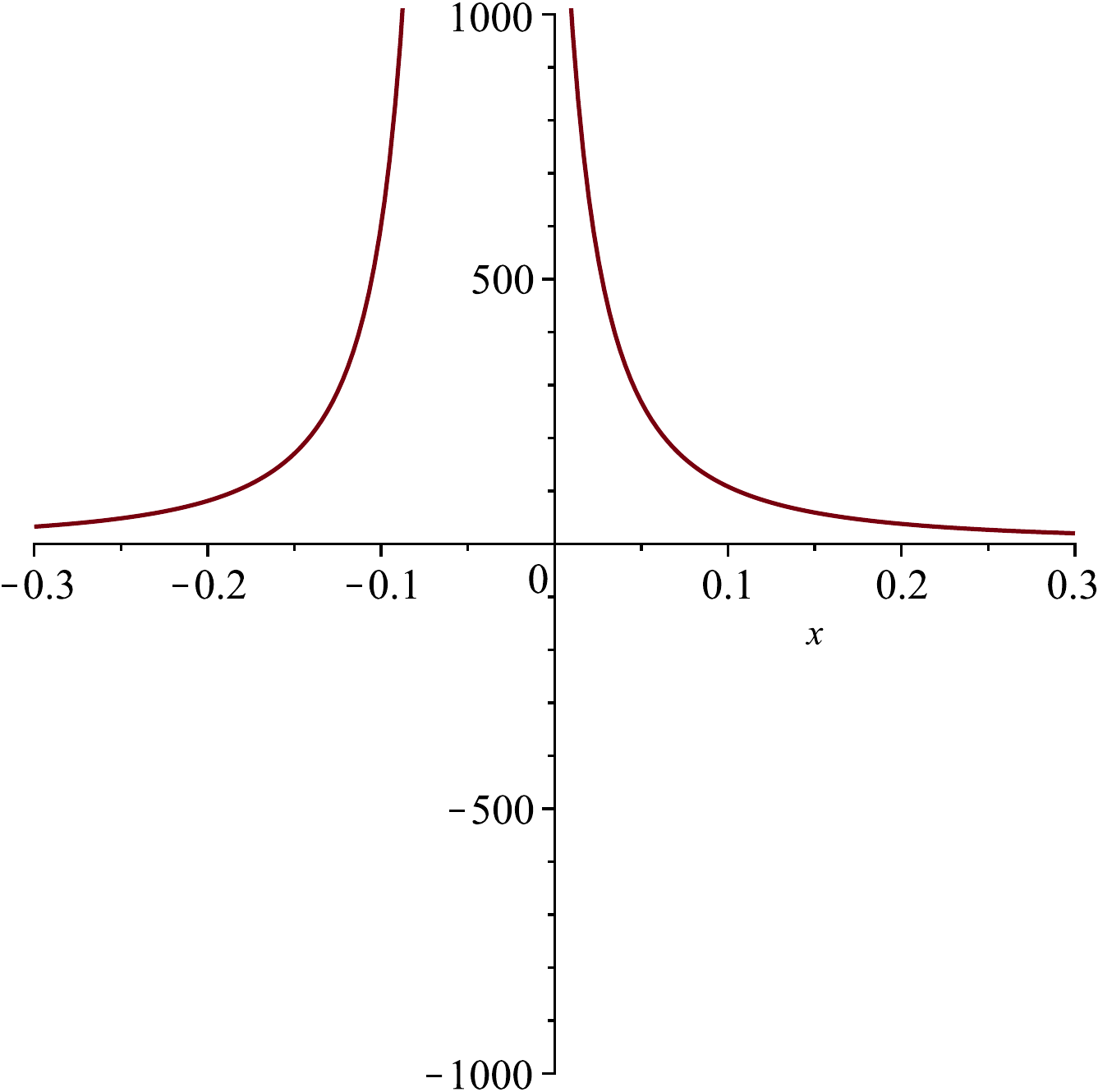} ~~~
        \includegraphics[width=3.6cm]{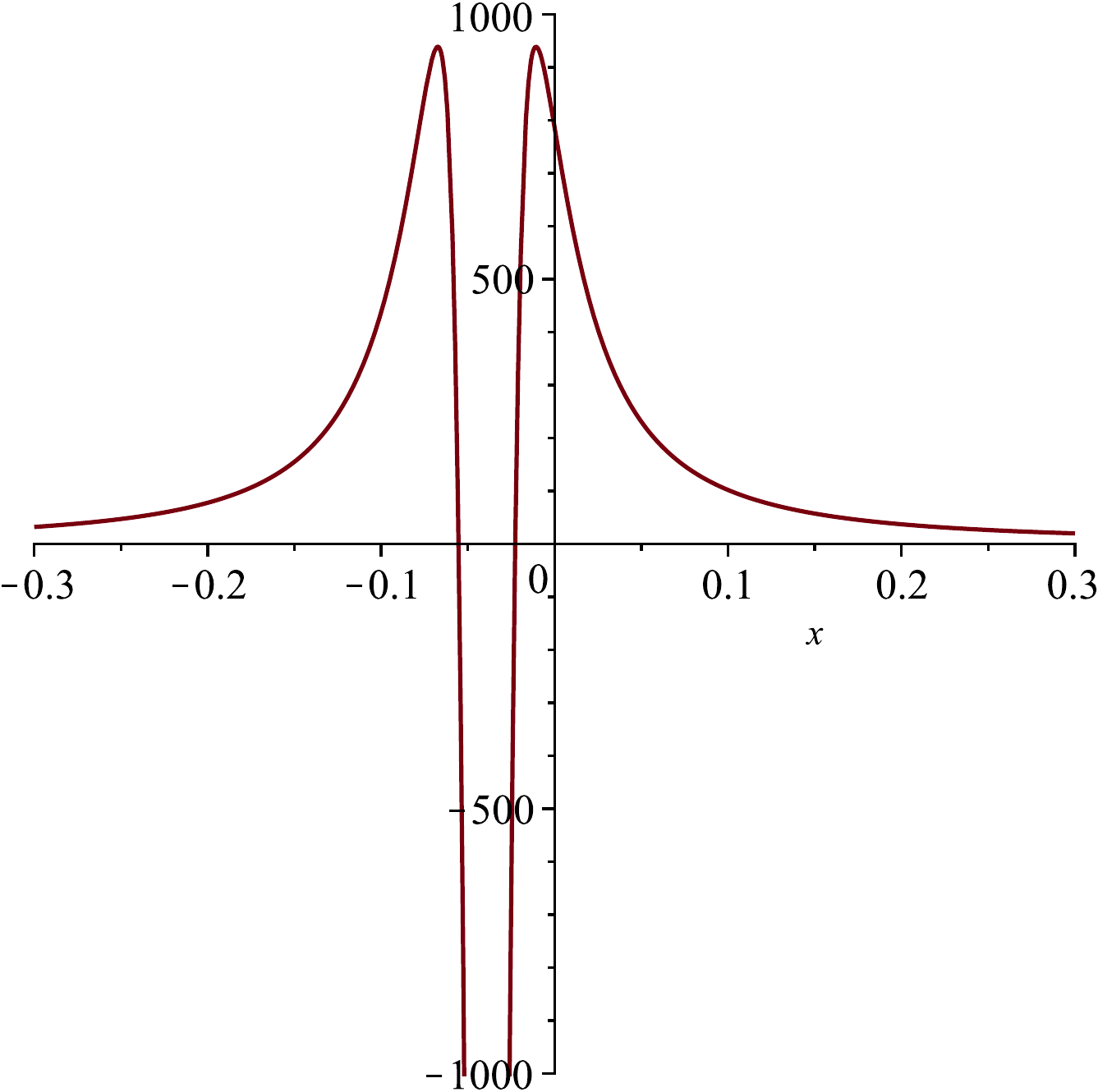} ~~~
        \includegraphics[width=3.6cm]{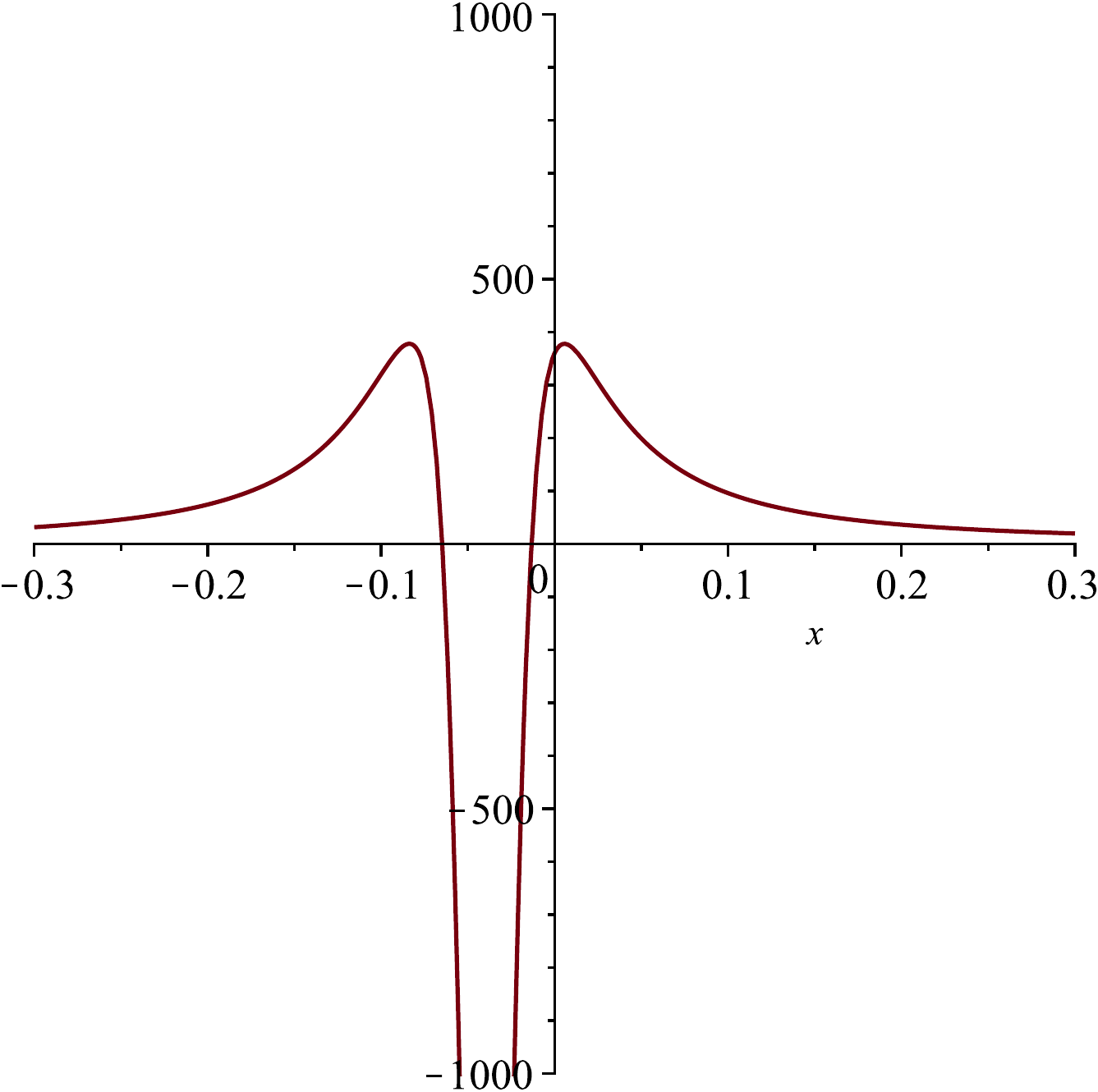} ~~~
        \includegraphics[width=3.6cm]{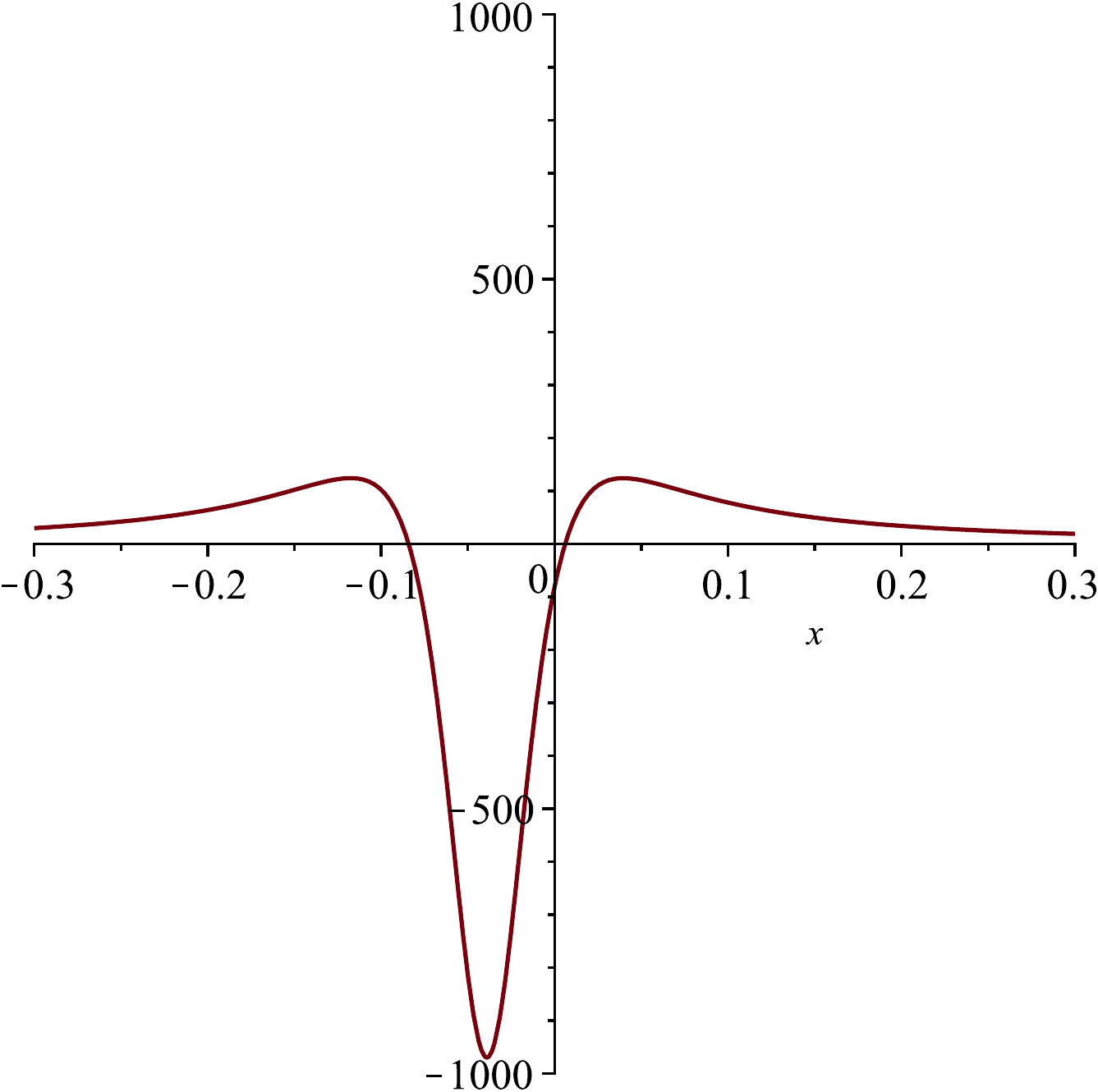} 
        }
      \caption{Singular soliton merger, more detail. Times $t=-0.0484,-0.0482,-0.0480,-0.0473$.}
\end{figure}

In summary, we have obtained $5$ types of solution by a single application of the BT to the starting
solution $f=\beta x$: standard solitons, singular solitons, a merging soliton, a singular soliton
absorbing a soliton, and the merger of a pair of singular solitons to a single soliton. We refer to these
solutions collectively as ``1 BT solutions''. 

We now consider superpositions of two 1 BT solutions of the form (\ref{yeq})
using the superposition principle (\ref{pl3}). The solution takes the simple form
\begin{equation}
  f_{12} = \beta x - \frac{y_1 y_{2xx} - y_2 y_{1xx}}{y_1 y_{2x} - y_2 y_{1x}}\ .  \label{2bt}
\end{equation}
We have not succeeded to give a complete (analytic) classification of these solutions, but
we report cases in which we have found superpositions without singularities: 
\begin{itemize}
\item  For certain parameter values, a pair of standard soliton solutions can be superposed to give a
  colliding 2-soliton solution.  The 2 solitons should be taken with velocities of differing signs;
  this is a necessary, but not sufficient, condition for such a superposition to be possible. 
\item  For certain parameter values, a standard soliton and a singular soliton can be superposed to give
  a 2-soliton solution. The resulting solutions include both colliding pairs and pairs moving in the
  same direction. 
\item  For certain parameter values, a standard soliton solution can be superposed with one of the two
  types of singular solution describing a merger, to give a solution with three solitary waves merging to two.
  See Figure 7, in which the soliton with parameters $\theta=8,C_1=1,C_2=1,C_3=0$ is superposed with
  the 1 BT solution with $\theta=-12,C_1=-1,C_2=1,C_3=1$ (for $\beta=5$). In the cases of this that we have
  found, the initial configuration always has two solitary waves moving in one direction, and the other in the
  opposite direction, but the final configuration can have two moving in one direction, or one in each
  direction. We have not found cases of mergers of 3 moving in the same direction to 2, but we cannot
  currently exclude this possibility. 
\end{itemize}
So far we have not found any cases of the merger of 4 solitary waves to 2, though we cannot currently
exclude this possibility. The superpositions of a pair of solutions describing a merger always seem to be
singular, describing, for example, the merger of 4 solitary waves to 2 singular solitons, or the absorption of
two singular solitons by two standard solitary waves. 

\begin{figure}
      \centerline{
        \includegraphics[width=3.6cm]{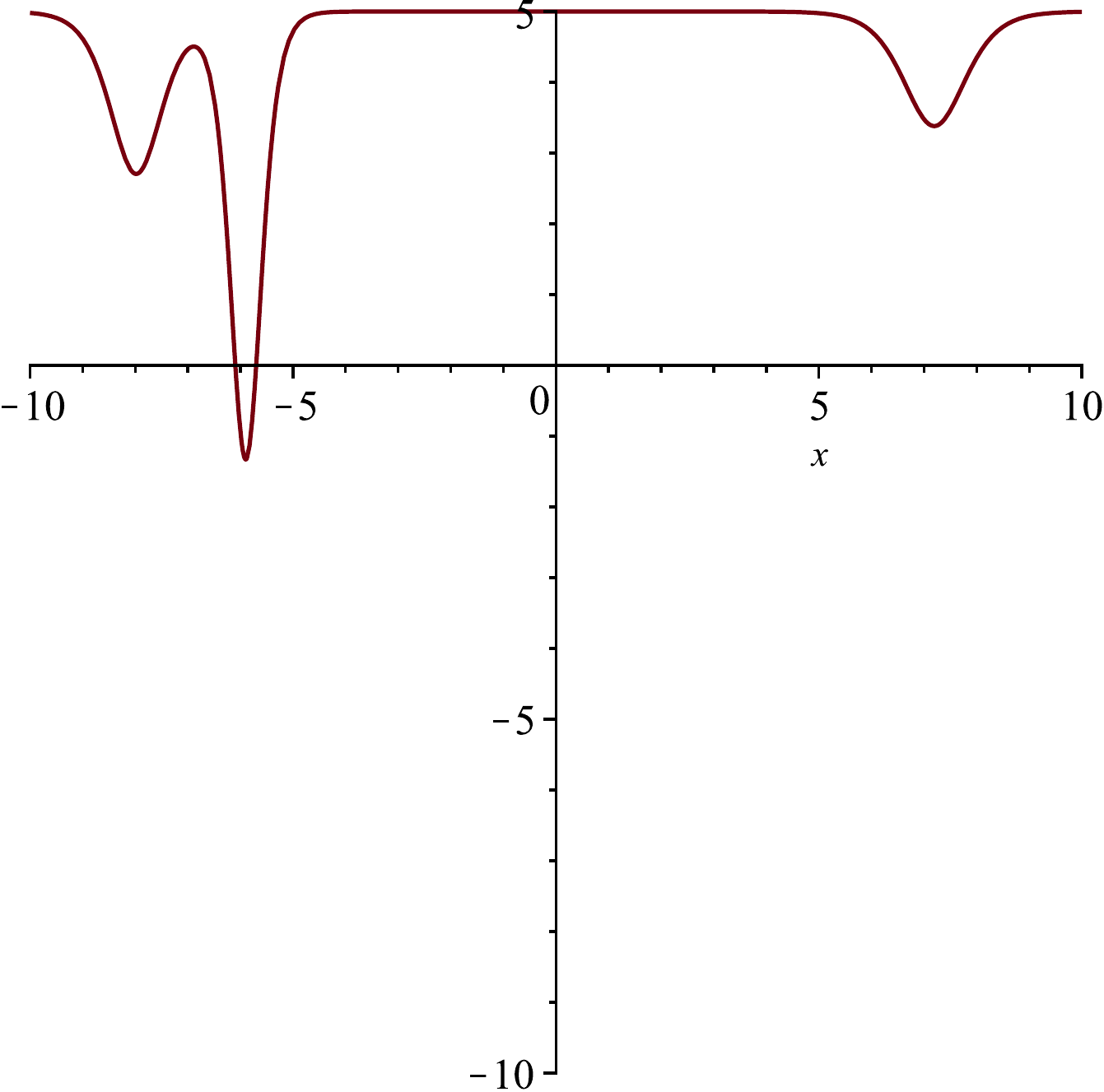} ~~~
        \includegraphics[width=3.6cm]{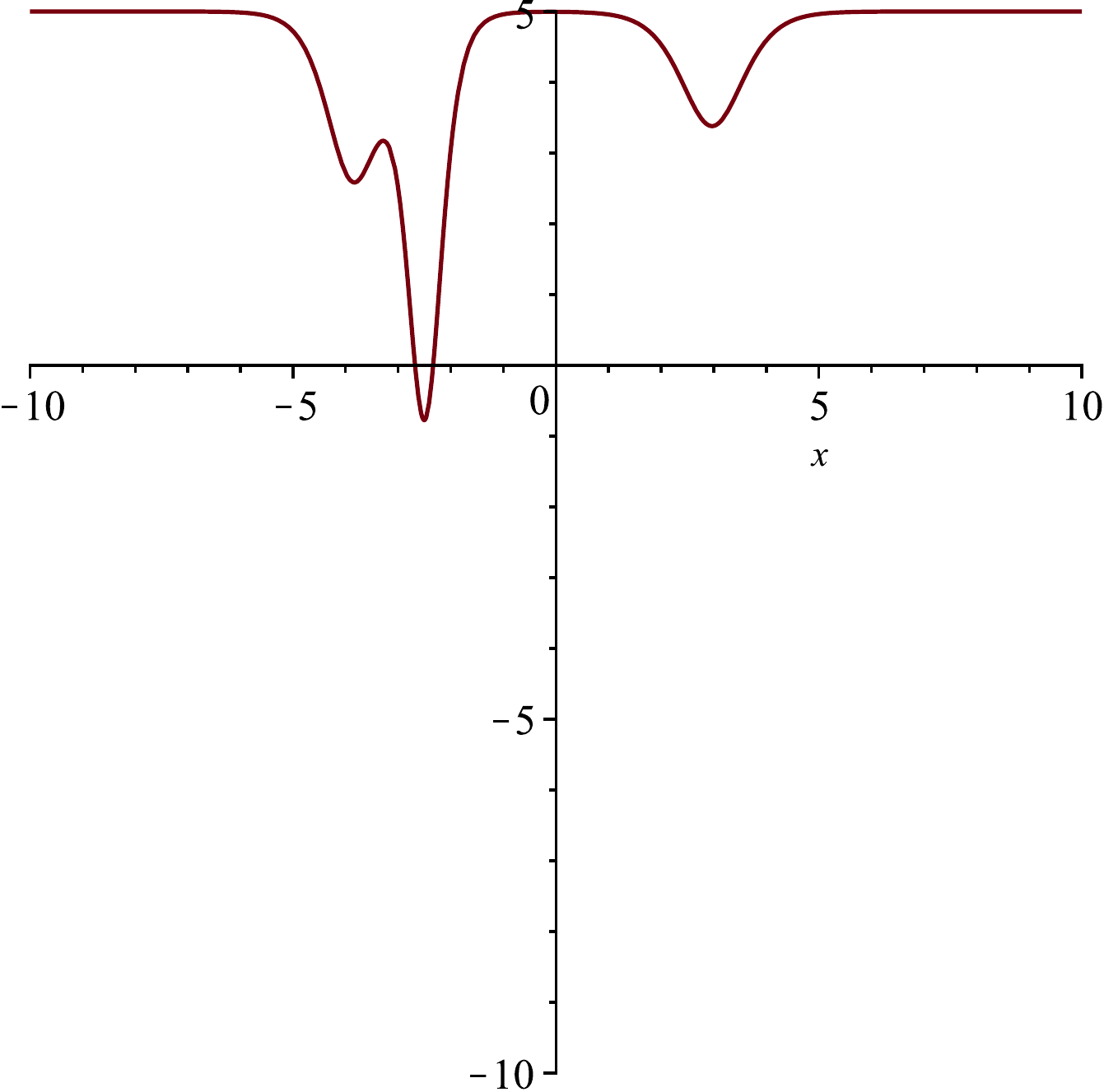} ~~~
        \includegraphics[width=3.6cm]{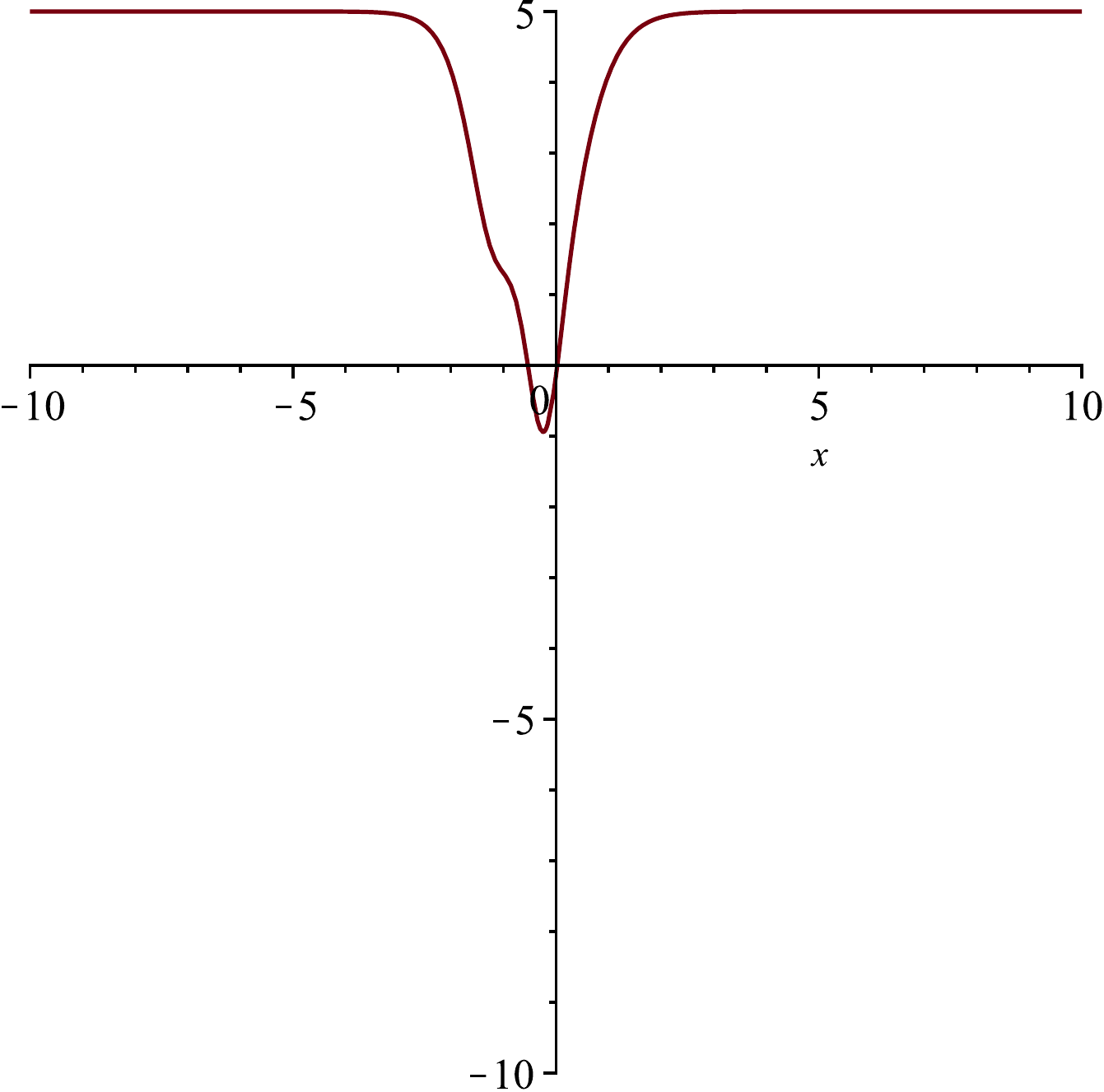} ~~~
        \includegraphics[width=3.6cm]{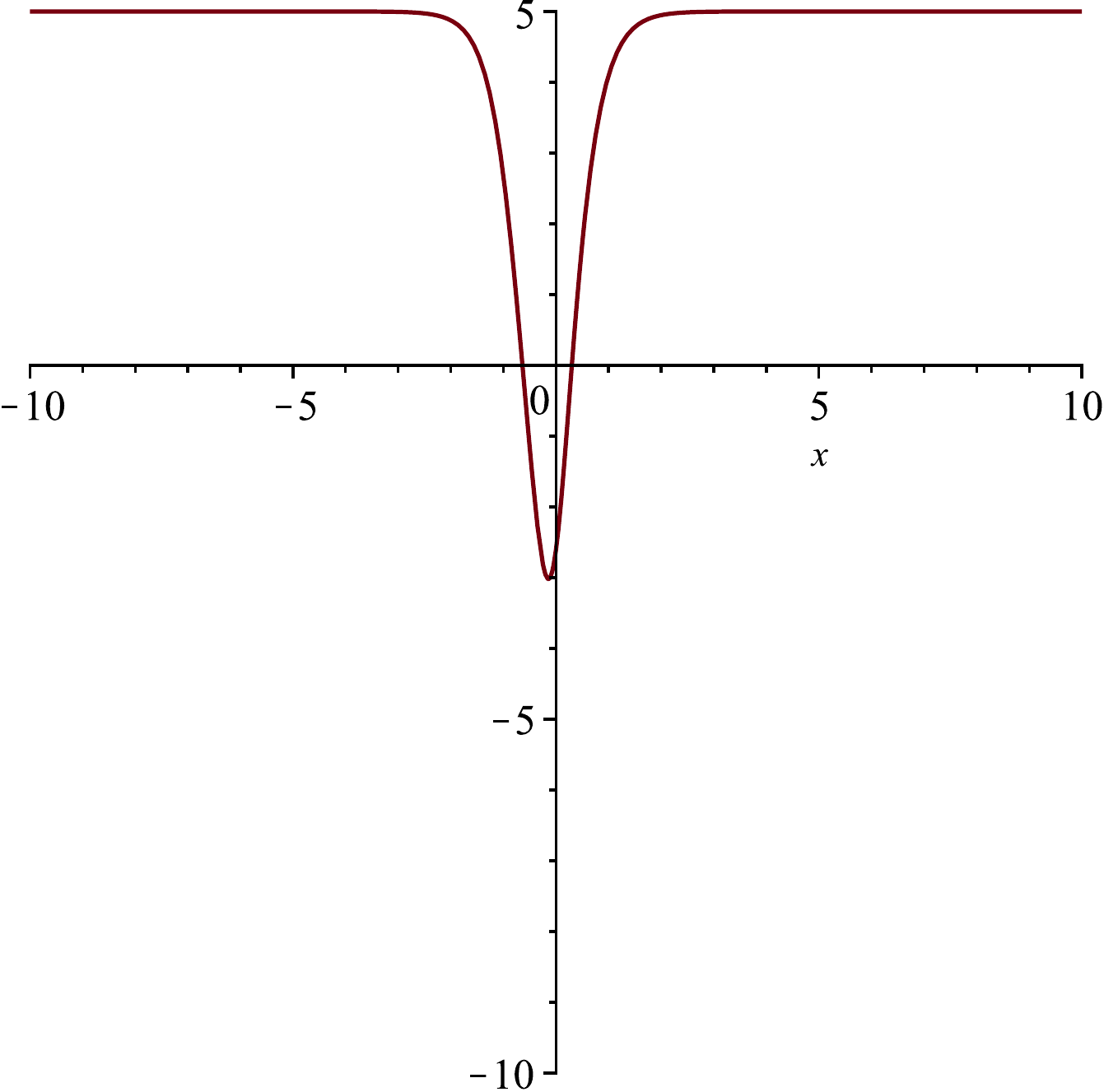} 
        }
      \centerline{
        \includegraphics[width=3.6cm]{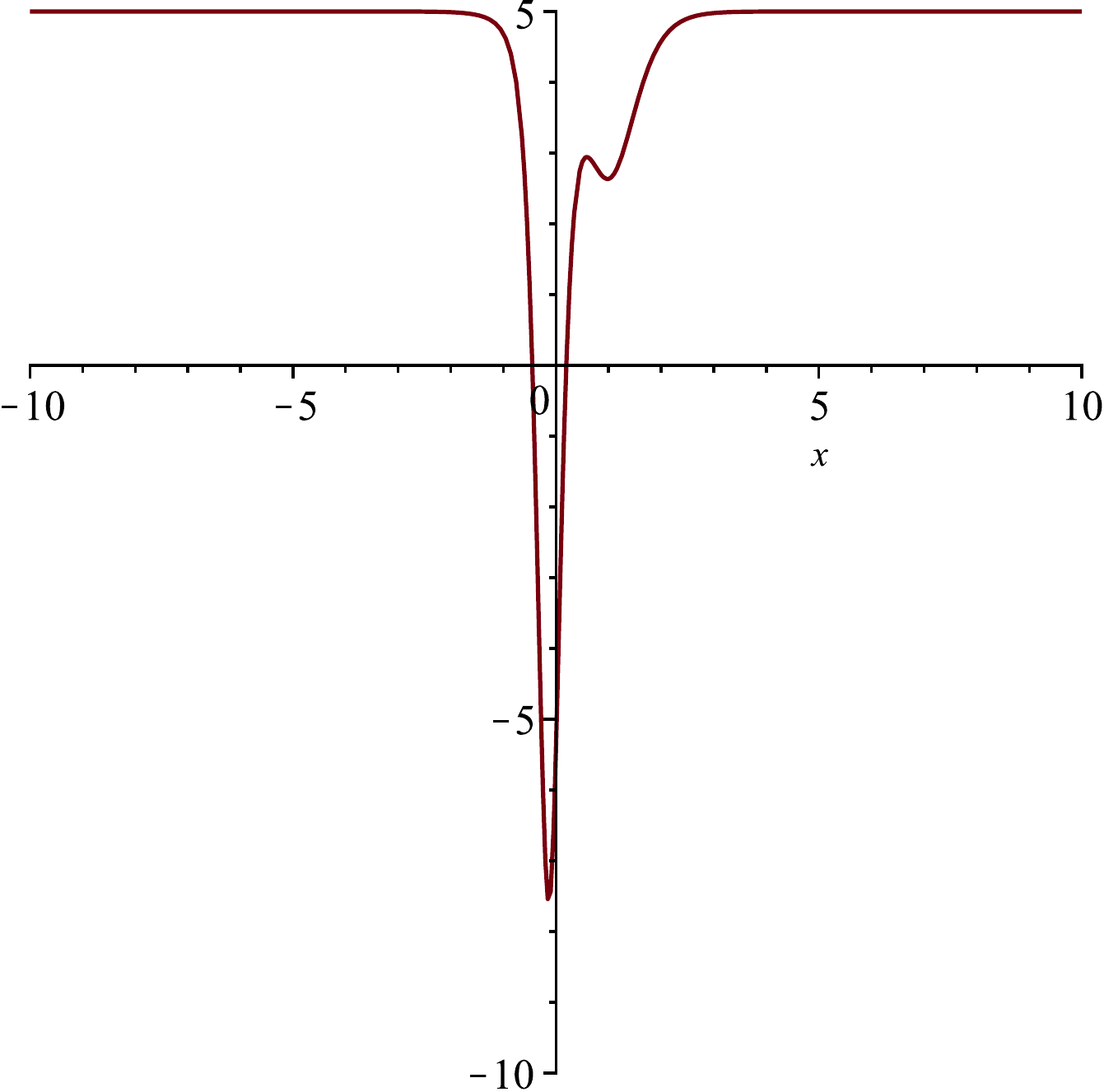} ~~~
        \includegraphics[width=3.6cm]{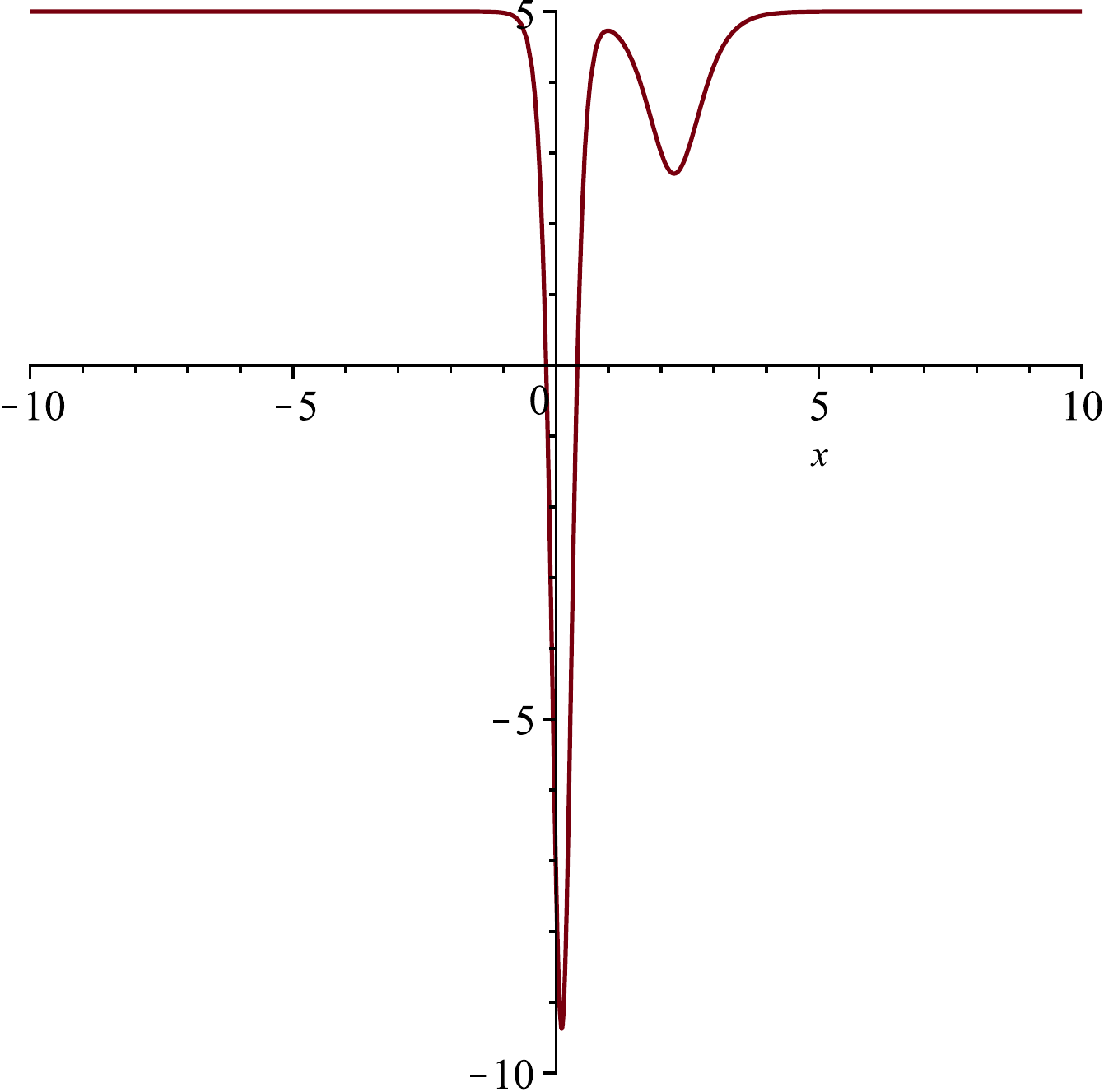} ~~~
        \includegraphics[width=3.6cm]{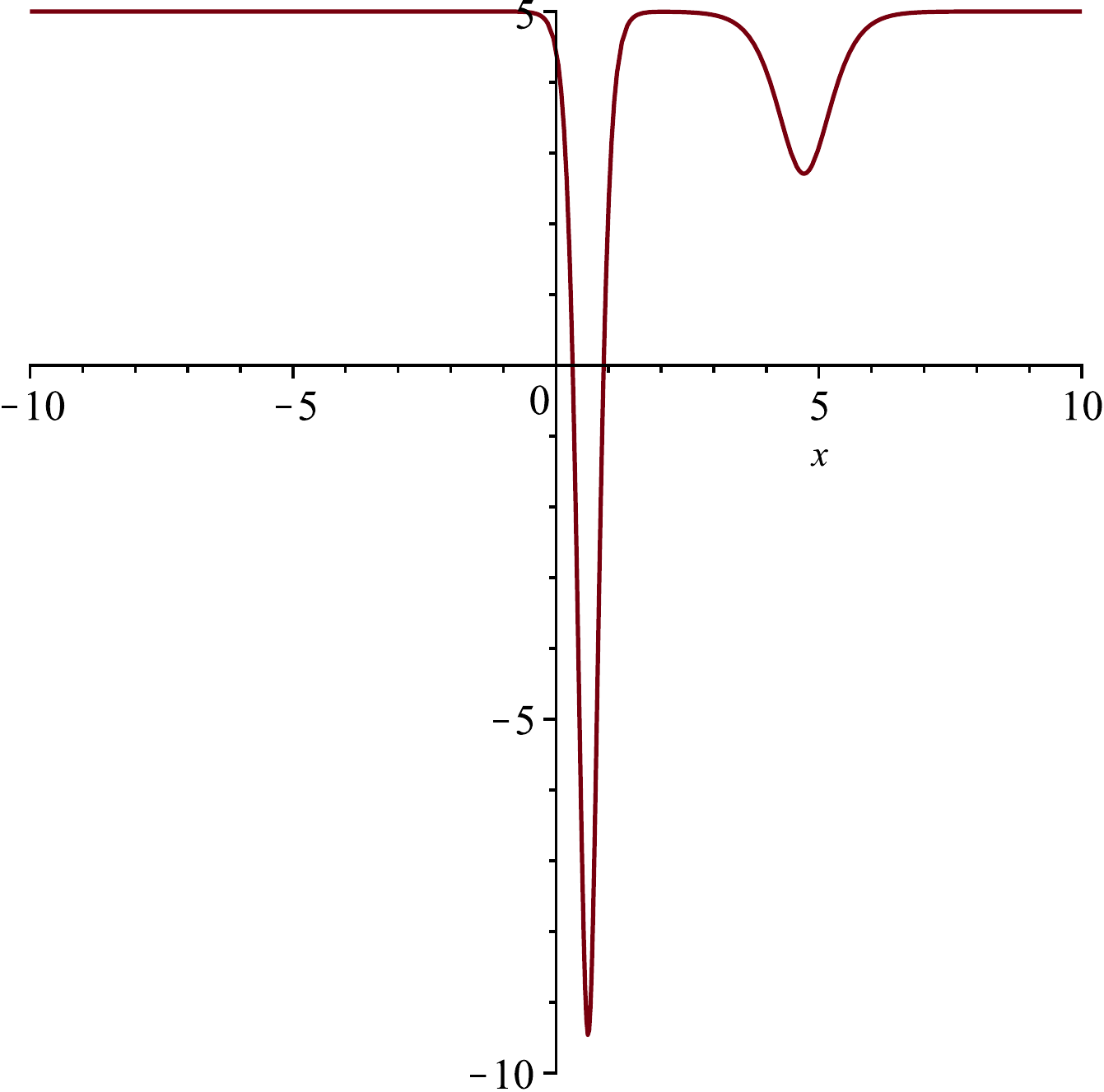} ~~~
        \includegraphics[width=3.6cm]{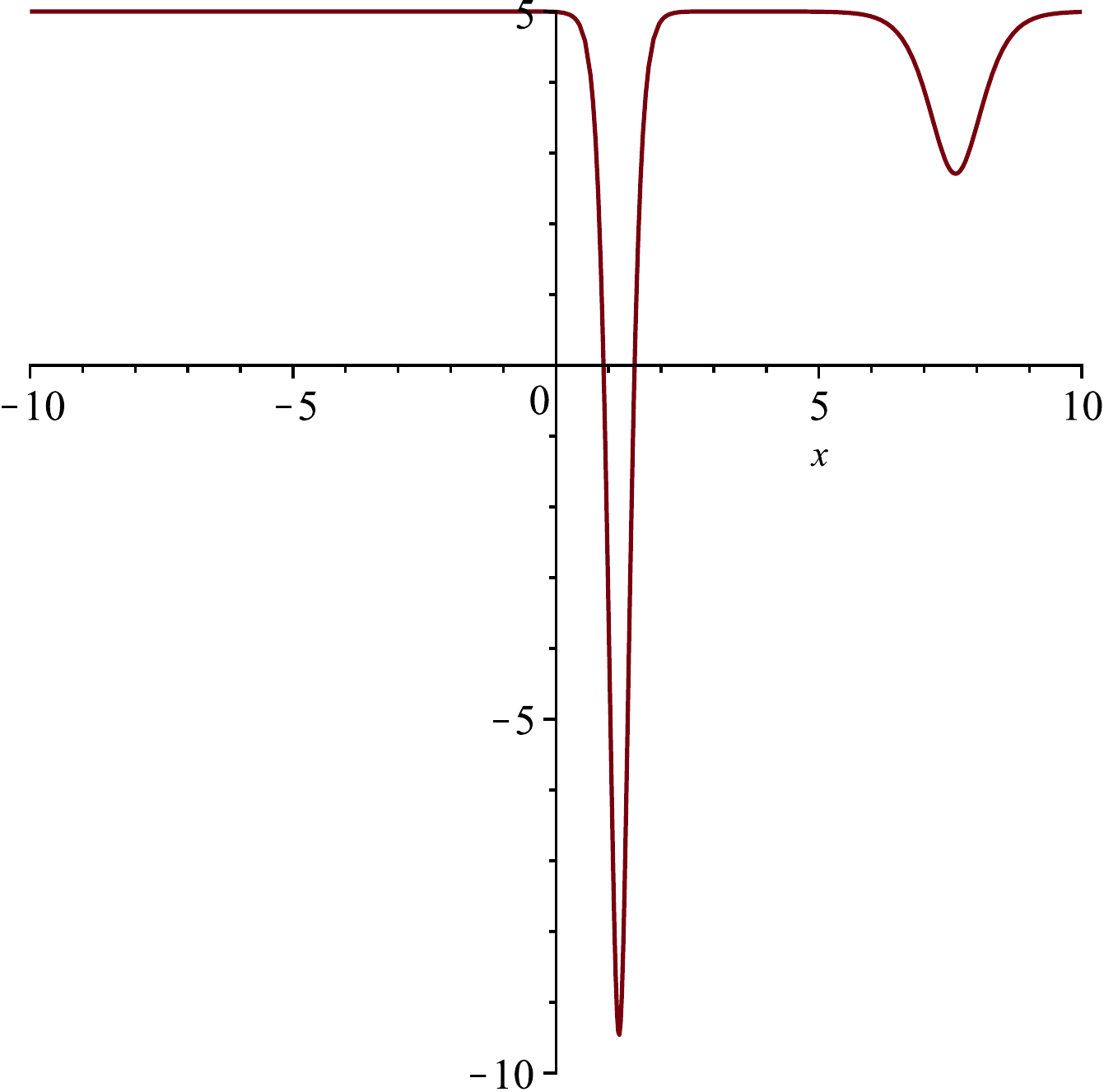} 
        }
      \caption{3 solitary waves merge to 2. Superposition of two solutions of type (\ref{yeq}) with
        $\theta=8,C_1=1,C_2=1,C_3=0$ and  $\theta=-12,C_1=-1,C_2=1,C_3=1$, for $\beta=5$. Plots
        of $u$ against $x$ for times $t=-1.8,-0.8,-0.16,0,0.2,0.5,1.1,1.8$. 
      }
\end{figure}





The superposition of $3$ 1 BT solutions of the form (\ref{yeq}), using equations (\ref{latt2}) and (\ref{pl3}),  
takes the form
\begin{equation}
f = \beta x  - 
\frac
{ \theta_1 y_1( y_2 y_{3x} - y_3 y_{2x}) 
+ \theta_2 y_2( y_3 y_{1x} - y_1 y_{3x})
+ \theta_3 y_3( y_1 y_{2x} - y_2 y_{1x} )
}
{
  y_1( y_{2x} y_{3xx} - y_{3x}y_{2xx} )
  + y_2( y_{3x} y_{1xx} - y_{1x}y_{3xx} ) 
  + y_3( y_{1x} y_{2xx} - y_{2x} y_{1xx}  ) 
} \ .  \label{3bt}
\end{equation}
Once again, we do not have a full analytic classification of solutions, but numerical experiments indicate that
this is simpler than for superpositions of 2 solutions. 3-soliton solutions  are obtained from (certain) superpositions of
2 standard solitons and 1 singular soliton. Solutions describing the merger of 4 waves to 3 are obtained from (certain)
superpositions of a standard soliton, a singular soliton, and a merging soliton. See
Figure 7 for an example. 

\begin{figure}
      \centerline{
        \includegraphics[width=3.6cm]{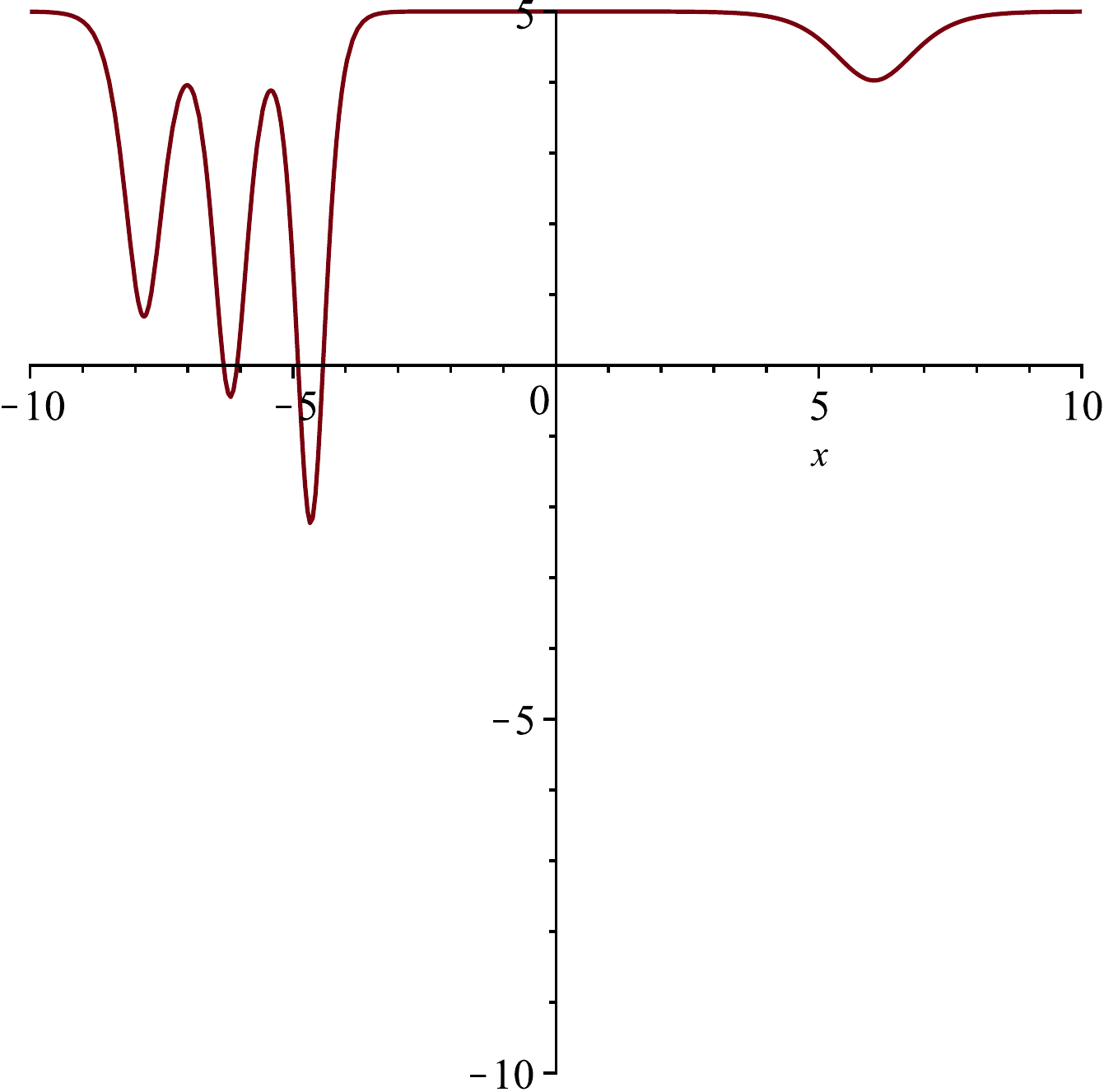} ~~~
        \includegraphics[width=3.6cm]{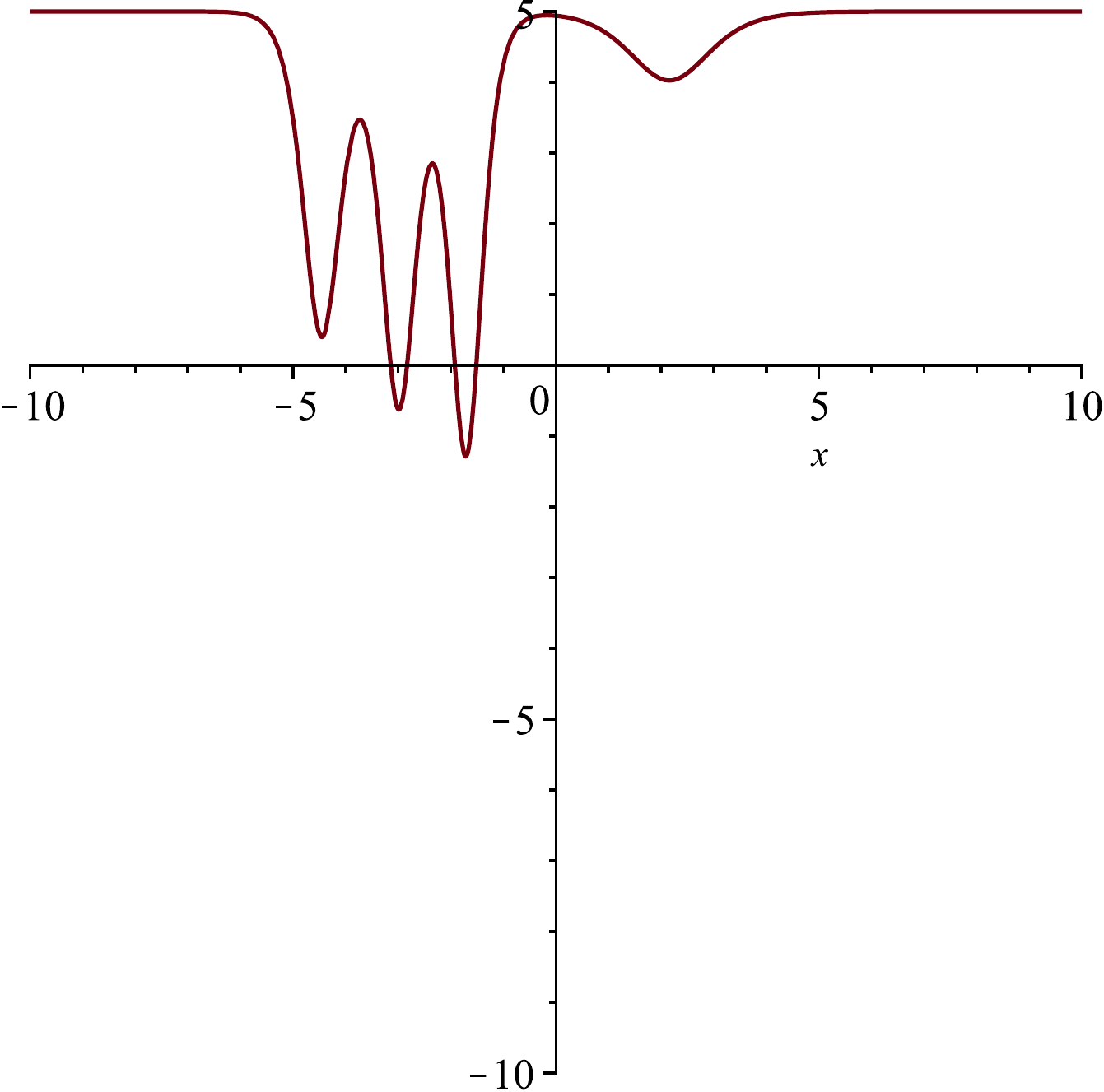} ~~~
        \includegraphics[width=3.6cm]{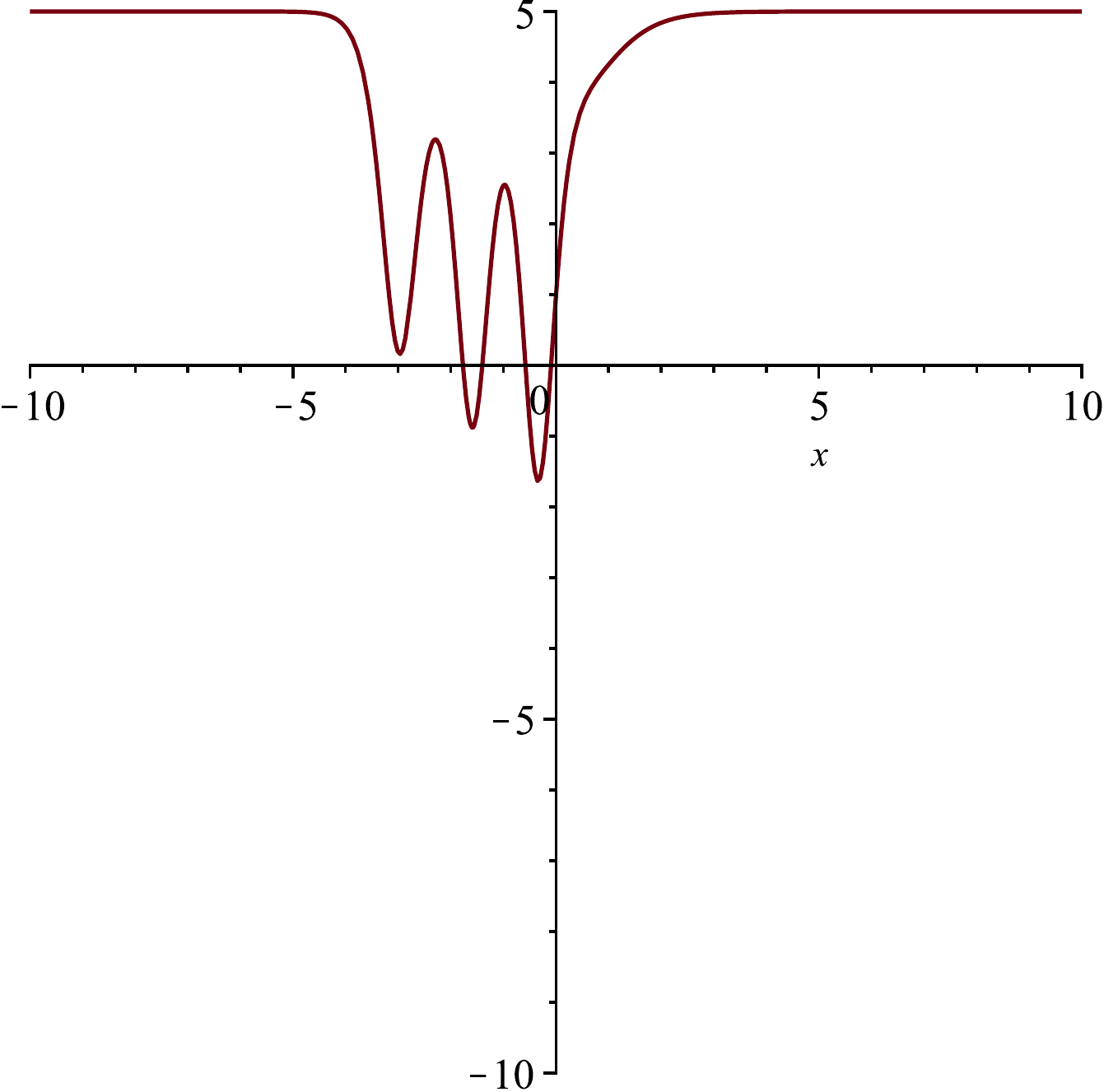} ~~~
        \includegraphics[width=3.6cm]{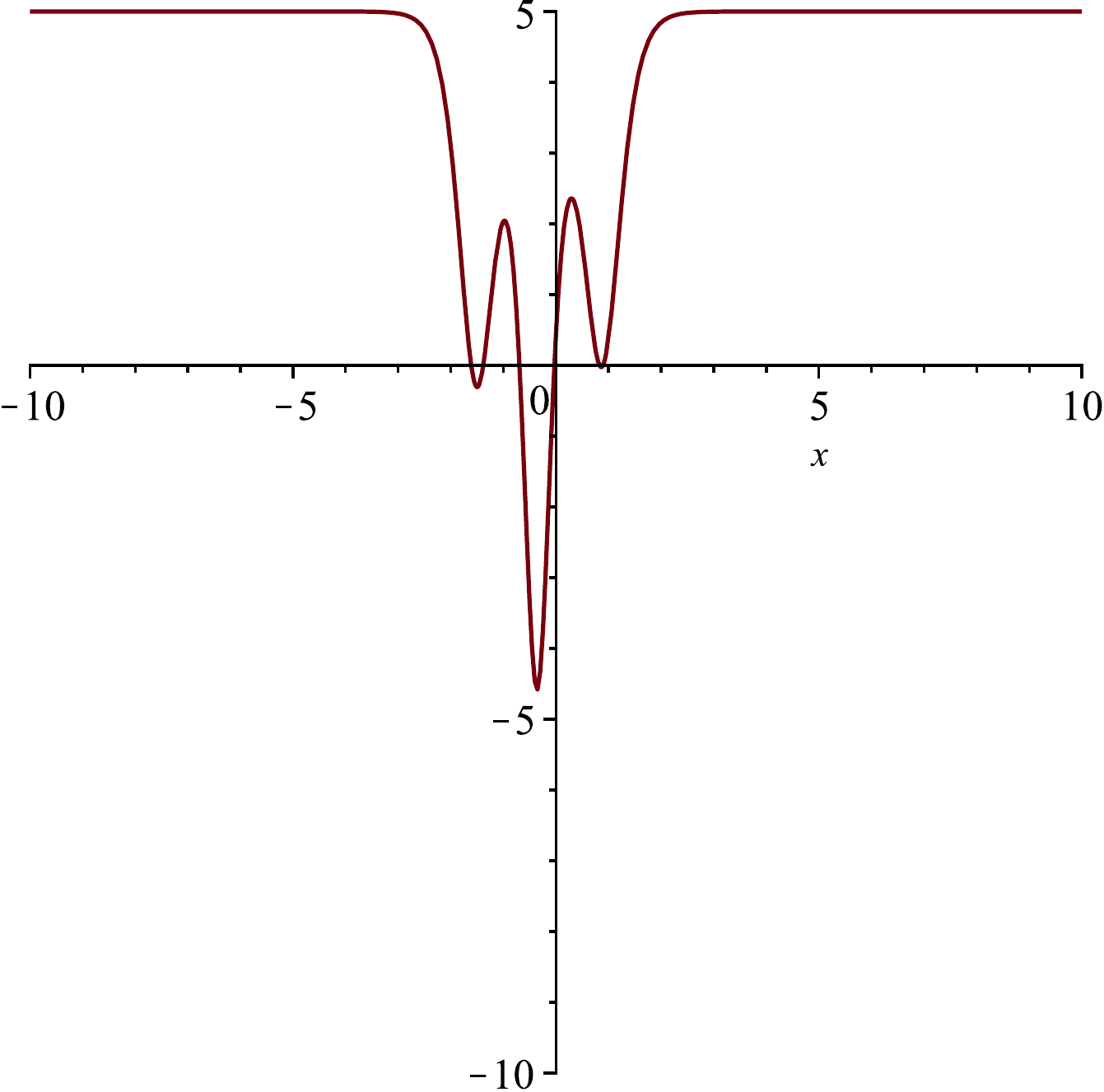} 
        }
      \centerline{
        \includegraphics[width=3.6cm]{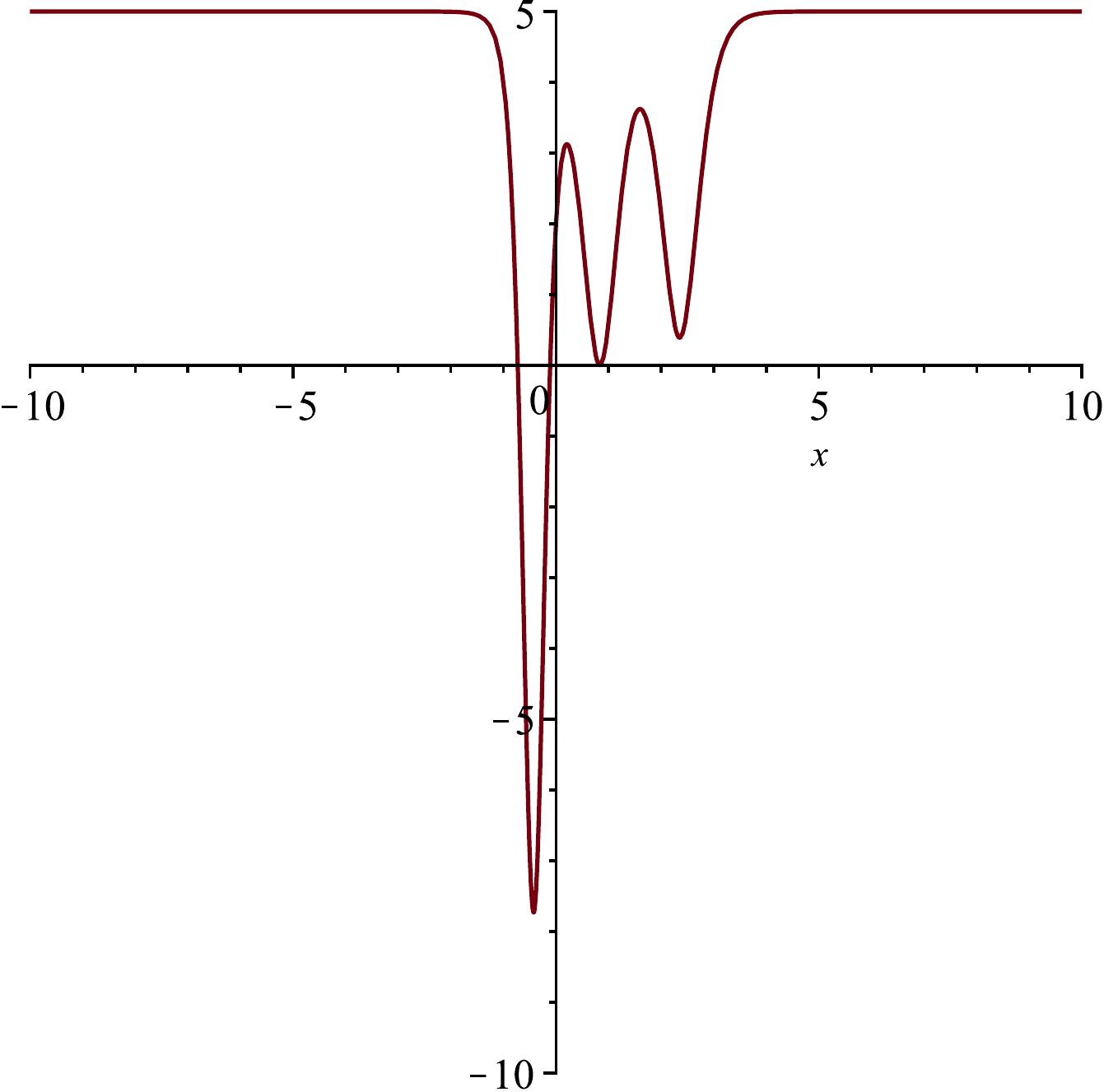} ~~~
        \includegraphics[width=3.6cm]{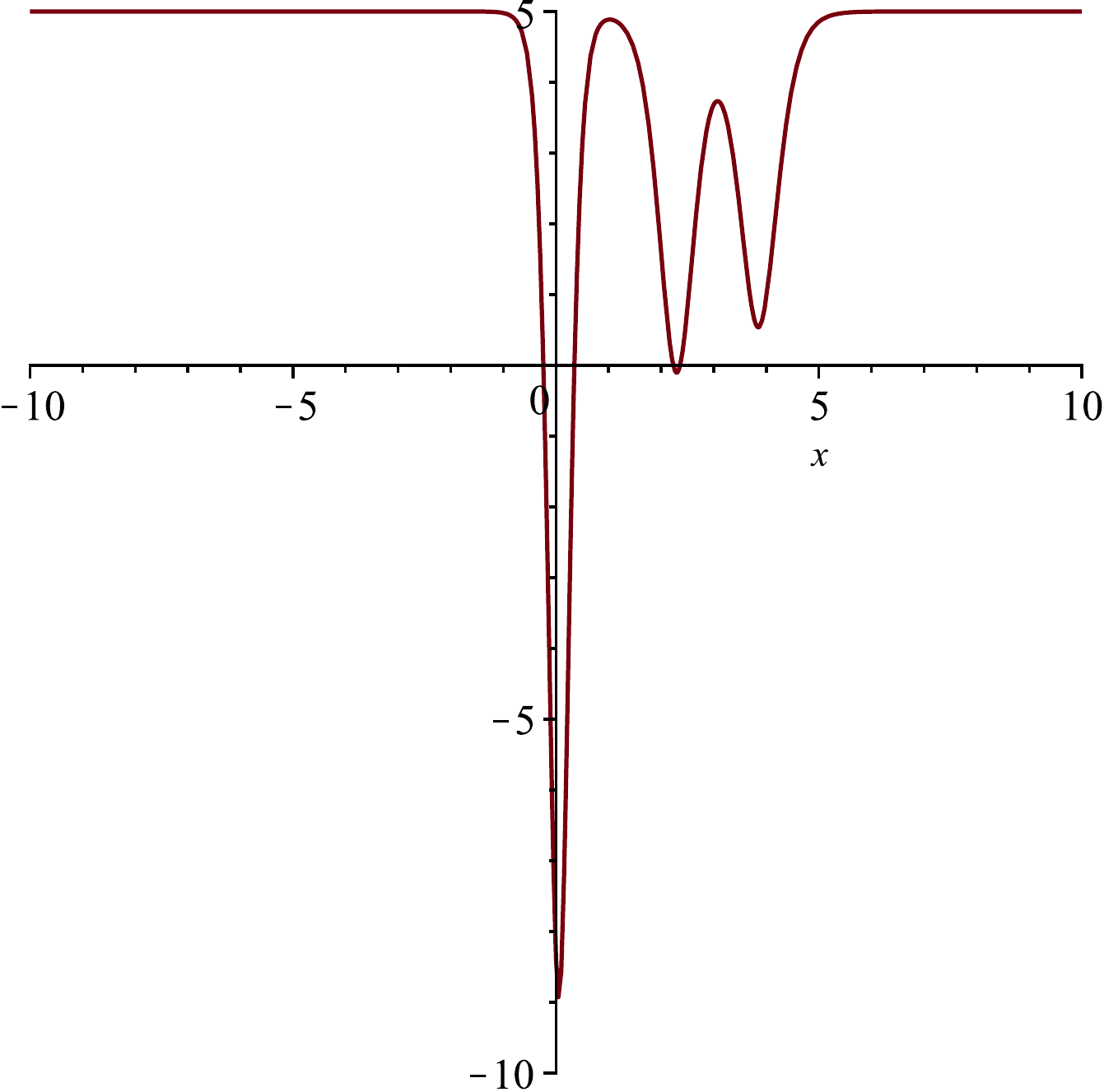} ~~~
        \includegraphics[width=3.6cm]{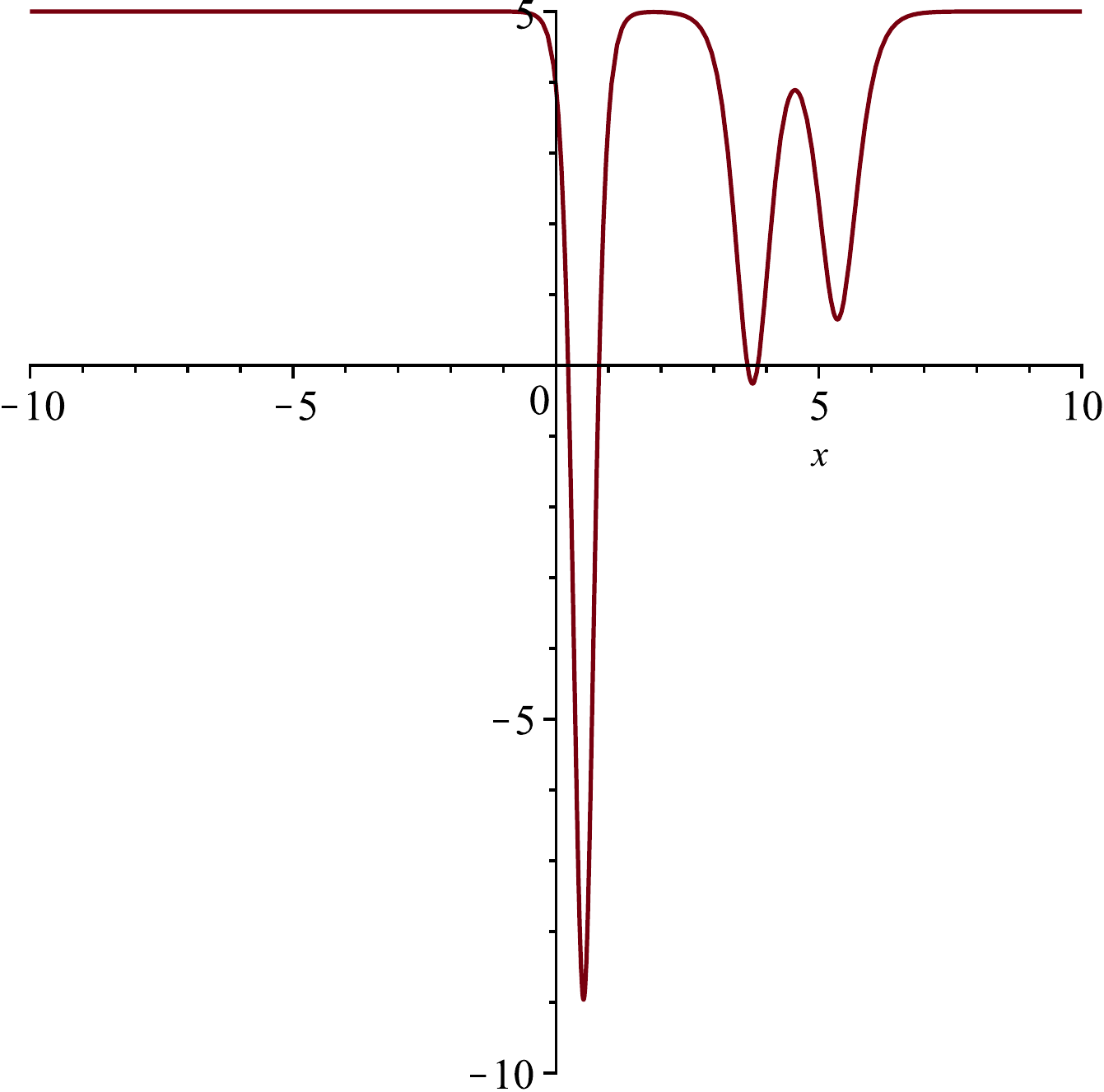} ~~~
        \includegraphics[width=3.6cm]{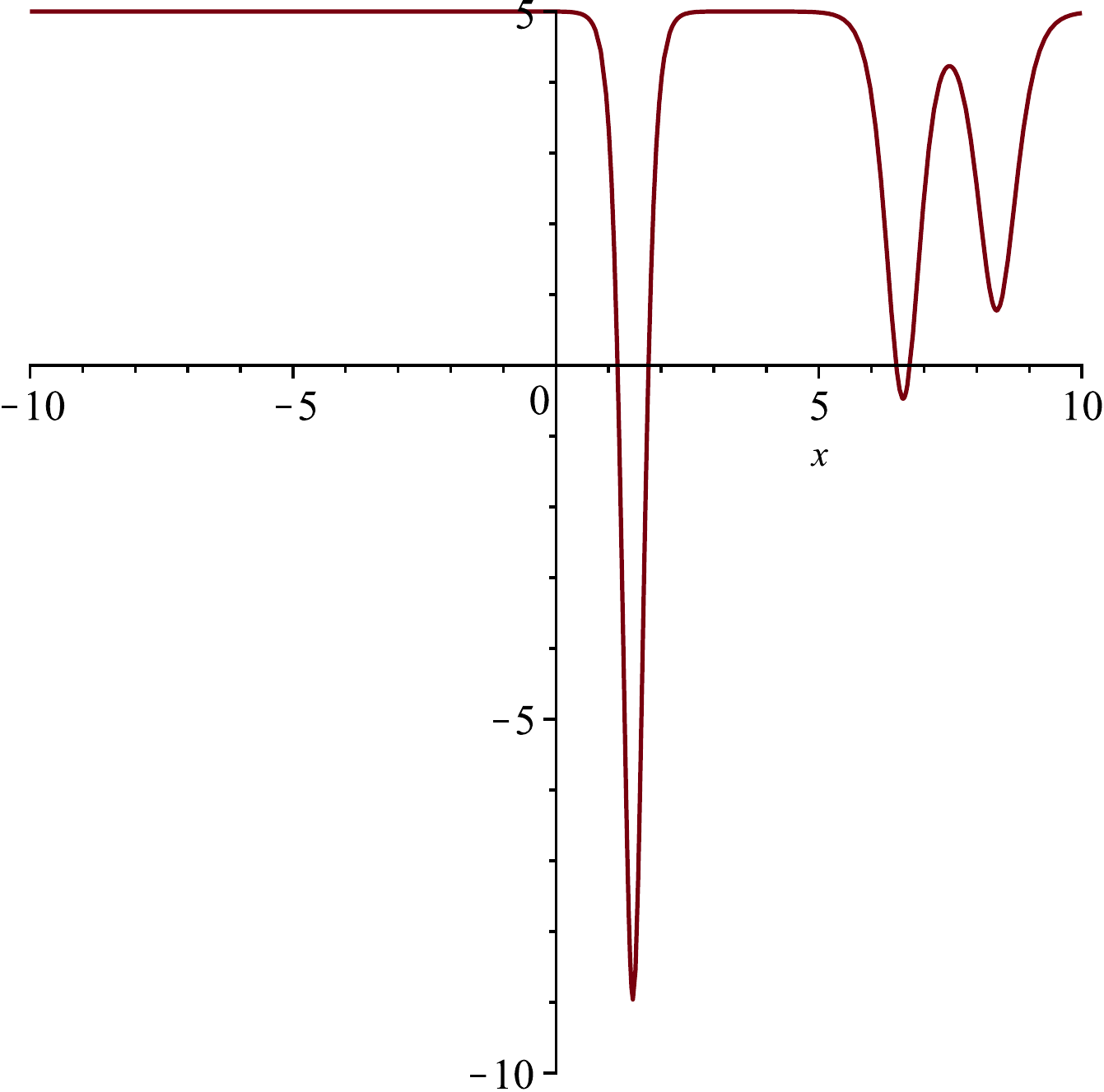} 
        }
      \caption{4 waves merge to 3. Superposition of three solutions of type (\ref{yeq}) with
        $\theta=-2,C_1=1,C_2=1,C_3=0$ (a soliton) 
        $\theta=-9,C_1=1,C_2=-1,C_3=0$ (a singular soliton) 
        and $\theta=-16,C_1=1,C_2=1,C_3=1$ (a merging soliton) 
        for $\beta=5$. Plots  of $u$ against $x$ for times $t=-1.7,-0.8,-0.4,0,0.4,0.8,1.2,2$. 
      }
\end{figure}

We have not succeeded in obtaining nonsingular solutions from a superposition using more than one merger-type
solution. However there are some remarkable singular solutions. In Figure 8 we present plots of the superposition of three
merger-type solutions, describing the evolution of 6 solitary waves into 3, via a brief, finite time duration 
singularity. The singularity forms after $t=-0.1$ and disappears before $t=0.3$. 

\begin{figure}
      \centerline{
        \includegraphics[width=3.6cm]{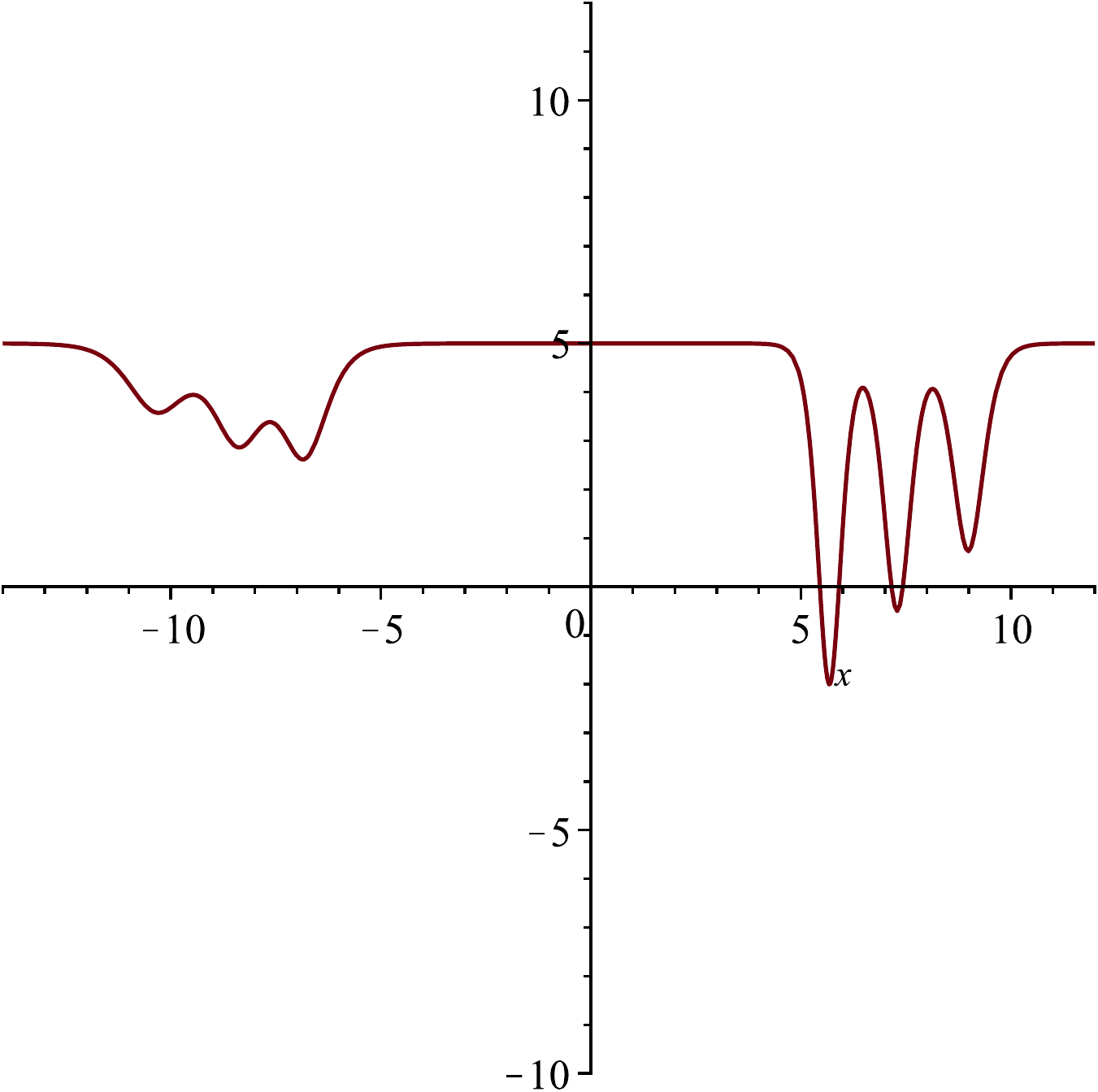} ~~~
        \includegraphics[width=3.6cm]{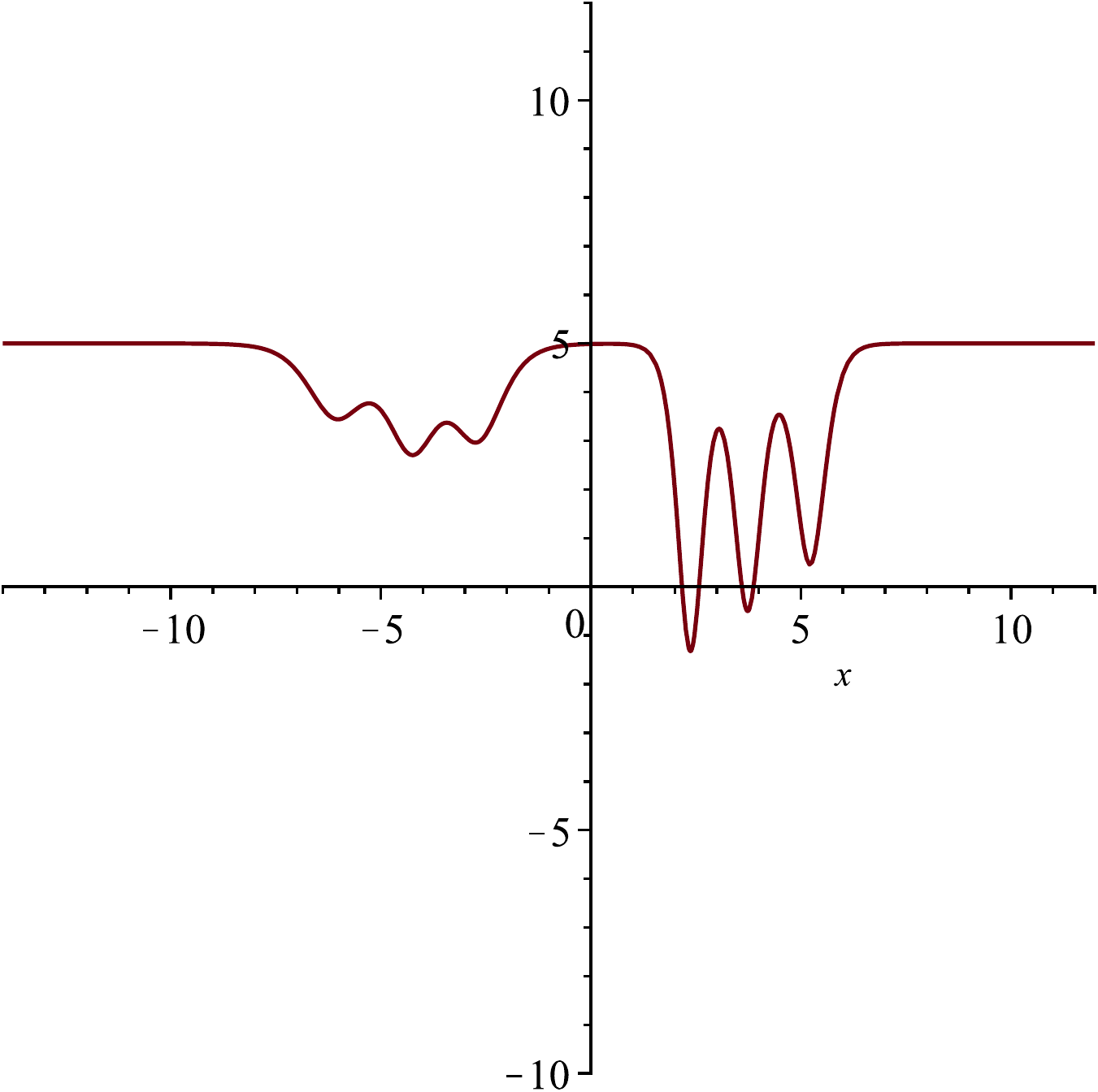} ~~~
        \includegraphics[width=3.6cm]{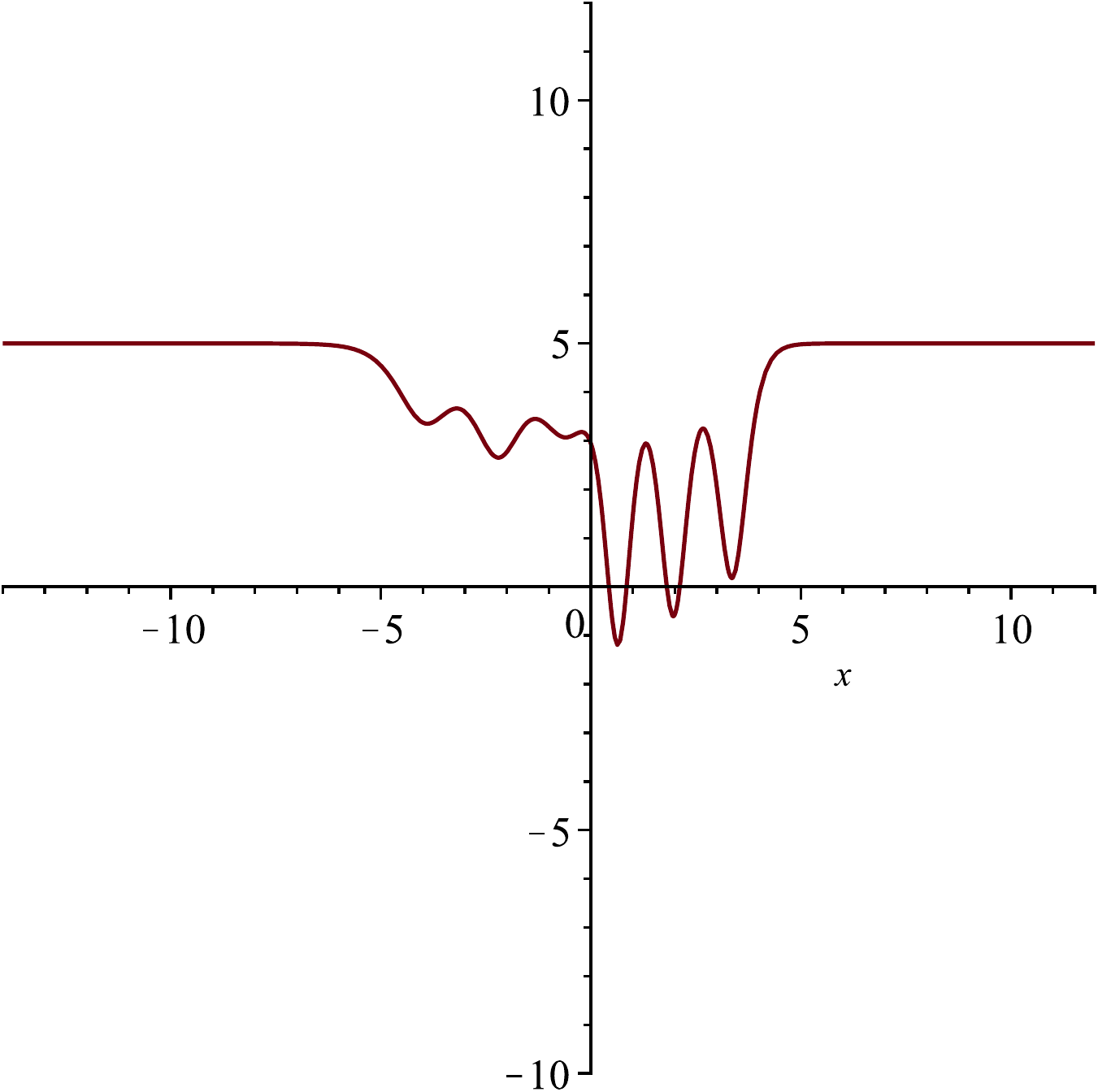} ~~~
        \includegraphics[width=3.6cm]{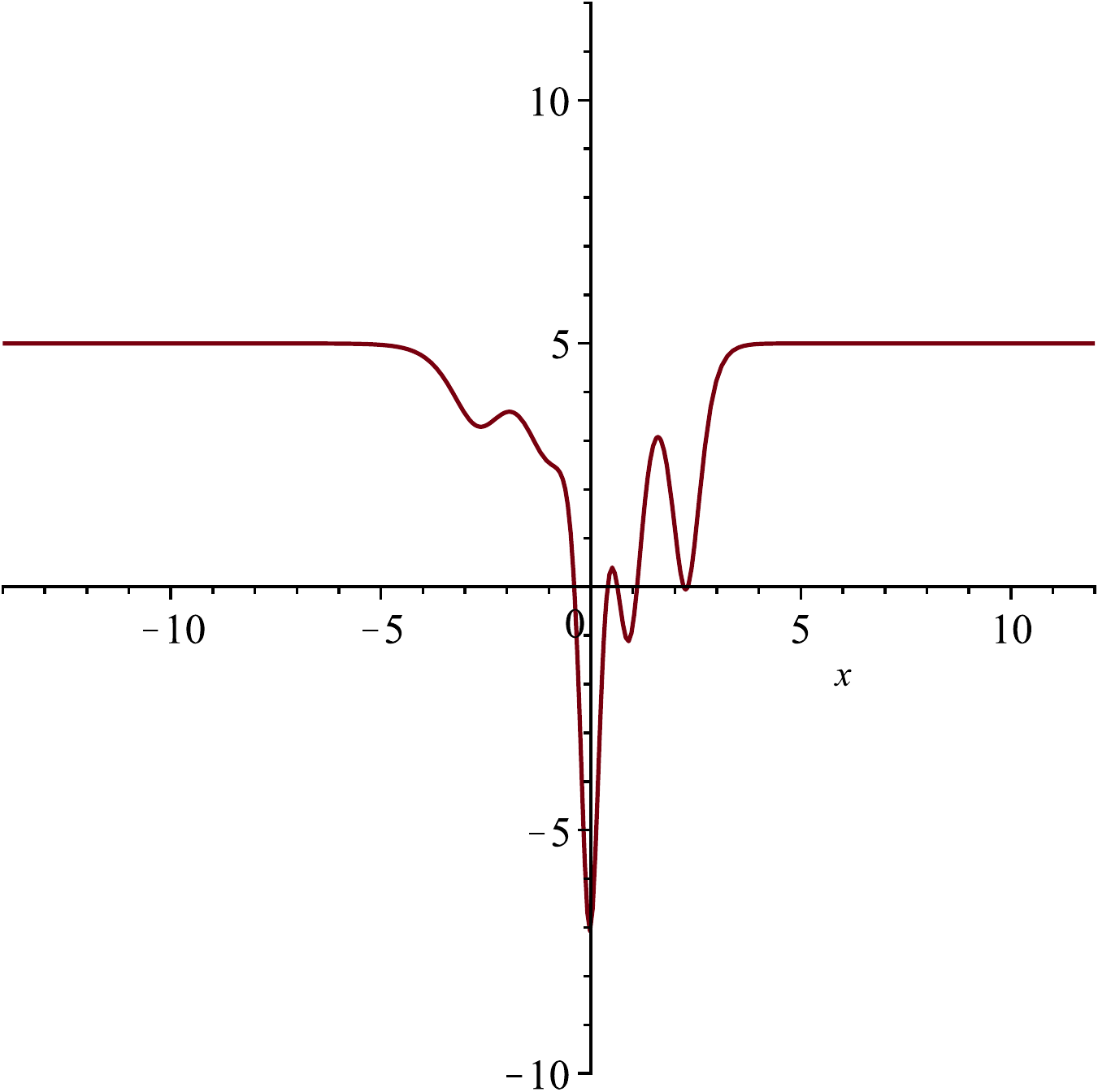} 
        }
      \centerline{
        \includegraphics[width=3.6cm]{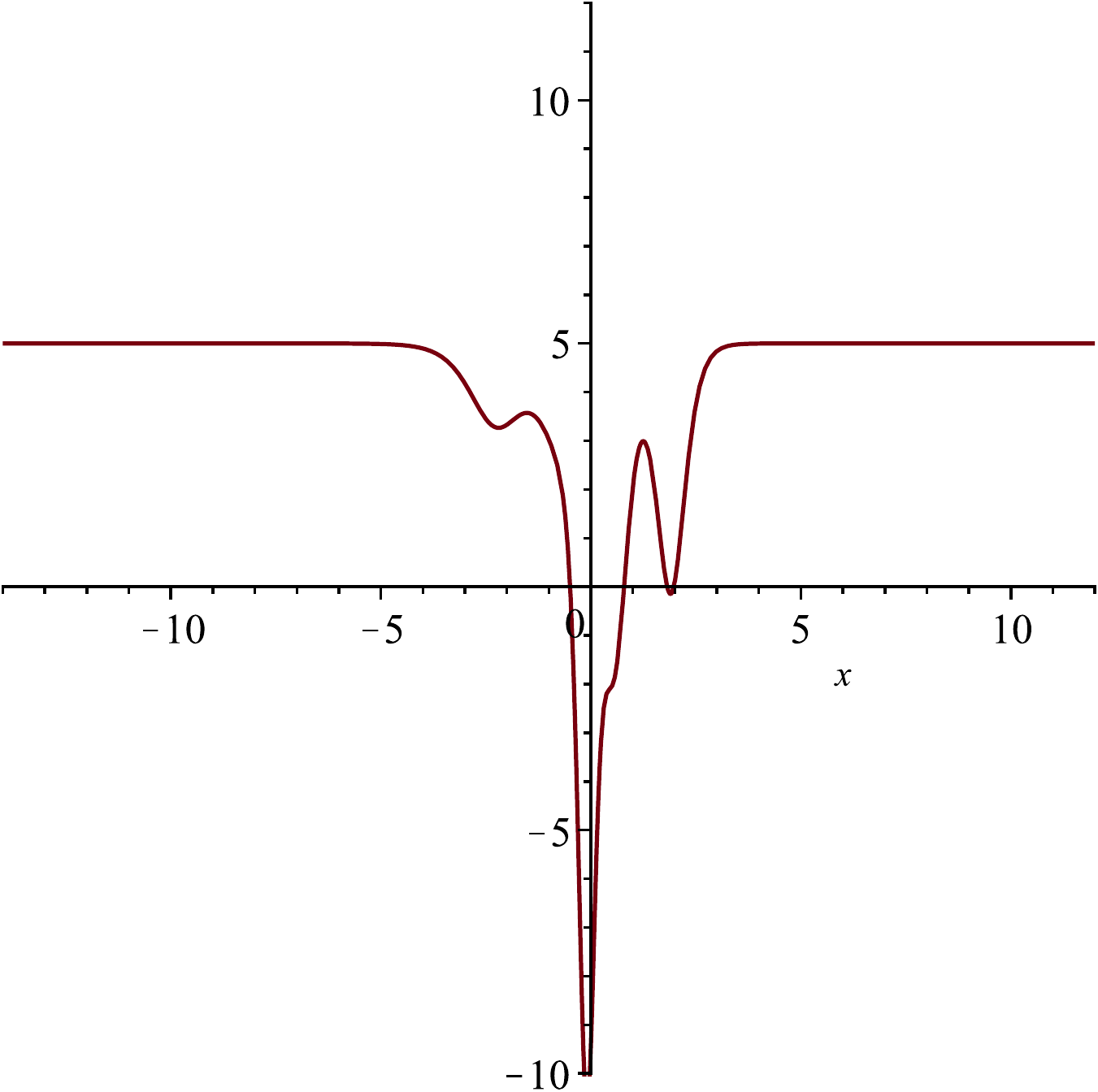} ~~~
        \includegraphics[width=3.6cm]{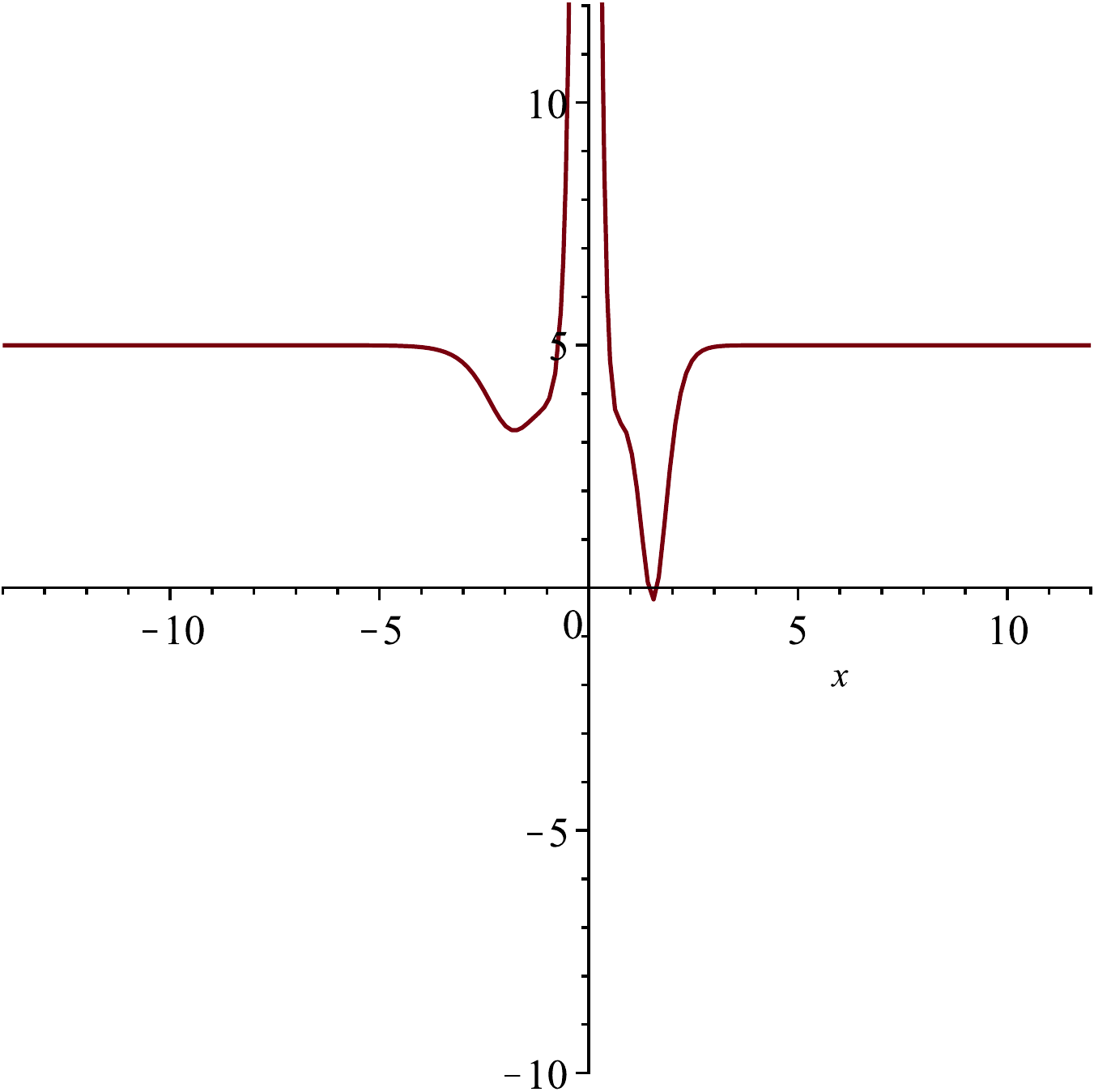} ~~~
        \includegraphics[width=3.6cm]{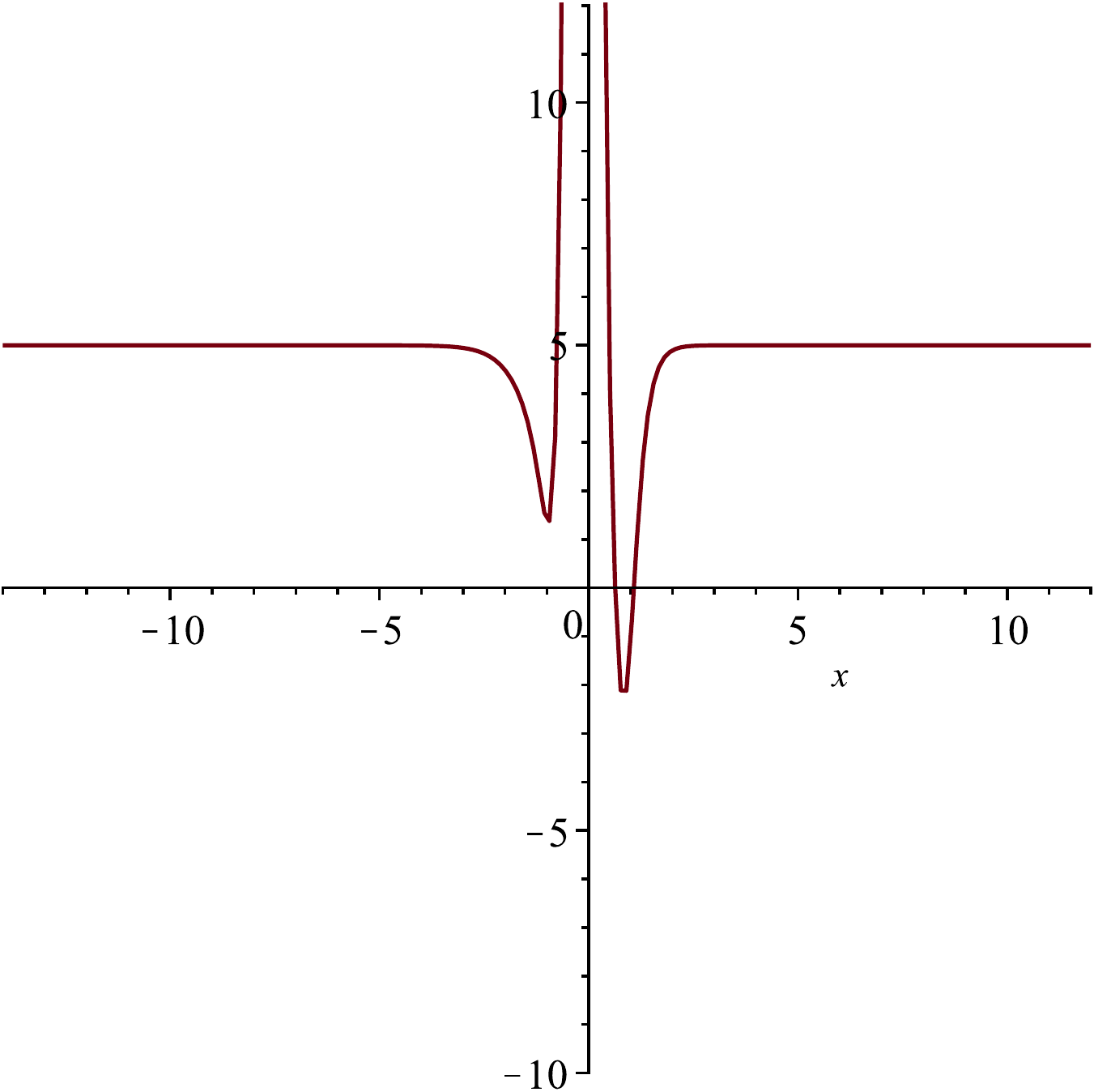} ~~~
        \includegraphics[width=3.6cm]{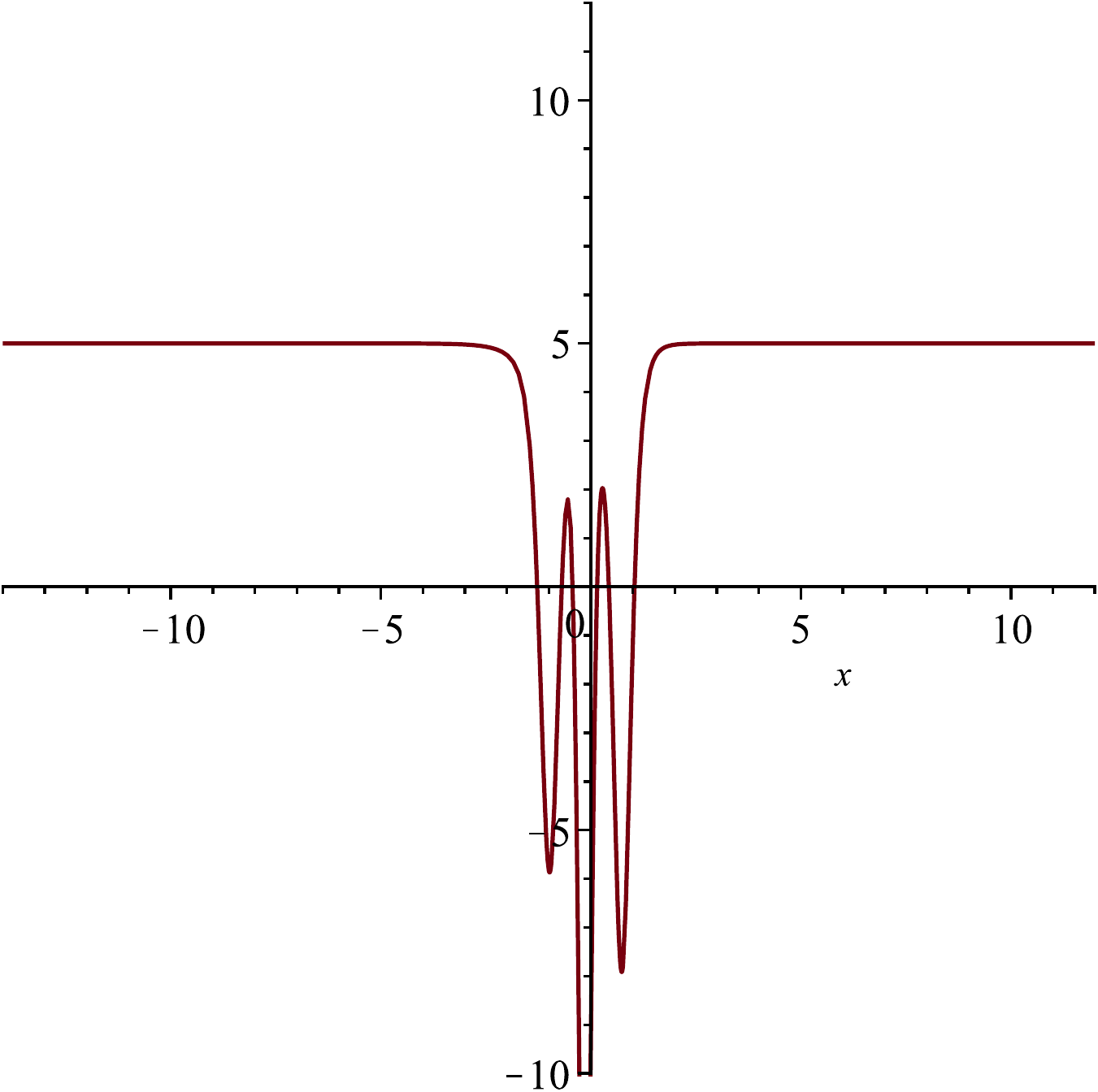} 
        }
      \centerline{
        \includegraphics[width=3.6cm]{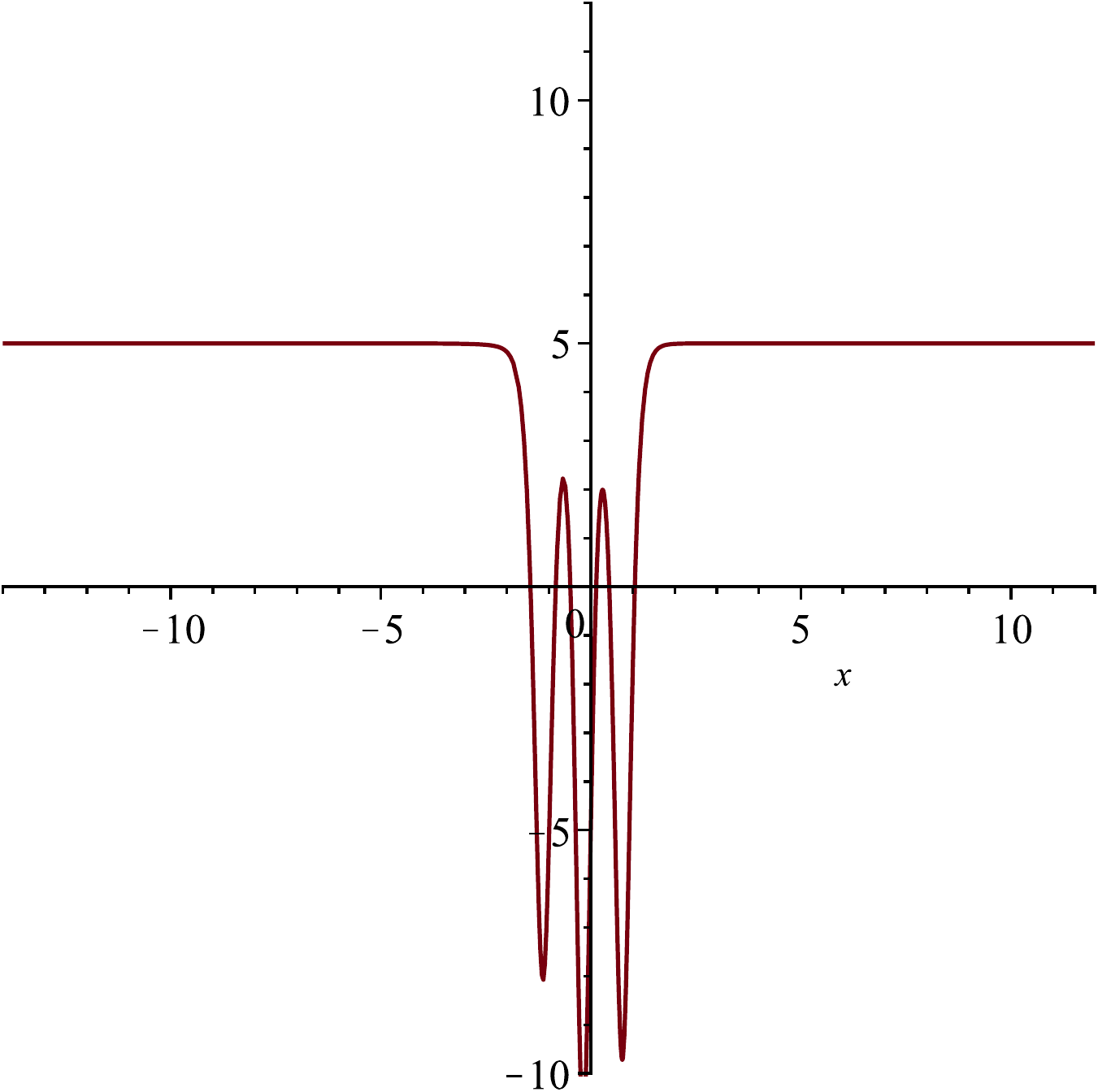} ~~~
        \includegraphics[width=3.6cm]{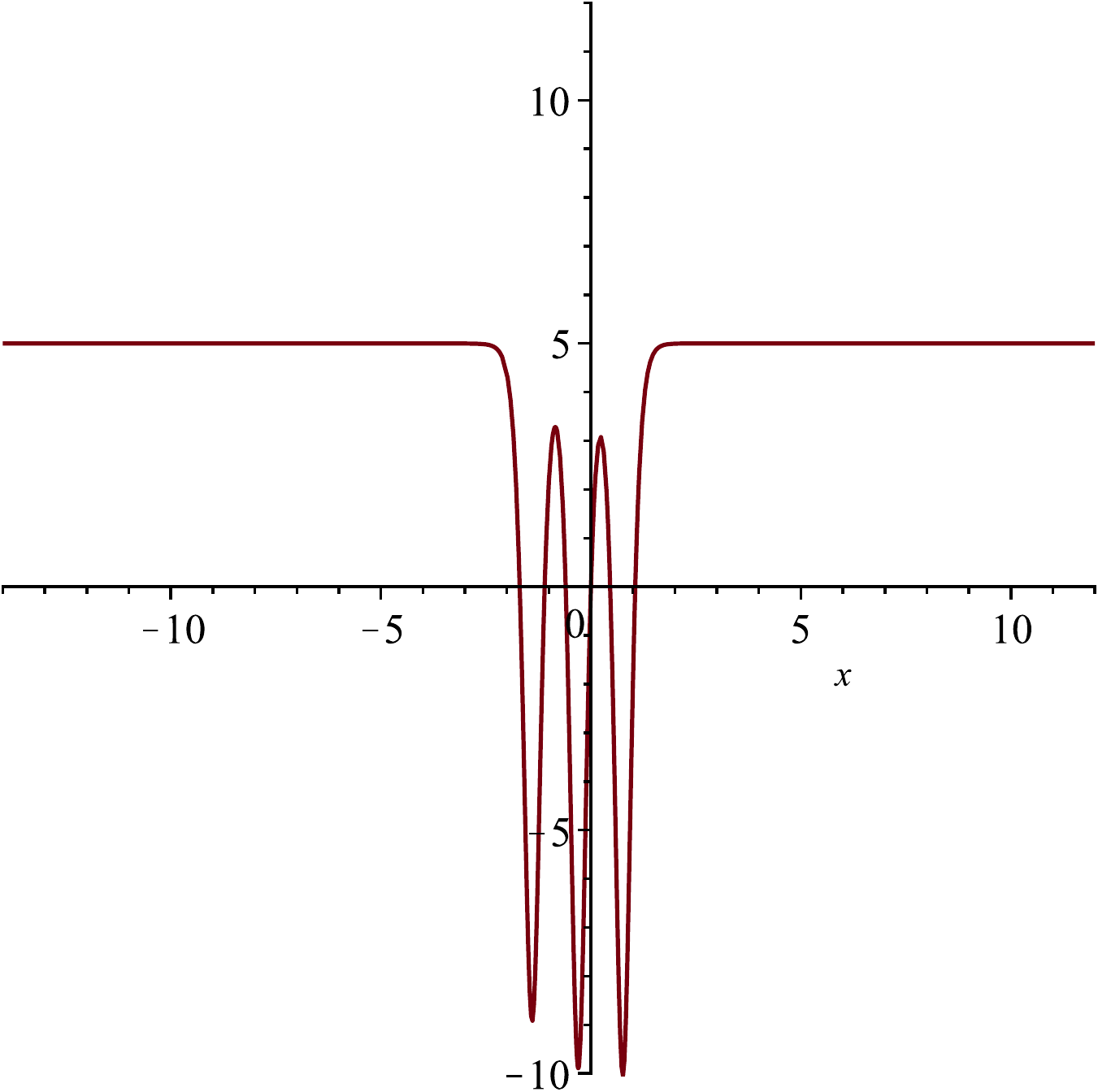} ~~~
        \includegraphics[width=3.6cm]{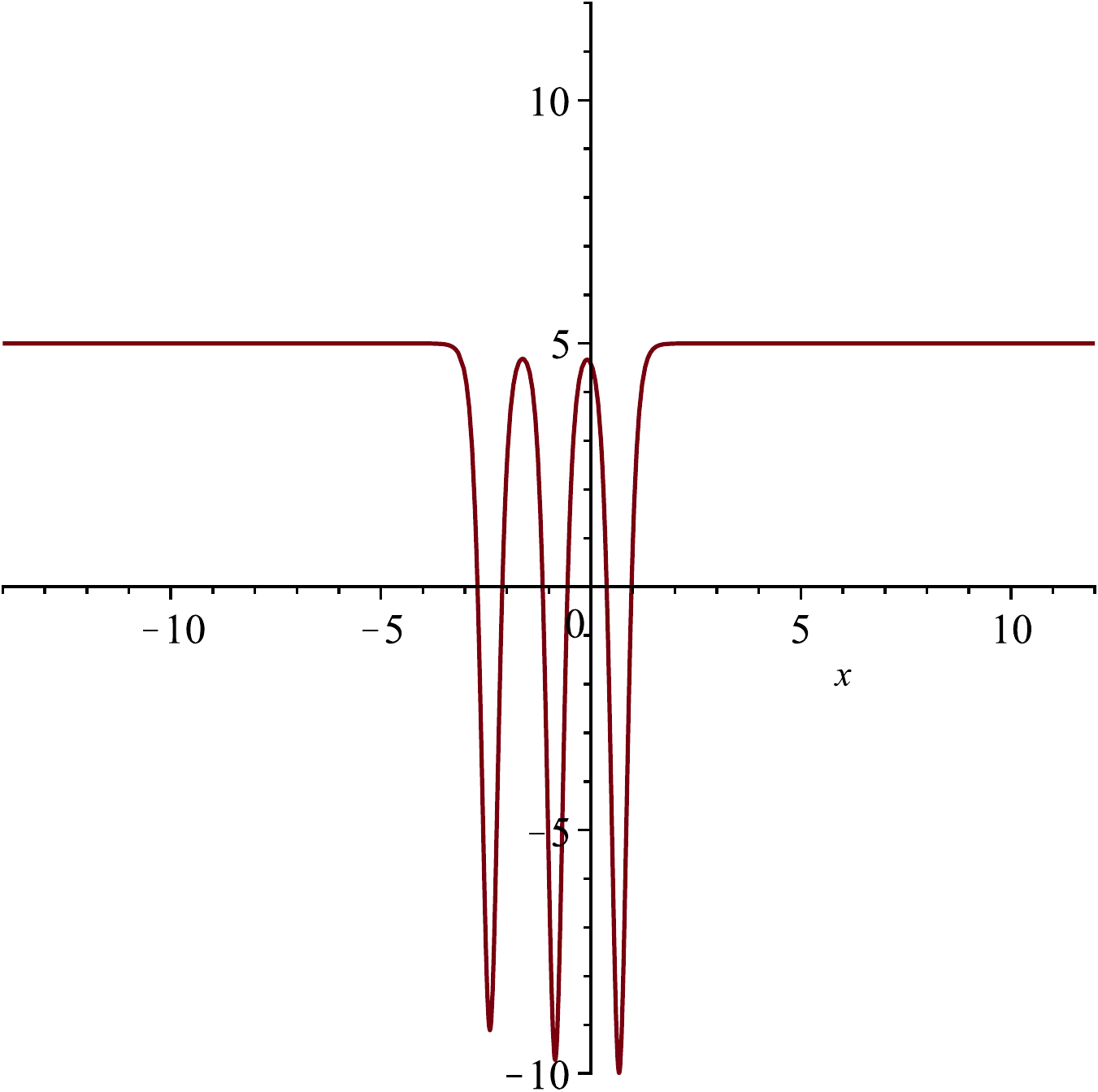} ~~~
        \includegraphics[width=3.6cm]{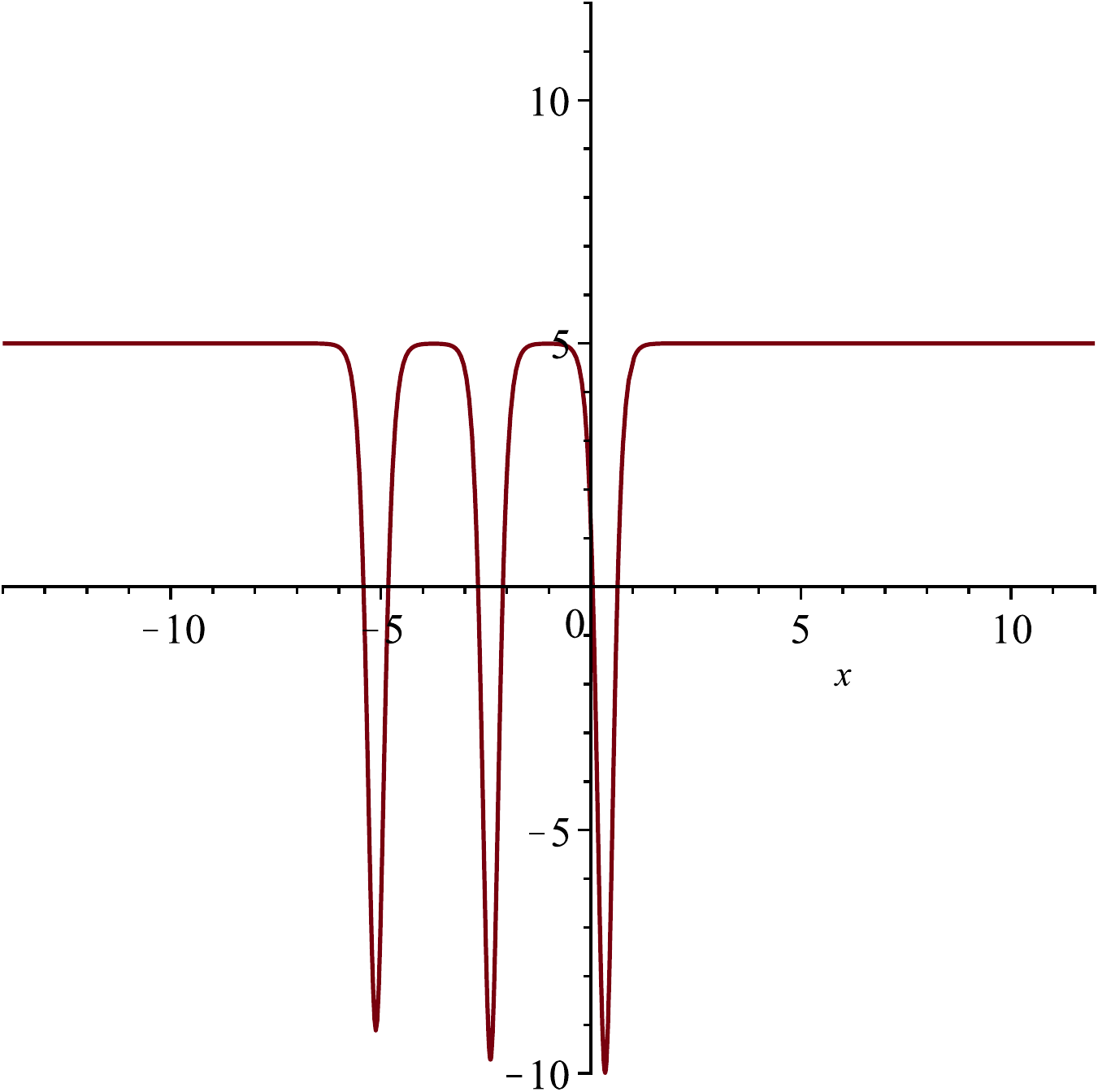} 
        }
      \caption{6 solitary waves merge to 3 via a singularity. Superposition of three solutions of type (\ref{yeq}) with
        $\theta=2,C_1=C_2=C_3=1$,  
        $\theta=9,C_1=1,C_2=-1,C_3=1$ 
        and $\theta=15,C_1=C_2=C_3=1$ 
        for $\beta=5$. Plots  of $u$ against $x$ for times $t=-2,-1,-0.5,-0.2,-0.1,0,0.2,0.3,0.4,0.6,1.5,4$. 
      }
\end{figure}


The forms (\ref{yeq}), (\ref{2bt}) and (\ref{3bt}) for the solutions obtain by $1,2$ or $3$ applications of the
BT suggest  that in general the form of the general solution obtained by $n$ applications of the BT to the starting
solution $f = \beta x$ 
should be
\begin{equation}
  f = \beta x- \frac{W_x}{W} \label{W}\end{equation}
where $W$ is the Wronskian
\begin{equation}
  W  = \det \left(
\begin{array}{cccc}
  y_1  & y_2  &   \ldots &  y_n  \\
  y_1'  & y_2'  &   \ldots &  y_n'  \\
  \vdots &   \vdots &  &   \vdots   \\
    y_1^{(n-1)}  & y_2^{(n-1)}  &   \ldots &  y_n^{(n-1)}  
\end{array}
\right) \label{W2}\end{equation}
and each of the functions $y_i$ are of the form in (\ref{yeq}), i.e. $y_i$ is a general solution of the
differential equation $y_i''' = 3 \beta y_i' + \theta_i y_i$. (In this paragraph we use primes to denote
differentiation with respect to $x$.) We prove this as follows. Assuming
$f=\beta x-\frac{W'}{W},
f_1=\beta x-\frac{W_1'}{W_1}, 
f_2=\beta x-\frac{W_2'}{W_2}, 
f_{12}= \beta x-\frac{W_{12}'}{W_{12}}$, 
substituting in (\ref{pl3}), simplifying and integrating once gives the requirement 
$$
W W_{12} = K ( W_1 W_2' - W_2 W_1' ) 
$$
where $K$ is an arbitrary constant (note that each of the $W$'s is only defined up to an overall constant). 
Now if $W,W_1,W_2,W_{12}$ all have the form of Wronskians, of dimensions $n-2,n-1,n-1,n$ respectively,
with $W_{12}$ being exactly the determinant of the matrix in (\ref{W2}), then
\begin{itemize}
\item $W$ is the determinant of the same matrix with the $(n-1)$'th and $n$'th rows and columns deleted\ ,
\item $W_1$ is the determinant of the same matrix with the  $n$'th row and $n$'th column deleted\ ,
\item $W_2$ is the determinant of the same matrix with the  $n$'th row and $(n-1)$'th column deleted \ ,
\item $W_1'$ is the determinant of the same matrix with the $(n-1)$'th row and $n$'th column deleted\ ,
\item $W_2'$ is the determinant of the same matrix with the $(n-1)$'th row and $(n-1)$'th column deleted\ .
\end{itemize}
The desired identity therefore follows from a case of Sylvester's theorem for determinants \cite{g}, that 
if $A$ is an arbitrary $n\times n$ matrix, 
$C$ is the same matrix with the $(n-1)$'th and $n$'th rows and columns deleted,
$B_1$ is the same matrix with the  $n$'th row and $n$'th column deleted,
$B_2$ is the same matrix with the $(n-1)$'th row and $n$'th column deleted,
$B_3$ is the same matrix with the  $n$'th row and $(n-1)$'th column deleted and 
$B_4$ is the same matrix with the $(n-1)$'th row and $(n-1)$'th column deleted, then 
$$ \det C \det A = \det B_1 \det B_2 - \det B_3 \det B_4 \ .  $$ 
The general result on the form of the solution follows by induction.  


A similar result appeared for the bad BEq in \cite{beq20}, however taking each $y_i$ to be the sum of only two exponentials. 
In \cite{beq34}, Hirota gave the general solution of the bad BEq using the ``Hirota method'' and it is interesting to see
how this works for the good BEq. For this paragraph we work directlty with the BEq in the form (\ref{be}), with
$\beta>0$. Writing $U = - (\log \tau)_{xx}$, the equation becomes
$$
\tau \tau_{tt} - \tau_t^2 - 4 \beta \left(\tau \tau_{xx} - \tau_x^2 \right)
+ \frac13\left(\tau \tau_{xxxx} - 4 \tau_{x}\tau_{xxx} + 3\tau_{xx}^2\right)
= 0 
$$
which is in ``Hirota bilinear form''. This has ``multisoliton'' solutions in the usual form 
$$\tau = 1 + \sum_i c_ie^{\eta_i} + \sum_{i<j} c_ic_j\phi_{ij} e^{\eta_i+\eta_j}
   + \sum_{i<j<k} c_ic_jc_k \phi_{ij}\phi_{jk}\phi_{ki} e^{\eta_i+\eta_j+\eta_k} + \ldots $$
where $\eta_i = a_i(x+b_it)$, $a_i,b_i,c_i$ constants, with $a_i^2 + 3b_i^2 = 12\beta$, and
$$ \phi_{ij} = - 
\frac{(a_i-a_j)^4 -12\beta(a_i-a_j)^2 + 3(a_ib_i-a_jb_j)^2}{(a_i+a_j)^4 -12\beta(a_i+a_j)^2 + 3(a_ib_i+a_jb_j)^2}\ . 
$$
To guarantee that all these solutions are nonsingular requires $\phi_{ij}>0$ for all choices of the constants
$a_i,a_j,b_i,b_j$ (in addition to choosing the constants $c_i>0$) 
and that is not the case here. Furthermore, it is possible to choose $a_i,a_j,b_i,b_j$ such
that $\phi_{ij}=0$. A straightforward calculation shows that if this happens then
$b_i^2 + b_ib_j + b_j^2 = 3\beta$, implying that there is some constant $\theta$ for which
$b_i,b_j$ are distinct solutions of the cubic equation $b^3 = 3\beta b + \theta$. This is the
origin of the merging soliton solutions in the the Hirota framework. 

\section{Conservation Laws and Symmetries}    

The remarkably simple method for finding conservation laws from a BT is very old,  see for example \cite{beq44}.
For the BEq, we simply need to observe that (\ref{sxx1})-(\ref{st1}) implies 
$$   s_t + (2 u - s_x - s^2)_x  = 0 \ . $$
Thus $s$, which depends on $\theta$,  provides a generating function for (densities of) conservation laws.
To obtain the standard conservation laws, 
observe that the solution $s$ to (\ref{sxx1}) can be written as an asymptotic series in $\theta$
in the form 
$$
s \sim  \sum_{i=-1}^{\infty}\theta^{-i/3}s_i \ . 
$$
Each of the coefficients $s_i$ is the density for a conservation law. 
The first few coefficients are given as follows: 
$$
s_{-1}^3 = 1 \ , \quad
s_{0}=0\ ,\quad
s_1=\frac{u}{s_{-1}}\ , \quad 
s_2=-\frac{v+u_x}{s_{-1}^2} \ .
$$
Further terms can be computed using the recurrence relation
$$
s_{k+2} =\frac{1}{3s_{-1}^2}
   \left(
   3us_{k}
  - \sum _{j=-1}^{k+1} \left( \sum_{i=\max(-1,-j-1)}^{\min(k+1,k-j+1)}s_{k-i-j}s_{i}s_{j} \right)
  -3\sum _{i=-1}^{k+1}s_{i}s_{(k-i)x}
  -s_{kxx} 
  \right)\ ,\quad k\geq1.
$$
So for example 
\begin{eqnarray*}
s_3&=&\frac{2}{3}u_{xx}+v_x\ , \\
s_4&=&\frac{1}{3s_{-1}}(3uv-u_{xxx}-2v_{xx})\ , \\
s_5&=&\frac{1}{9s_{-1}^2}(u_{xxxx}+3v_{xxx}-3u^3+3uu_{xx}-9uv_x-9v^2-18u_xv) \ . 
\end{eqnarray*}
For each $i=1,2,\ldots$, $s_i$ is the density $F$ of a conservation law $F_t + G_x = 0$. 
For $i=3$ the conservation law is evidently trivial ($F=H_x$, $G=-H_t$ for some $H$). Indeed we will
shortly show that all the conservation laws for $i=3,6,9,\ldots$ are trivial. The associated
flux $G$ is the  coefficient of $\theta^{-i/3}$ in $2u-s_x-s^2$.  Thus for $i=1,2,4,5$ we have fluxes 
\begin{eqnarray*}
  G_1&=&\frac{1}{s_{-1}}(u_x+2v) \ ,  \\ 
  G_2&=&-\frac{1}{s_{-1}^2}\left(u^2+v_x+\frac{1}{3}u_{xx}\right)\ , \\  
  G_4&=&\frac{1}{s_{-1}}\left(\frac{2}{3}u^3-uv_x-2uu_{xx}-u_x^2+u_xv+v^2+\frac{1}{9}u_{xxxx}\right)\ ,\\  
  G_5&=&\frac{1}{9s_{-1}^2}\left(18u^2v-9u^2u_x-3uu_{xxx}-21uv_{xx} -30u_{xx}v -36vv_x-9u_xu_{xx}-45u_xv_x\right.\\
     && \left. +3v_{xxxx}+u_{xxxxx}\right) \ . 
\end{eqnarray*}

We note there are $3$ possible series for $s$, corresponding to the $3$ possible choices of $s_{-1}$.
The dependence of $s_1,s_2,\ldots$ on the choice of $s_{-1}$ is clear, and can be verified to be
consistent with the recursion relation. We denote the three solutions of (\ref{sxx1}) with
these three asymptotic series by $s^{(1)},s^{(2)},s^{(3)}$. If we
define $\sigma =  s^{(1)} + s^{(2)} + s^{(3)} $ then $\sigma$ has 
asymptotic series $\sum_{i=1}^\infty 3s_{3i} \theta^{-i} $.  However, if we define 
\begin{equation}
A=(s^{(2)}-s^{(3)})s^{(1)}_x+(s^{(3)}-s^{(1)})s^{(2)}_x+(s^{(1)}-s^{(2)})s^{(3)}_x
  - (s^{(1)}-s^{(2)}) (s^{(2)}-s^{(3)})(s^{(3)}-s^{(1)})\ , 
\label{Aref}\end{equation}
then it can be verified (using (\ref{sxx1}) for each of the functions $s^{(1)},s^{(2)},s^{(3)}$)
that
\begin{equation}
\sigma = s^{(1)}+s^{(2)}+s^{(3)} = -(\log A)_x \ . 
\label{Aid}\end{equation}
It follows that $s_{3i}$ is a total $x$ derivative for all $i$, and the associated
conservation laws are trivial. 

The use of a BT to generate symmetries is rather newer \cite{rs1}. The critical observation made in
\cite{rs1} for the KdV, Sine Gordon and Camassa Holm equations, was that while individual BTs are
not  ``small'' transformation (and thus not directly related to symmetries, which are transformations
of solutions that are infinitesimally close to the identity), the composition of two BTs can be small 
in this sense. For the BEq this is not the case, and it is necessary to consider the composition
of 3 BTs. This has its origins in the fact that the Lax pair for the BEq is a $3\times 3$ matrix Lax pair. 
Equation (\ref{latt2}) for a triple BT can be written, using (\ref{pl2}), in the
form 
\begin{eqnarray}
f_{123} &=& f + \frac{(\theta_3-\theta_2)f_1+(\theta_1-\theta_3)f_2+(\theta_2-\theta_1)f_3}
{(f_2^2-f_3^2-f_{2x}+f_{3x})f_1+(f_3^2-f_1^2-f_{3x}+ f_{1x})f_2+(f_1^2 -f_2^2-f_{1x}+f_{2x})f_3} \nonumber \\
&=& f 
-\frac{(\theta_3-\theta_2)s_1+(\theta_1-\theta_3)s_2+(\theta_2-\theta_1)s_3}
     {(s_2-s_3)s_{1x}+(s_3-s_1)s_{2x}+(s_1-s_2)s_{3x}-(s_1-s_2)(s_2-s_3)(s_3-s_1)}\ .  \label{symgen}
\end{eqnarray}
The critical observation is that as $\theta_2,\theta_3$ tend to $\theta_1$, the numerator of the
second term becomes small, but the denominator can remain large by taking $s_1,s_2,s_3$ 
to be distinct solutions of (\ref{sxx1})-(\ref{st1}). Thus, 
writing $\theta_1=\theta,\theta_2 = \theta + b\epsilon, \theta_3 = \theta + a\epsilon$ and
taking the limit $\epsilon\rightarrow0$, we obtain the following generator for infinitesimal symmetries acting on $f$
(via $f \rightarrow f + \epsilon Q_f(\theta)$: 
\begin{equation}
Q_f(\theta) = \frac{a(s^{(1)}-s^{(2)}) +b  (s^{(3)}-s^{(1)}) } 
{(s^{(2)}-s^{(3)})s^{(1)}_x+(s^{(3)}-s^{(1)})s^{(2)}_x+(s^{(1)}-s^{(2)})s^{(3)}_x- (s^{(1)}-s^{(2)})(s^{(2)}-s^{(3)})(s^{(3)}-s^{(1)})}\ .
\label{Qf}\end{equation}
Here $s^{(1)},s^{(2)},s^{(3)}$ are distinct solutions of (\ref{sxx1})-(\ref{st1}).
The generator for the field $h$ can be written down but is long and complicated. The generator for $w$ (see Section 2)
takes a  simpler form: 
\begin{equation}
Q_w(\theta) =
\frac{as^{(3)}(s^{(2)}-s^{(1)})+bs^{(2)}(s^{(1)}-s^{(3)})}
     {(s^{(2)}-s^{(3)})s^{(1)}_x+(s^{(3)}-s^{(1)})s^{(2)}_x+(s^{(1)}-s^{(2)})s^{(3)}_x - (s^{(1)}-s^{(2)})(s^{(2)}-s^{(3)})(s^{(3)}-s^{(1)})
     }\ .
\label{Qw}\end{equation}
The  generators for the fields $u$ and $v$ are  $x$-derivatives of the generators for $f$ and $w$ respectively. In computing
these derivatives, it is useful to notice that the quantity in the denominator $Q_f(\theta)$ and $Q_w(\theta)$ is the quantity
$A$ introduced above in the discussion of conservation laws, see (\ref{Aref}), which satisfies  $A_x = -A(s^{(1)}+s^{(2)}+s^{(3)})$. 
Using the asymptotic expansions for $s^{(1)},s^{(2)},s^{(3)}$
obtained in the discussion of conservation laws we obtain the first few local 
symmetries: 
\begin{eqnarray*}
X_1&=&\frac{\partial}{\partial w} \ , \\
X_2&=&\frac{\partial}{\partial f} \ , \\
X_4&=&u\frac{\partial}{\partial f}+v\frac{\partial}{\partial w} \ , \\
X_5&=&(-2v-u_x)\frac{\partial}{\partial f}+\left(v_x+\frac{2}{3}u_{xx}-u^2\right)\frac{\partial}{\partial w} 
   \qquad      \left( = f_t\frac{\partial}{\partial f}+w_t\frac{\partial}{\partial w} \right)  \ , \\
X_7&=&3(6uu_x+12uv-u_{xxx}-2v_{xx})\frac{\partial}{\partial f}\\
  && +(12u^3-18u_{xx}u-18uv_x -9u_x^2+18v^2+2u_{xxxx}+3v_{xxx})\frac{\partial}{\partial w}\ , \\
X_8&=&(15u^3-15uu_{xx}+45u_xv+45v^2+u_{xxxx})\frac{\partial}{\partial f}  \\
&& +(18u^2u_x+45vu^2-6u_{xxx}u-12v_{xx}u-6u_{xx}u_x-12u_{xx}v-9u_xv_x-18vv_x)\frac{\partial}{\partial w}\ .
\end{eqnarray*}
Here we have taken, without loss of generality, $a=1,b=0$. The index $i$ on the vector field $X_i$ indicates that it
is obtained from the coefficient of $\theta^{-i/3}$ in the expansions of the generators. Note that the vector fields
$X_3,X_6,\ldots$ vanish, in analog of the situation for conservation laws. 

The local symmetries listed above are generated from the (\ref{Qf})-(\ref{Qw}) by expansion in powers
of $\theta$. Using the identity (\ref{Aid}) the full symmetry can be written (for the case $a=1,b=0$) in the form 
$$
X =  (s^{(1)}-s^{(2)})e^{\int s^{(1)}+s^{(2)}+s^{(3)}dx} \left(
\frac{\partial}{\partial f} -  s^{(3)} \frac{\partial}{\partial w} \right) \  . 
$$
This is a symmetry provided $s^{(1)},s^{(2)},s^{(3)}$ are solutions of (\ref{sxx1})-(\ref{st1}). 
Taking, for example, $s^{(3)}=s^{(1)}$ we obtain the {\em nonlocal} symmetry 
$$
X =  (s^{(1)}-s^{(2)})e^{\int 2s^{(1)}+s^{(2)} dx} \left(
\frac{\partial}{\partial f} -  s^{(1)} \frac{\partial}{\partial w} \right) \  . 
$$
There are 6 distinct versions of this symmetry arising from permutations of 
$s^{(1)},s^{(2)},s^{(3)}$. Nonlocal symmetries are useful as it is possible to construct invariant solutions
with respect to nonlocal, as well as local, symmetries \cite{Lou}. 

Returning to local symmetries, 
it is straightforward to verify, using just (\ref{sxx1}),  that the symmetry generators
$$ Q_f(\theta) =  (s^{(1)}-s^{(2)})e^{\int s^{(1)}+s^{(2)}+s^{(3)}dx} \ ,\qquad 
Q_w(\theta)= -s^{(3)}Q_f(\theta)  $$
satisfy the linear differential equations 
\begin{eqnarray*}
&& \left( \begin{array}{cc}
- \frac{2}{3}D^3 + Du + uD   &
-\frac{1}{3}D^4 + D^2u + 2Dv + vD   \\
 \frac{1}{3}D^4 - uD^2 + 2vD + Dv    & 
\frac{2}{9}D^5 - \frac{2}{3} (uD^3 +D^3u) + u^2D + Du^2 - (v_xD + Dv_x) 
\end{array} \right) 
\left( \begin{array}{c}
 Q_w \\
 Q_f 
\end{array} \right)  \\
&=&
\theta
\left( \begin{array}{cc}
 0 & -D  \\
 D  &  0 
\end{array} \right) 
\left( \begin{array}{c}
 Q_w \\
 Q_f 
\end{array} \right)\ .
\end{eqnarray*}
(Here $D$ denotes differentiation with respect to $x$.) Denoting the matrix differential operator on the LHS of this
equation as $P_2$, and the one on the RHS as $P_1$, we have
$$ P_1^{-1} P_2 
\left( \begin{array}{c}
 Q_w \\
 Q_f 
\end{array} \right)
= \theta
\left( \begin{array}{c}
 Q_w \\
 Q_f 
\end{array} \right)  $$
implying that the operator $P_1^{-1} P_2$ can be identified as the  recursion operator \cite{Olver}
for the potential BEq.
Since $Q_u(\theta)=Q_f(\theta)_x$ and $Q_v(\theta)=Q_w(\theta)_x$, 
the operator $P_2 P_1^{-1}$ can be identified as the recursion operator of the BEq. 
(Note that the recursion operator differs from the standard one for the BEq,
as given, for example, in \cite{olverbook}, as our form of the BEq (\ref{ueq})-(\ref{veq}) is
slightly different.)




\section{Conclusion}
The theme of this paper has been how the B\"acklund transformation, and particularly its superposition
principles, can give so much insight into the properties of the Boussinesq equation. Specifically, we
have obtained two systems of lattice equations associated with the superposition principles, we have
used the superposition principle to study the soliton solutions of the equation, which have a rich
structure that has not yet been full explored, and we have given a concise and complete account of the
theory of conservation laws and symmetries of the equation, using a generating function for symmetries
derived immediately from the superposition principle of 3 BTs.

The novelty in this work in the context of the theory of B\"acklund transformations, in comparison,
say, to our recent work on the BT for the Camassa-Holm equation \cite{rs2}, 
is in the need to look at the superposition principle for 3 BTs. For the BEq, the superposition principle of
2 BTs is not purely algebraic, whereas for 3 BTs it is. We expect this structure to be shared by the many
interesting equations associated with the Lie algebra $SL(3)$ (i.e. with $3\times 3$ matrix Lax pairs). 

A number of open questions have emerged in the course if this paper. In our work on lattice systems,
we arrived at the system of equations (\ref{latt})-(\ref{latt2}) on a cube, and it would be interesting
to have a characterization of the integrability of this system. In our work on soliton solutions, we have seen
that although we have a formula for the general multisoliton solution, we still lack much in the physical
understanding of these solutions. In particular, the question of whether there exists a nonsingular solution describing
the merger of 4 solitons to 2 is open, and there is much work to be done understanding the changes in 
speeds (and also phase shifts) between  initial and final solitons in the merger solutions. 

Another open direction is to understand the action of the BT and application of the superposition
principles to rational solutions \cite{beq41,cl1,cl2,cld,beq35}  and symmetry reductions of the BEq 
\cite{beq15,beq19,beq40,beq39}, and in particular to investigate the possible
reductions associated with the nonlocal symmetries given in Section 5. 


\bibliographystyle{acm}
\bibliography{P}
\end{document}